\documentclass[twocolumn]{pasj02}

\usepackage{graphicx}
\usepackage{multirow}
\usepackage{makecell}
\usepackage{url}
\usepackage{lineno}

\jyear{2026}

\newcommand{\okina}{\kern-0em \textquoteleft \kern-0em}
\newcommand{\citealt}[1]{\cite{#1}}

\begin{document}

\title{PEGASUS (Pfs Emission-line GAlaxy SUrvey with Subaru): II. Metal Poor AGNs in High Equivalent-Width Narrow-Band Emitters} 

\author{Hisakazu \textsc{Uchiyama}\altaffilmark{1,2}\altemailmark,
Tohru \textsc{Nagao}\altaffilmark{3},
Mariko \textsc{Kubo}\altaffilmark{4,5},
Tadayuki \textsc{Kodama}\altaffilmark{5},
Rhythm \textsc{Shimakawa}\altaffilmark{6},
Akio \textsc{Inoue}\altaffilmark{7,8},
Masayuki \textsc{Tanaka}\altaffilmark{2},
Jialai \textsc{Wang}\altaffilmark{9,10,11,12},
Ronaldo \textsc{Laishram}\altaffilmark{2},
Yuma \textsc{Sugahara}\altaffilmark{7,8,13},
Masato \textsc{Onodera}\altaffilmark{14},
Kazuki \textsc{Daikuhara}\altaffilmark{15},
Ko \textsc{Ishida}\altaffilmark{5},
Ryo Albert \textsc{Sutanto}\altaffilmark{5},
Haruka \textsc{Kusakabe}\altaffilmark{16},
Yoshiki \textsc{Matsuoka}\altaffilmark{3},
Alberto \textsc{Rodriguez-Ardila}\altaffilmark{17},
Yuta \textsc{Suzuki}\altaffilmark{18},
Yoshiki \textsc{Toba}\altaffilmark{19,3,20},
Xinfeng \textsc{Xu}\altaffilmark{21},
Yongquan \textsc{Xue}\altaffilmark{22},
Yongming \textsc{Liang}\altaffilmark{23}}

\email{hisakazu.uchiyama.86@hosei.ac.jp}

\altaffiltext{1}{Department of Advanced Sciences, Faculty of Science and Engineering, Hosei University, Koganei, Tokyo 184-8584, Japan}
\altaffiltext{2}{National Astronomical Observatory of Japan, Mitaka, Tokyo 181-8588, Japan}
\altaffiltext{3}{Research Center for Space and Cosmic Evolution, Ehime University, Bunkyo-cho 2-5, Matsuyama, Ehime 790-8577, Japan}
\altaffiltext{4}{Department of Physics and Astronomy, School of Science, Kwansei Gakuin University, 1 Gakuen Uegahara, Sanda, Hyogo 669-1330, Japan}
\altaffiltext{5}{Astronomical Institute, Tohoku University, 6-3, Aramaki, Aoba-ku, Sendai, Miyagi 980-8578, Japan}
\altaffiltext{6}{Waseda Institute for Advanced Study (WIAS), Waseda University, 1-21-1, Nishi-Waseda, Shinjuku, Tokyo 169-0051, Japan}
\altaffiltext{7}{Waseda Research Institute for Science and Engineering, Faculty of Science and Engineering, Waseda University, 3-4-1 Okubo, Shinjuku, Tokyo 169-8555, Japan}
\altaffiltext{8}{Department of Physics, School of Advanced Science and Engineering, Faculty of Science and Engineering, Waseda University, 3-4-1 Okubo, Shinjuku, Tokyo 169-8555, Japan}
\altaffiltext{9}{Department of Astronomy, University of Science and Technology of China, Hefei 230026, China}
\altaffiltext{10}{School of Astronomy and Space Science, University of Science and Technology of China, Hefei 230026, China}
\altaffiltext{11}{Kavli Institute for the Physics and Mathematics of the Universe (Kavli IPMU, WPI), The University of Tokyo Institutes for Advanced Study (UTIAS), The University of Tokyo, Kashiwa, Chiba 277-8583, Japan}
\altaffiltext{12}{Center for Data-Driven Discovery, Kavli IPMU (WPI), UTIAS, The University of Tokyo, Kashiwa, Chiba 277-8583, Japan}
\altaffiltext{13}{Department of Natural Sciences, Faculty of Science and Engineering, Tokyo City University, 1-28-1 Tamazutsumi, Setagaya, Tokyo 158-8557, Japan}
\altaffiltext{14}{Subaru Telescope, National Astronomical Observatory of Japan, National Institutes of Natural Sciences (NINS), 650 North A'ohoku Place, Hilo, HI 96720, USA}
\altaffiltext{15}{Institute of Space and Astronautical Science, Japan Aerospace Exploration Agency, 3-1-1, Yoshinodai, Chuou-ku, Sagamihara, Kanagawa 252-5210, Japan}
\altaffiltext{16}{Department of General Systems Studies, Graduate School of Arts and Sciences, The University of Tokyo, 3-8-1 Komaba, Meguro-ku, Tokyo, 153-8902, Japan}
\altaffiltext{17}{Observatório Nacional, Rua Gen. José Cristino 77, CEP 20921-400, São Cristovão, Rio de Janeiro, Brazil}
\altaffiltext{18}{Center for General Education, Shinshu University, 3-1-1 Asahi, Matsumoto, Nagano 390-8621, Japan}
\altaffiltext{19}{Department of Physical Sciences, Ritsumeikan University, 1-1-1 Noji-higashi, Kusatsu, Shiga 525-8577, Japan}
\altaffiltext{20}{Academia Sinica Institute of Astronomy and Astrophysics, 11F of Astronomy-Mathematics Building, AS/NTU, No.1, Section 4, Roosevelt Road, Taipei 10617, Taiwan}
\altaffiltext{21}{Department of Physics and Astronomy, Northwestern University, 2145 Sheridan Road, Evanston, IL, 60208, USA}
\altaffiltext{22}{Department of Astronomy, University of Science and Technology of China, Hefei, China}
\altaffiltext{23}{Institute for Cosmic Ray Research, The University of Tokyo, 5-1-5 Kashiwanoha, Kashiwa, Chiba 277-8582, Japan}

\KeyWords{galaxies: active --- galaxies: abundances --- techniques: spectroscopic --- surveys}

\maketitle

\begin{abstract}
We present Subaru {\okina}Ōnohi{\okina}ula Prime Focus Spectrograph (PFS) optical--NIR spectroscopic follow-up of high-equivalent-width (EW)
narrow-band (NB) emission-line galaxies selected from Hyper Suprime-Cam Subaru Strategic Program, 
to assess whether high-EW sources preferentially host low-metallicity AGNs. 
The PFS spectroscopy yielded secure spectroscopic identifications for 583 targets, including 581 sources with reliable PFS redshift solutions and two manually identified $z\sim4$ interlopers.
We construct BPT diagnostics for the 70 objects with significant detections in H$\alpha$, H$\beta$, [O\thinspace{\sc iii}], and [N\thinspace{\sc ii}]. 
Among them, $\sim34$ \% are located in the AGN region of the BPT diagram, suggesting that AGNs constitute a significant fraction of the sample.
A non-negligible fraction of the sample does not populate the classical AGN branch; instead, it lies in the intermediate ``valley'' between the star-forming and AGN sequences. 
Comparison to the \textsc{Cloudy} model suggests that the valley population is consistent with sub-solar metallicities.
Three He\thinspace{\sc ii}/[Ne\thinspace{\sc v}]-detected sources are also found in the star-forming region of the BPT diagram, and they extend to lower-metallicity solutions, consistent with $Z/Z_{\odot}\approx0.25-0.5$ ($12+\log({\rm O/H})\approx8.1-8.4$). 
One object shows a clear broad H$\alpha$ component; a multi-component fit yields $M_{\rm BH}\approx8.8\times10^{6}\,M_{\odot}$ and $\lambda_{\rm Edd}\approx0.035$, and this source is consistent with near-solar narrow-line region metallicity. 
We further find that the fraction of point-like sources in our parent target sample ($22\%\pm2\%$) is intermediate between those of star-forming galaxies ($12\%\pm5\%$) and AGN ($33\%\pm10\%$), consistent with a contribution from compact components. 
Stacked [O\thinspace{\sc iii}] line profiles do not show clear evidence for strong outflow signatures. 
Our findings suggest that high-EW NB selection efficiently identifies systems including candidates for metal-poor AGNs.
\end{abstract}


\section{Introduction}

Identifying early-stage supermassive black holes (SMBHs) and characterising their properties is crucial for understanding the growth history of SMBHs. 
In the local universe, tight scaling relations have been established between SMBH mass ($M_{\rm BH}$) and the properties of their host galaxies, particularly bulge components, suggesting a co-evolution between black hole growth and galaxy formation \citep{HaringRix2004, McConnellMa2013, KormendyHo2013, GrahamSahu2023}. 
However, it remains unclear when and through what physical processes such correlations were established over cosmic time. 
Identifying SMBHs in their early growth stages is therefore essential for probing the origin of these correlations, motivating systematic searches for such objects.

The origin of these correlations can be addressed by directly probing the relative growth of SMBHs and their host galaxies at high redshift, where the scaling relations may still be in the process of forming. Recent James Webb Space Telescope (JWST) observations have begun to constrain the stellar components of quasar host galaxies at $z \gtrsim 6$. 
\citet{Ding2023}, for example, reported that two $z>6$ quasars may be broadly consistent with the local $M_{\rm BH}$--host galaxy stellar mass relation. 
In contrast, other studies have suggested that SMBHs at $z \sim 5-7$ may be overmassive relative to their hosts \citep{Stone2024,Pacucci2024}, indicating that the picture of co-evolution in the early universe remains debated. However, high-redshift studies are often biased towards luminous AGN and are limited by sample size and measurement uncertainties, making it difficult to probe low-luminosity SMBHs in their early growth phases \citep{Lauer2007}. 
Consequently, a complementary approach is important to search systematically for SMBHs that are at early growth stages in the low-redshift universe.

Various methods have been employed to identify candidate early-stage SMBHs at low redshift (see \citealt{Greene2020} for review). 
These include the identification of low-mass black hole candidates through optical spectroscopy (particularly broad H$\alpha$ emission) \citep{Greene2004}, X-ray selection of low-luminosity AGN \citep{Miller2015}, mid-infrared searches for dust-obscured activity \citep{Satyapal2014}, radio observations of low-mass AGNs \citep{Nyland2017}, and variability-based techniques in optical bands \citep{Baldassare2018, Morokuma2016}. 
These efforts have revealed signatures of nuclear activity even in dwarf and low-mass galaxies \citep{Reines2013}, providing important samples for investigating black hole seed formation and early growth. 
Nevertheless, each method is subject to its own selection effects. 
Broad-line selection tends to favour unobscured, actively accreting type 1 AGN, potentially missing low-Eddington or dust-obscured systems. X-ray selection may overlook heavily absorbed AGN (particularly those with high column densities), while radio selection preferentially identifies jet-dominated sources. 
Mid-infrared diagnostics can be contaminated by dust emission from star formation, and variability-based methods are limited by observational cadence and time baselines. 
Identifying early-stage SMBHs and understanding their connection to host galaxies remain challenging, motivating the development of new observational approaches. 

AGNs residing in low-metallicity environments may provide an important clue in this context. 
Since galactic chemical enrichment proceeds over time, low metallicity may reflect relatively unevolved systems, at least from the perspective of chemical evolution. 
\citet{Matsuoka2011} showed that the broad-line region metallicity of Sloan Digital Sky Survey (SDSS) quasars correlates strongly with black hole mass, suggesting that AGN metallicity primarily reflects the chemical evolution of their host galaxies (see also \citealt{Xu2018}). 
\citet{Groves2006} used photoionisation models and SDSS Seyfert 2 galaxies to show that sub-solar-metallicity narrow-line AGN are rare, identifying only $\sim40$ candidates among $\sim 2.3\times10^4$ Seyfert 2 galaxies. 
\citet{Kawasaki2017} selected sources in the transition region (``valley'') of the BPT diagram from $\sim2.1\times10^5$ SDSS emission-line galaxies and investigated low-metallicity narrow-line region (NLR) AGN candidates. 
They found that 43 of these sources show evidence for AGN activity and are unlikely to be explained by star-forming galaxies with an unusually high gas density or ionisation parameter, suggesting that the transition region of the BPT diagram provides an efficient way to identify low-metallicity AGN candidates. 

Broad H$\alpha$ emission has been reported in extremely metal-poor dwarf galaxies with a gas metallicity of $12+\log(\mathrm{O/H}) \sim 7.4-8.0$ \citep{Izotov2008}, suggesting that black hole growth can occur in such primitive environments. 
For comparison, typical AGN host galaxies often exhibit near-solar or super-solar gas metallicities ($12+\log(\mathrm{O/H}) \gtrsim 8.7$; see \citealt{Groves2006}), highlighting the unusual nature of AGN activity in extremely metal-poor environments. 
These extremely metal-poor systems are also characterised by strong emission lines, including large equivalent widths (EWs) in [O\thinspace{\sc iii}] and H$\alpha$. 
Large emission-line EWs are commonly observed in metal-poor galaxies with young stellar populations, where harder ionising radiation fields and weaker stellar continua enhance line emission relative to the underlying continuum. 
If obscured AGN are embedded in such systems, their narrow-line emission may provide an additional contribution to the emission-line flux while contributing little to the optical continuum, potentially producing even more high EWs. 
This suggests that galaxies with unusually large emission-line EWs may provide a promising avenue for identifying low-metallicity AGN candidates. 
Such objects appear to be extremely rare in the local universe (e.g., four candidates among $\sim 7\times10^5$ SDSS galaxies, corresponding to $\sim 0.0006\%$; \citealt{Izotov2008}), and systematic spectroscopic surveys targeting this population have so far been scarce. 

One practical way to target high-EW systems efficiently is narrow-band (NB) selection, where emission-line galaxies are identified via a significant NB flux excess relative to an adjacent broadband (BB). 
This technique preferentially selects sources with large observed-frame equivalent widths and can be applied homogeneously over wide fields, enabling efficient searches for rare populations across cosmologically representative volumes \citep{Hayashi2018,Hayashi2020}. 
The Hyper Suprime-Cam Subaru Strategic Program (HSC-SSP) provides an ideal dataset, combining a wide field-of-view with deep imaging and a suite of NB filters \citep{Aihara2018}.
In the COSMOS UltraDeep field, deep NB imaging from HSC-SSP (e.g., NB816 and NB921, and NB1010) is complemented by the CHORUS programme, which adds additional deep NB coverage such as NB718 and NB973 \citep{Hayashi2018,Hayashi2020,Inoue2020}. 
Because NB selection preferentially identifies strong emission-line galaxies within a well-defined redshift window, follow-up spectroscopy can efficiently provide high signal-to-noise measurements of key nebular lines. 
This enables constraints on nebular conditions, including gas-phase metallicity through photoionisation modelling.

In this paper, we spectroscopically follow up these high-EW NB selected targets to statistically assess what physical properties high-EW sources can exhibit, and in particular whether they preferentially include low-metallicity AGN candidates. 
The spectroscopy is carried out with {\okina}Ōnohi{\okina}ula Prime Focus Spectrograph (PFS), a wide-field, massively multiplexed, fibre-fed optical--near-infrared spectrograph on the 8.2-m Subaru Telescope. 
PFS can obtain spectra for $\sim$2400 objects simultaneously over a $\sim$1.3-degree diameter field of view, with broad wavelength coverage from $\sim$380 to 1260 nm in a single exposure \citep{Takada2014,Tamura2016}.  
This combination is particularly well suited to our goal: the wide field enables efficient follow-up of rare NB-selected sources over large areas, while the wavelength coverage captures the key rest-frame optical nebular lines used to separate AGN and star-forming galaxies via standard diagnostic diagrams (e.g., BPT), such as H$\beta$, [O\thinspace{\sc iii}]$\lambda5007$, H$\alpha$, and [N\thinspace{\sc ii}]$\lambda6583$ (and, when available, higher-ionisation lines such as He\thinspace{\sc ii}$\lambda4686$ and [Ne\thinspace{\sc v}]$\lambda3426$). 
PFS therefore provides an optimal, uniform spectroscopic data set for simultaneously constraining ionisation mechanism and chemical properties of high-EW emitters.
These observations are conducted as part of \textit{PEGASUS} (Pfs Emission-line GAlaxy SUrvey with Subaru; Kubo et al., submitted), a programme designed to build large spectroscopic samples of emission-line galaxies selected from wide-field NB imaging. 
Here we present the initial results of PEGASUS, focusing on a pilot search for AGN and their physical properties in high-EW sources in the COSMOS region.

The paper is organised as follows. 
In section~2, we describe the data and target selection. 
In section~3, we present the PFS spectroscopic observations. 
In section~4, we describe the spectroscopic measurements. 
In section~5, we present the results.
In section~6, we discuss the implications of our findings.
Throughout this work, we adopt a standard $\Lambda$CDM cosmology with $H_0=70~{\rm km\,s^{-1}\,Mpc^{-1}}$, $\Omega_{\rm m}=0.3$, and $\Omega_\Lambda=0.7$, and use AB magnitudes for photometry.

\section{Photometric Data and Target Selection}

\subsection{Photometric Data}

\subsubsection{HSC-SSP}

The HSC-SSP is a deep, wide-field imaging survey conducted with the Hyper Suprime-Cam on the Subaru Telescope, which provides a $1.5^{\circ}$-diameter field of view \citep{Aihara2018}. 
HSC-SSP consists of three layers (Wide, Deep, and UltraDeep) observed in five BBs ($grizy$) and several NB filters.
The survey design and the filter information are given in \citet{Aihara2018} and \citet{Kawanomoto2018}, respectively. 
The HSC camera system and CCD dewar are described in \citet{Komiyama2018}. 
An on-site quality assurance system was used during the HSC observations to monitor data quality in real time \citep{Furusawa2018}.

In this study, we use imaging data in the COSMOS field from the HSC-SSP UltraDeep layer, primarily based on Public Data Release~2 (PDR2; \citealt{Aihara2019}), which corresponds to the internal processing version commonly referred to as S18A. 
For the NB1010 filter, which is not available in PDR2, we additionally use imaging from HSC-SSP Public Data Release~3 (PDR3; \citealt{Aihara2022}), where NB1010 observations in the COSMOS field become available.
The UltraDeep layer is designed to reach extremely faint flux limits by focusing on HSC 2 pointings, enabling the efficient selection of rare populations such as extreme equivalent-width emission-line galaxies.
In COSMOS, the UltraDeep data provide deep $grizy$ imaging with typical 5$\sigma$ limiting magnitudes of
$g\simeq28.1$, $r\simeq27.7$, $i\simeq27.4$, $z\simeq26.8$, and $y\simeq26.3$ mag ($2$ arcsec-diameter apertures), with point-source depths typically $\sim$0.3 mag deeper \citep{Aihara2018}. 
For NB718, NB816, and NB921, the typical 5$\sigma$ limiting magnitudes in the COSMOS field are 25.61, 25.58, and 25.39 mag, respectively \citep{Hayashi2020}.
The data were reduced using the HSC pipeline \texttt{hscPipe} \citep{Bosch2018}, and the astrometric and photometric calibrations are tied to the Pan-STARRS1 system \citep{Schlafly2012,Tonry2012,Magnier2013}.

\begin{table*}
\caption{Photometric quality cuts applied to the HSC-SSP/CHORUS catalogs.}
\label{tab:qualitycuts}
\centering
\begin{tabular}{p{2cm} p{4.6cm} p{2.5cm} p{6.5cm}}
\hline\hline
Category & Flag / criterion & Applied to $^{1}$ & Description \\
\hline
Primary object selection
& \texttt{isprimary = True}
& all objects
& Select primary objects without deblending children in the inner tract/patch region \\
\hline
Input exposure requirement
& \texttt{inputcount\_flag\_noinputs = False}
& all $grizy$ + selection NB
& Require valid input images in each band used for the selection \\
& \texttt{inputcount\_value > 2}
& all $grizy$ + selection NB
& Require at least three contributing visits in each band used for the selection \\
\hline
Pixel-level quality cuts
& \texttt{pixelflags\_bad = False}
& selection NB + associated BB
& Exclude bad pixels at the source position \\
& \texttt{pixelflags\_edge = False}
& selection NB + associated BB
& Exclude edge-affected sources \\
& \texttt{pixelflags\_saturatedcenter = False}
& selection NB + associated BB
& Exclude sources affected by central saturation \\
\hline
Measurement quality cuts
& \texttt{merge\_peak = True}
& selection NB
& Select objects identified as peaks in the deblending process \\
& \texttt{sdssshape\_flag\_badcentroid = False}
& selection NB
& Require successful centroiding in the NB image \\
& \texttt{cmodel\_flag = False}
& selection NB
& Require valid cmodel photometry in the NB image \\
& \texttt{psfflux\_flag = False}
& selection NB
& Require valid PSF photometry in the NB image \\
& \texttt{psfflux\_flag\_badcentroid = False}
& selection NB
& Require successful centroiding for PSF photometry \\
\hline
NB detection
& $\mathrm{S/N} \geq 5$
& selection NB
& Require significant detection in the NB used for the selection for PSF photometry  \\
\hline
Bright object mask
& \texttt{mask\_pdr2\_bright\_objectcenter = False}
& all $grizy$ for PDR2
& Exclude sources affected by bright objects in the $grizy$ bands \\
& \texttt{mask\_brightstar\_halo = False}
& all $grizy$ for PDR3
& Exclude sources affected by bright-star halos \\
& \texttt{mask\_brightstar\_ghost = False}
& all $grizy$ for PDR3
& Exclude sources affected by optical ghosts \\
& \texttt{mask\_brightstar\_blooming = False}
& all $grizy$ for PDR3
& Exclude sources affected by blooming artefacts \\
\hline
\end{tabular}
\vspace{1mm}
\begin{minipage}{0.98\linewidth}
\footnotesize
$^{1}$ Associated BB denotes the broad-band used for the colour selection with each NB filter: 
$i$ for NB718 and NB816, $z$ for NB921 and NB973, and $y$ for NB1010.\\
\end{minipage}

\end{table*}

\subsubsection{CHORUS}

In the COSMOS UltraDeep field, the available HSC-SSP NB imaging is complemented by the CHORUS programme that provides additional deep HSC observations in several NB and intermediate-band filters \citep{Inoue2020}. 
CHORUS extends the HSC-SSP UltraDeep imaging by adding NB observations that are not included in the standard HSC-SSP releases, most notably NB718 and NB973, as well as IB945.

In this work, we specifically use the CHORUS Public Data Release 1 (PDR1) catalogs for the NB718 and NB973 samples. 
The CHORUS imaging in COSMOS covers $\sim$1.7~deg$^2$, with an effective area of $\sim$1.4~deg$^2$ after applying masking and quality cuts. 
These data reach typical 5$\sigma$ limiting magnitudes of $\sim$25.6 and $\sim$24.6 mag for NB718 and NB973, respectively \citep{Hayashi2020}.

\begin{table*}
\caption{Summary of spectroscopic outcomes for the high-EW AGN candidate sample.} 
\centering
\label{table1}
\begin{tabular}{llccccccc}
\hline
Filter & Emitter &
\makecell{Targeted$^{1}$} &
\makecell{Observed$^{2}$} &
\makecell{Confirmed$^{3}$} &
\makecell{Redshift\\ mismatch$^{4}$} &
\makecell{Unconfirmed$^{5}$} &
\makecell{Purity$^{6}$} \\
\hline
\multirow{2}{*}{NB1010}
 & HAE@$z_{\rm exp}=0.54$     & 111 & 75  & 68  & 5  & 7(7)  & 0.926 \\
 & O3E@$z_{\rm exp}=1.02$   & 216 & 154 & 125 & 11 & 29(19) & 0.912\\
\hline
\multirow{2}{*}{NB973}
 & HAE@$z_{\rm exp}=0.48$     & 24  & 18  & 18  & 0  & 0(0) & 1.000  \\
 & O3E@$z_{\rm exp}=0.94$   & 42  & 32  & 29  & 20 & 3(1)  & 0.310 \\
\hline
\multirow{2}{*}{NB921}
 & HAE@$z_{\rm exp}=0.40$     & 32  & 26  & 26  & 4  & 0(0) & 0.846  \\
 & O3E@$z_{\rm exp}=0.84$   & 266 & 217 & 217 & 17 & 0(0)  & 0.922 \\
\hline
NB816
 & O3E@$z_{\rm exp}=0.63$   & 86  & 57  & 57  & 8  & 0(0)  & 0.860 \\
\hline
NB718
 & O3E@$z_{\rm exp}=0.43$   & 50  & 45  & 43  & 4  & 2(2)  & 0.907 \\
\hline
\hline
\multicolumn{2}{l}{Total}   & 827 & 624 & 583 & 69 & 41(29) &  0.882 \\
\hline
\end{tabular}
\vspace{1mm}
\begin{minipage}{0.98\linewidth}
\footnotesize
$^{1}$ Number of intended targets.\\
$^{2}$ Number of targets observed.\\
$^{3}$ Number of targets with either a reliable spectroscopic redshift solution ($Q\ge2$) or a manually identified $z\sim4$ interloper.\\
$^{4}$ Number of spectroscopically confirmed objects whose $z_{\rm spec}$ lies outside the expected narrowband redshift range defined by the filter bandpass. This is a subset of the confirmed sample. The two manually identified $z\sim4$ interlopers are included in this category.\\
$^{5}$ Number of targets for which no reliable spectroscopic redshift solution could be determined. The number in parentheses indicates cases rejected based on manual inspection, including spectra severely affected by sky-subtraction residuals. The remaining unconfirmed targets are cases with no consistent redshift solution from the fitting procedure.\\
$^{6}$ Purity of the NB selection, defined as the fraction of spectroscopically confirmed sources that are consistent with the expected NB redshift range: $(N_{\rm confirmed}-N_{\rm mismatch})/N_{\rm confirmed}$. 
\end{minipage}
\end{table*}

\subsection{NB-AGN candidate selection in COSMOS}

Our NB selection strategy closely follows that of \citet{Hayashi2020}, which was developed for NB-selected emission-line samples based on HSC-SSP PDR2 and CHORUS PDR1 imaging. 
We adopt essentially the same photometric measurements and quality/masking criteria, and extend the selection to the NB1010 filter using HSC-SSP PDR3. 
The selection is based on HSC and CHORUS forced photometry, using aperture-matched BB and NB flux measurements to identify sources with significant NB excesses.
To ensure consistency with the HSC-SSP-based samples, we cross-match the CHORUS catalogs with the HSC-SSP catalogs, and apply identical photometric quality and masking criteria. 

To ensure reliable photometry, we apply a set of quality cuts that are equivalent to those adopted in \citet{Hayashi2020}. 
The HSC-SSP survey area is divided into tracts, each of which is further subdivided into smaller patches for data processing. 
We first select primary objects located in the unique inner tract/patch regions and exclude sources with deblended children.
We exclude sources flagged as saturated, affected by bad pixels, or located near image edges, and require a sufficient number of input exposures. 
Concretely, for each object we require the following conditions in the bands used for the selection, i.e., the relevant NB filter and its associated BB(s): (i) no missing inputs and at least three contributing visits,  (ii) no pixel-level issues at the source position, and (iii) successful centroiding and photometric measurements. 
We do not impose any morphology-based selection (e.g., point-source criteria). 
We require a significant NB detection (S/N$\ge 5$), evaluated using PSF photometry in the NB band, following \citet{Hayashi2020}, to ensure a consistent definition of detection significance. 
In addition, we exclude sources affected by bright stars using the standard HSC bright-star mask flags (halo, ghost, and blooming) in the $grizy$ bands. 
The detailed flag definitions and masking criteria are summarised in Table~\ref{tab:qualitycuts}. 

Applying these criteria, we select NB emitters using five NB filters (NB718, NB816, NB921, NB973, and NB1010) and construct eight NB emitter populations summarised in Table~\ref{table1}.
The corresponding redshift ranges are determined from the emission lines falling within each NB filter, as described below: 
H$\alpha$ emitters (HAEs) in the redshift ranges
$0.392<z<0.410$,
$0.475<z<0.493$, and
$0.530<z<0.548$,
and [O\thinspace{\sc iii}] emitters (O3Es) in the redshift ranges
$0.427<z<0.441$,
$0.619<z<0.637$,
$0.829<z<0.851$,
$0.935<z<0.953$, and
$1.009<z<1.027$.
The selection is based on the colour excess between the NB and BB filters. 
Following \citet{Hayashi2020}, for the colour-excess measurement, we use the \texttt{undeblended\_convolvedflux\_2\_11} photometry measured on PSF-matched images, corresponding to a 1\farcs1 aperture. 
The other photometric measurements are based on \texttt{convolvedflux\_2\_20\_mag}, i.e., the 2\arcsec-aperture photometry measured on PSF-matched images. 
The HSC \texttt{convolvedflux} measurements are corrected for aperture losses using the corresponding aperture corrections. 
The continuum flux density at the NB wavelength is estimated as
$f_{\nu,c}(\mathrm{NB}) = w_1 f_\nu(\mathrm{BB}_1) + w_2 f_\nu(\mathrm{BB}_2)$ with $w_1+w_2=1$. 
Here, $\mathrm{BB}_1$ and $\mathrm{BB}_2$ denote the two BB filters adjacent to each NB filter. 
The weights $w_1$ and $w_2$ were determined from the effective wavelengths of the two adjacent BB filters and the NB filter, following the linear interpolation scheme of \citet{Hayashi2020} and are summarised in Table~\ref{tab:nb_selection}. 
For NB1010, however, we approximate the continuum using the $y$-band flux density alone (i.e. $(w_1,w_2)=(0,1)$) to avoid extrapolation beyond the $z$--$y$ wavelength range.
Following \citet{Hayashi2020}, we apply colour cuts to remove bright continuum sources with intrinsically red spectra, and impose a magnitude cut of $>18.5$ to eliminate saturated bright objects.

\begin{table}
\centering
\caption{Selection criteria for NB-selected emission-line candidates.}
\begin{tabular}{lccccc}
\hline
NB & BBs & Weights $(w_1,w_2)$ &  colour cut  \\
\hline
NB1010 & $z, y$ & $(0.000,\,1.000)$  & $>0.20$ \\
NB973  & $z, y$ & $(0.052,\,0.948)$  & $>0.20$ \\
NB921  & $z, y$ & $(0.643,\,0.357)$  & $>0.20$ \\
NB816  & $i, z$ & $(0.631,\,0.369)$  & $>0.25$ \\
NB718  & $r, i$ & $(0.079,\,0.921)$  & $>0.25$ \\
\hline
\end{tabular}
\label{tab:nb_selection}
\end{table}

We further restrict the sample to emission-line candidates in redshift slices corresponding to each NB filter, ensuring that the observed NB excess is associated with a specific target emission line within the filter bandpass. 
First, objects with secure spectroscopic redshifts ($z^{\rm HSC}_{\rm spec}$) consistent with the expected NBE redshift windows are included. 
Existing spectroscopy mostly lacks the NIR coverage required for the BPT analysis used in this study. 
We adopt $z^{\rm HSC}_{\rm spec}$ from the spectroscopic-redshift compilation provided in the HSC-SSP database,\footnote{\url{https://hsc-release.mtk.nao.ac.jp/doc/index.php/specz-2/}} which combines various public spectroscopic surveys (e.g., SDSS, GAMA, zCOSMOS, and DEEP2) and homogenises the quality flags for convenient selection of secure measurements \citep{Aihara2022}. 
Sources with secure spectroscopic redshifts that are inconsistent with the expected redshift of each NB selection are regarded as contaminants and excluded from the sample. 
For sources without spectroscopic redshifts, we restrict our sample to those for which all three photometric redshift estimates---\textsc{Mizuki}, \textsc{DEmP}, and the COSMOS2015 $Z_{\rm PDF}$---are available \citep{Tanaka2015,Tanaka2018,HsiehYee2014,Laigle2016}. 
Following the selection strategy of \citet{Hayashi2020}, we require that all three estimates satisfy a fractional redshift uncertainty of $\sigma_z/(1+z) < 0.2$ and are consistent with the expected redshift slice of the corresponding NB selection. 
Sources that do not satisfy this redshift-consistency criterion but still have reliable photometry are instead subjected to the two-colour cuts defined in \citet{Hayashi2020} to select additional candidate emitters for each NB selection. 
For the NB1010 selection, we apply colour-colour criteria to identify candidate emitters, as described in Appendix~\ref{sec:nb1010_colour}.

\begin{figure}
\begin{center}
\includegraphics[width=1.0\linewidth]{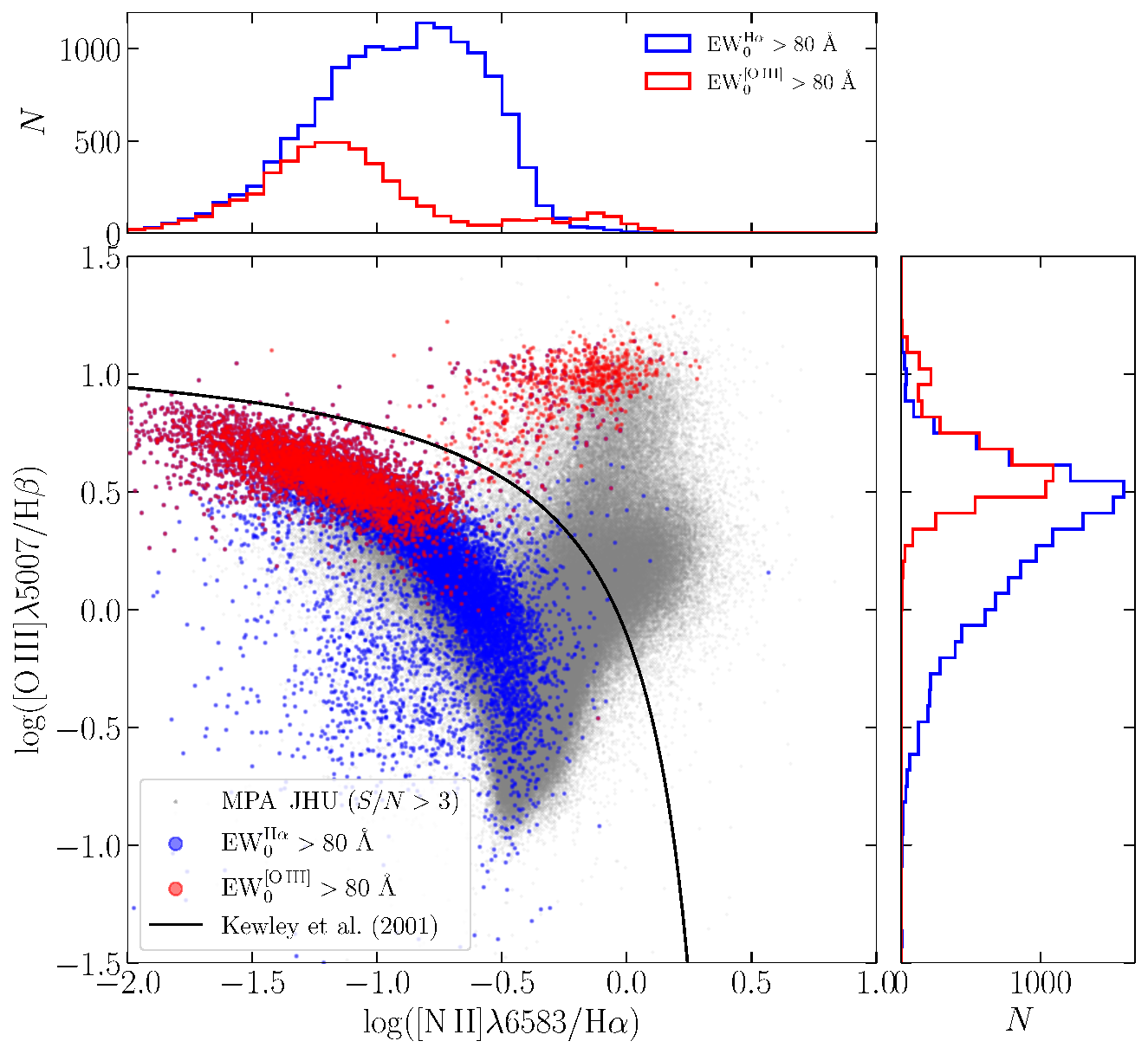}
\end{center}
  \caption{
BPT diagram for local emission-line galaxies from the MPA--JHU SDSS catalog.
The grey dots show the distribution of all emission-line galaxies with $SNR>3$
in the [O\thinspace{\sc iii}]$\lambda5007/{\rm H\beta}$ versus
[N\thinspace{\sc ii}]$\lambda6583/{\rm H\alpha}$ plane.
The solid curve indicates the demarcation line of \citet{Kewley2001}.
Galaxies with ${\rm EW_{0}^{\rm [O\,III]}} > 80$ \AA\ are highlighted
in red, while those with ${\rm EW_{0}^{\rm H\alpha}} > 80$ \AA\ are shown in blue.
The top and right panels show the corresponding number distributions
along the [N\thinspace{\sc ii}]$\lambda6583/{\rm H\alpha}$ and
[O\thinspace{\sc iii}]$\lambda5007/{\rm H\beta}$ axes, respectively,
for the two EW-selected samples.

Alt text: BPT diagram of local SDSS emission-line galaxies. High-EW [O III] emitters are concentrated towards higher [O III]/H$\beta$ and lower [N II]/H$\alpha$ than high-EW H$\alpha$ emitters, and extend into the valley and AGN regions above the Kewley et al. demarcation. }
\label{fig:bpt_mpa_jhu}
\end{figure}

To isolate systems with strong ionising radiation fields, we select emitters with ${\rm EW_{0}^{\rm [O\,III]}} > 80$ \AA\ or ${\rm EW_{0}^{\rm H\alpha}} > 80$ \AA\  in the rest frame. 
To assess the nature of such systems, we construct a local reference sample of emission-line galaxies from the MPA--JHU SDSS catalog  \citep{Brinchmann2004} and examine their position on the BPT diagram. 
Figure~\ref{fig:bpt_mpa_jhu} shows the distribution of the MPA--JHU SDSS galaxies in the [O\thinspace{\sc iii}]$\lambda5007/{\rm H\beta}$ versus [N\thinspace{\sc ii}]$\lambda6583/{\rm H\alpha}$ plane. 
High-EW [O\thinspace{\sc iii}] objects tend to lie above high-EW H$\alpha$ ones on the BPT diagram, while both populations often extend towards the low-[N\thinspace{\sc ii}]/H$\alpha$ side. 
This reflects ionisation conditions that favour O$^{++}$ over O$^{+}$ and N$^{+}$, thereby enhancing [O\thinspace{\sc iii}]/H$\beta$ and suppressing [N\thinspace{\sc ii}]/H$\alpha$. 
In contrast, strong H$\alpha$, as a recombination line, does not by itself imply a similarly high ionisation state. 
We find that $13.9$ \% (1.4 \%) of these high-EW [O\thinspace{\sc iii}] (H$\alpha$) emitters lie in the AGN region above the \citet{Kewley2001} demarcation curve, yielding a combined fraction of 5.1 \%. 
Interestingly, the high-EW [O\thinspace{\sc iii}] or H$\alpha$ sources satisfying the AGN criteria do not simply follow the main locus of optically selected AGNs. 
Instead, they are preferentially offset towards higher [O\thinspace{\sc iii}]$/H\beta$ and lower [N\thinspace{\sc ii}]$/H\alpha$, populating the ``valley'' between the star-forming sequence and the bulk of the AGN branch.  
Here, we define the "valley" as the relatively underpopulated region between the star-forming sequence and the main AGN branch in the BPT diagram. 
This region of the BPT diagram is known to be associated with low-metallicity, high-ionisation AGNs \citep{Groves2006, Kewley2013, Kawasaki2017}. 
The location of these high-EW sources suggests that this selection may efficiently identify systems occupying the BPT valley, where low-metallicity AGNs have previously been found.
These properties suggest that high-EW NB-selected emitters may provide a useful sample for spectroscopic searches for metal-poor AGN candidates. 
We hereafter refer to these sources as ``high-EW AGN candidates''. 

We further impose a magnitude cut of ${\rm NB} < 23.3$ to ensure sufficient signal-to-noise ratios for follow-up spectroscopy. 
This selection yields a statistically powerful sample for studying AGN demographics at intermediate redshifts. 
The final target list consists of 827 high-EW AGN candidates, as summarised in table~\ref{table1}. 
Among them, 60 have $z^{\rm HSC}_{\rm spec}$. 
We note that NB921 emitters identified as [O\thinspace{\sc iii}] emitters at $z\simeq0.84$ that have already been observed as part of the PFS-SSP survey are excluded from the present sample. 
These objects are removed to avoid duplication with the ongoing PFS-SSP emission-line galaxy programme and to ensure that our analysis focuses on a homogeneous set of targets newly observed within the PEGASUS framework.

\section{PFS spectroscopy}

The PFS observations presented here are part of the broader PEGASUS programme.
A comprehensive description of the survey design, target prioritisation, observing strategy, and data reduction will be presented in Kubo et al., submitted.
We summarise here the observational information essential for the present analysis. 

\subsection{Instrumental setup and observations}

The PFS provides wavelength coverage of $\lambda \simeq 380$--1260 nm using the blue, red, and near-infrared spectrograph arms \citep{Sugai2015, Tamura2016, Tamura2022, Tamura2024}. The resolving power in the low-resolution mode is $R\sim2500$ at 500 nm (blue), $R\sim3000$ at 800 nm (red), and $R\sim4500$ at 1100 nm (NIR), corresponding to typical spectral resolutions of $\sim2.1$, 2.7, and 2.4 \AA, respectively.

The initial PEGASUS survey area covers $\sim 1.5~{\rm deg}^2$, corresponding to the COSMOS--UDF region of the HSC-SSP (${\rm RA} = 150.1192^\circ$, ${\rm Dec} = +2.2058^\circ$), observed with the Subaru Telescope in the S25A semester (S25A-058QN, PI: M. Kubo). 
The observations were conducted in queue mode. 
Individual exposures were obtained with a unit exposure time of 450 s.
A total of 8,718 NBE targets were prepared for the fibre assignment, yielding a source density of $\sim 5,800~{\rm deg^{-2}}$. 
The targets consist of H$\alpha$, [O\thinspace{\sc iii}], and [O\thinspace{\sc ii}] emitters selected from the HSC-SSP and CHORUS narrowband data over $0.4 \lesssim z \lesssim 1.6$ \citep{Hayashi2020}, with an emphasis on sources with line fluxes above $2\times 10^{-17}~{\rm erg~s^{-1}~cm^{-2}}$. 
Among these, 827 targets were selected for the present project (table~\ref{table1}), of which 624 were successfully observed with PFS.
We confirmed that all 827 targets in the final target sample (table \ref{table1}) satisfy the adopted line-flux criterion, and therefore no additional objects are removed by this threshold after the NB and EW selections. 
The flux-limited NB emitter sample includes high-EW objects (rest-frame EW$_{0} > 80$ \AA) as candidate rare low-metallicity AGNs (table \ref{table1}). 

The target allocation followed a priority scheme designed to maximise the scientific return for both environmental and AGN-oriented goals.
The highest priorities were assigned to NB emitters associated with known clusters in COSMOS and to high-EW AGN candidates, followed by double-NB detections and other NB emitter classes ordered by redshift and expected line detectability. 
The observations were executed in the PFS low-resolution mode.

Within the PEGASUS project, the NB1010 targets are primarily associated with our AGN subproject and therefore typically received long integrations ($>7200$ s) to enable BPT-quality line diagnostics. 
In contrast, targets in the other NB samples were often shared among multiple subprojects with different science goals, including programmes requiring shorter exposures (e.g., 1800 s). As a result, the achieved integration times differ significantly among the NB samples.

\begin{figure}
\begin{center}
\includegraphics[width=1\linewidth]{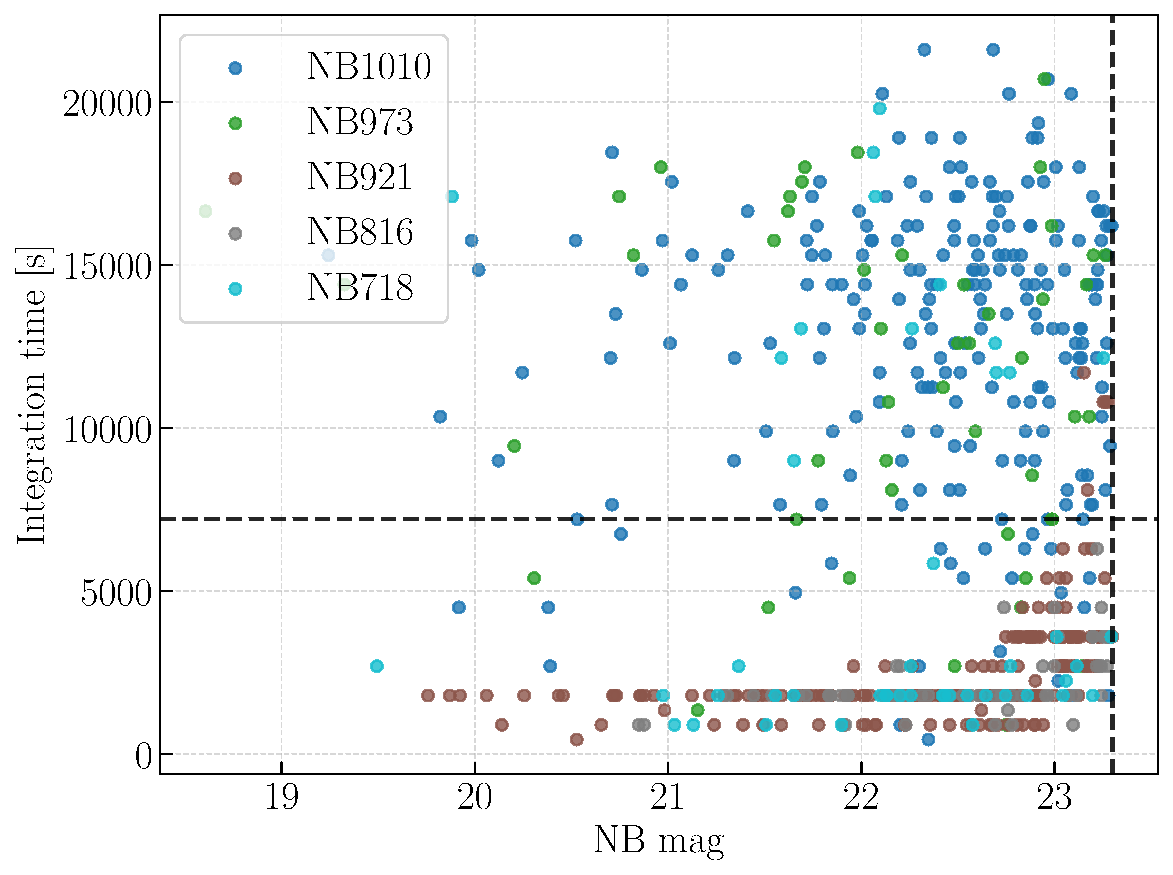}
\end{center}
\caption{
Integration time as a function of NB magnitude for our targets.
The horizontal axis shows the NB magnitude corrected by the attenuation factor, 
while the vertical axis represents the total integration time. 
The horizontal dashed line indicates the required integration time of 7200~s,
and the vertical dashed line marks the NB magnitude cut at ${\rm NB}=23.3$.
Points are colour-coded by NB filter. 
Sources above the horizontal dashed line satisfy the required exposure time. 

Alt text: Scatter plot of PFS integration time versus NB magnitude for targets selected with five NB filters. NB1010 and NB973 targets generally have longer integrations, frequently exceeding 7200 s, whereas NB718, NB816, and NB921 targets are dominated by shorter integrations near 1800 s.
} \label{fig:nbmag_vs_integration_time}
\end{figure}

The exposure times for individual targets were determined primarily from the NB line-flux estimates. For the high-EW, metal-poor AGN candidates, we adopted a representative template scaled to ${\rm NB}\simeq 23.3$ mag and required 7200 sec of on-source exposure to enable BPT-quality line diagnostics.
 As shown in figure~\ref{fig:nbmag_vs_integration_time}, the NB1010 and NB973 samples generally received longer integrations than the other NB samples as a consequence of the observing strategy described above.  
 The median integration times are 13725 s and 12600 s for NB1010 and NB973, respectively. In contrast, the NB718, NB816, and NB921 samples are dominated by short integrations, all with median integration times of 1800 s. 
The NB921 sample, which constitutes the largest number of observed targets, also shows predominantly short exposures with a low fraction of sources meeting the required integration time (2\%). These trends are summarised in table~\ref{tab:nb_integration_time}. The fractions of sources satisfying the required integration time are high for NB1010 and NB973 (0.88 and 0.80, respectively), but much lower for NB718, NB816, and NB921 (0.29, 0.00, and 0.02, respectively).

\begin{table}
\caption{Integration-time statistics for each NB}
\centering
\begin{tabular}{lcccc}
\hline
NB & $N$  & Median [s] & $f_{\rm req}^{a}$ \\
\hline
NB1010 & 229  & 13725 & 0.88 \\
NB973  & 50   & 12600 & 0.80 \\
NB921  & 243  & 1800  & 0.02 \\
NB816  & 57   & 1800  & 0.00 \\
NB718  & 45   & 1800  & 0.29 \\
\hline
\end{tabular}
\vspace{2pt}
\begin{flushleft}
\footnotesize
$^{a}$ Fraction of sources meeting the required integration time ($t_{\rm int} \geq 7200\,\mathrm{s}$).
\end{flushleft}
\label{tab:nb_integration_time}
\end{table}

\subsection{Data reduction}
The PFS data used in this study are based on the ``S25A\_April2026'' data release and were processed with the standard PFS data reduction pipelines (DRPs)\footnote{https://subaru-pfs.github.io/pfs\_helpdesk\_tutorial}, which adopt a two-stage framework: the 2D-DRP that converts raw detector images into flux- and wavelength-calibrated one-dimensional spectra, and the 1D-DRP that measures spectroscopic quantities from the extracted spectra \citep{Shimono2016}. 

In the 2D-DRP, detector-level calibrations and instrumental corrections were automatically applied by the pipeline, including bias subtraction, flat-fielding, identification and tracing of fibre profiles, and extraction of one-dimensional spectra for each exposure.
Wavelength calibration was performed using calibration-lamp exposures and, when appropriate, sky features within the same framework.
Sky subtraction was carried out using the dedicated sky fibres and/or local sky modelling implemented in the DRP, and relative and absolute flux calibration was performed using standard stars observed in the same configurations.
Multiple exposures were then coadded to produce the final one-dimensional spectra for each target. 

\begin{figure*}
\begin{center}
\includegraphics[width=1\linewidth]{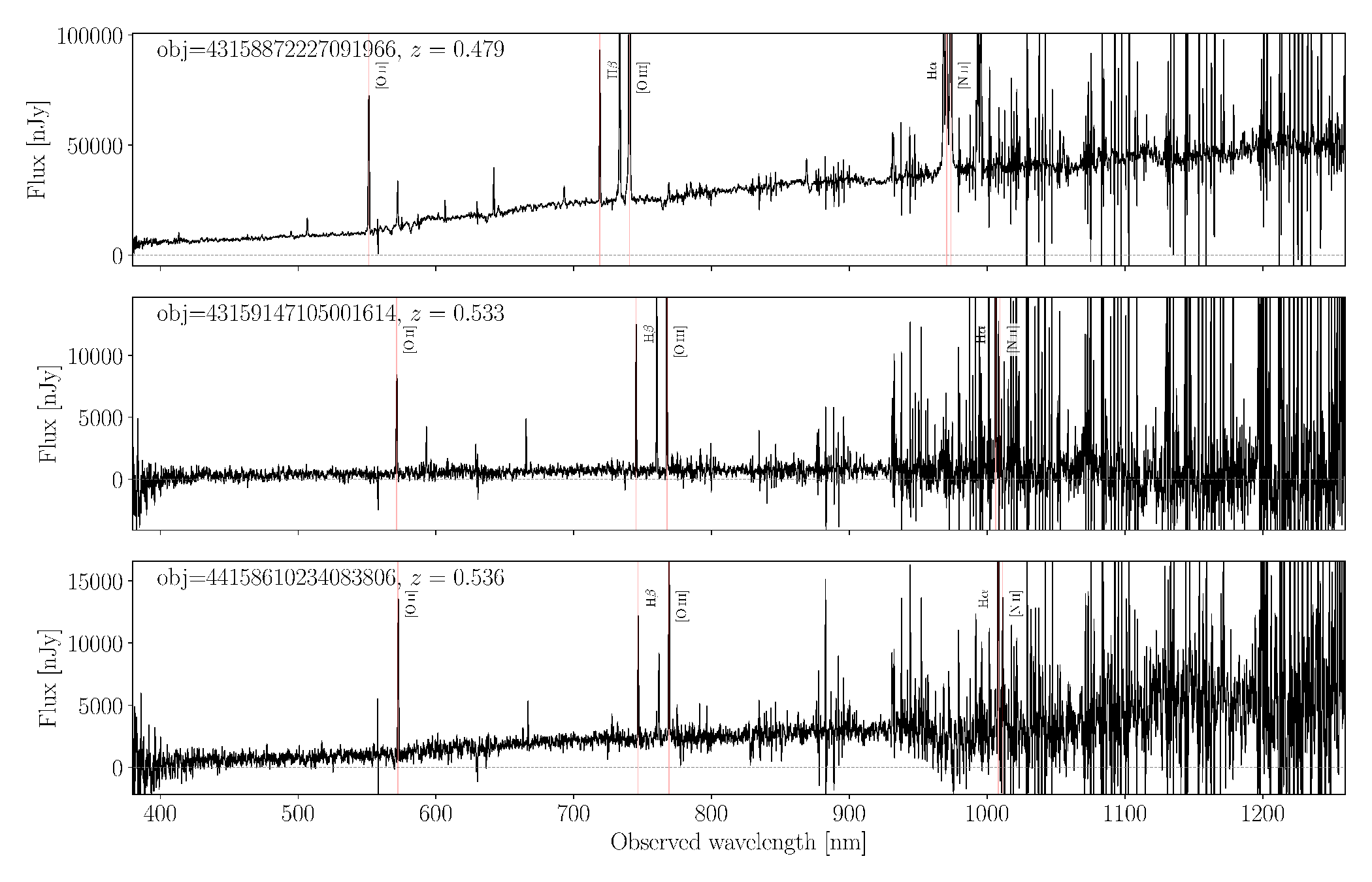}
\end{center}
\caption{
Representative examples of full PFS spectra for three objects in our sample. 
The spectra cover the full observed wavelength range of $\sim380-1260$ nm obtained in a single exposure with the PFS low-resolution mode. 
Major emission lines, including [O\thinspace{\sc ii}], H$\beta$, [O\thinspace{\sc iii}], H$\alpha$, and [N\thinspace{\sc ii}], are marked by vertical red lines at their expected observed wavelengths based on the spectroscopic redshift of each object. 
The spectra are lightly smoothed for display purposes. 

Alt text: Full PFS spectra of three representative objects covering approximately $380-1260$ nm. Multiple emission lines, including [O\thinspace{\sc ii}], H$\beta$, [O\thinspace{\sc iii}], H$\alpha$, and [N\thinspace{\sc ii}], are detected at the expected wavelengths, demonstrating the broad simultaneous wavelength coverage of PFS. 
} \label{fig:full_spectrum_examples}
\end{figure*}

Representative examples of the resulting spectra are shown in figure~\ref{fig:full_spectrum_examples}, illustrating the broad wavelength coverage of PFS and the simultaneous detection of multiple diagnostic emission lines. 
Redshifts and emission-line measurements are derived from the coadded one-dimensional spectra in our own analysis; details of the measurements are described in section~\ref{sec:spec_measurements}.

\section{Spectroscopic Measurements} \label{sec:spec_measurements}

\subsection{Construction of the spectroscopic sample}

\begin{figure*}[t]
\begin{center}
\includegraphics[width=1\linewidth]{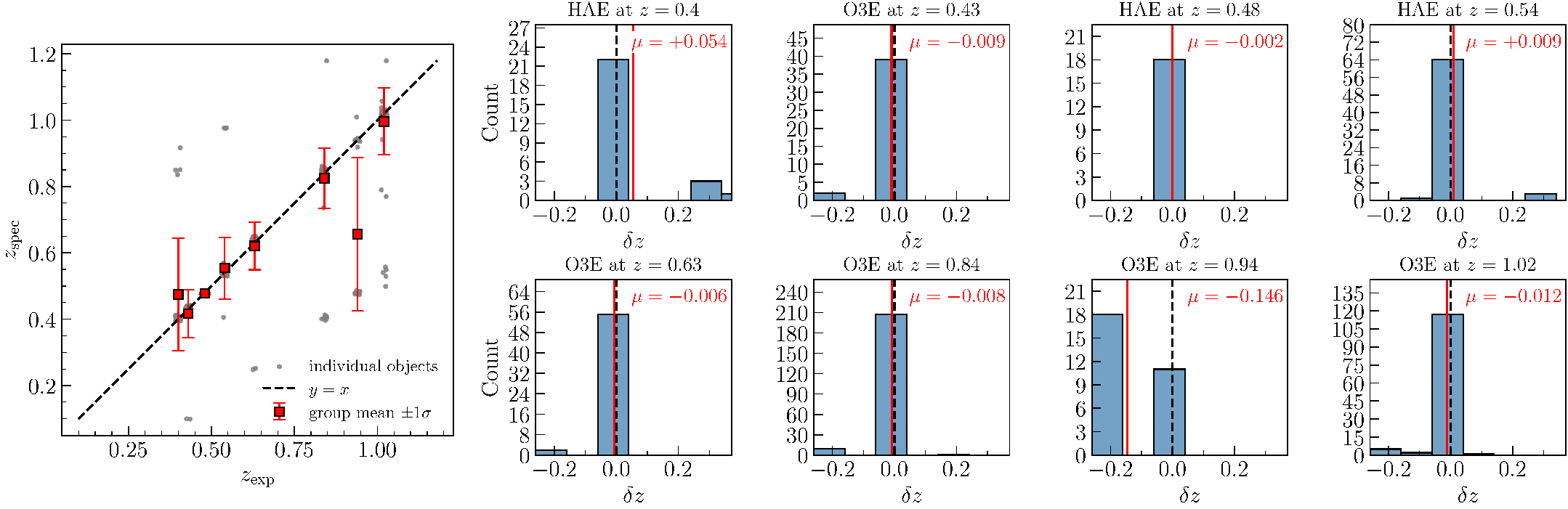}
\end{center}
\caption{ Validation of our spectroscopic redshifts measured from PFS spectra.
\textit{Left:} measured spectroscopic redshift ($z_{\rm spec}$) as a function of the redshift expected from the narrowband (NB) target selection ($z_{\rm exp}$; assuming H$\alpha$ for HAEs and [O\thinspace{\sc iii}] for O3Es).
grey circles show individual objects (with a small random offset applied for visibility), the dashed line indicates $y=x$, and red squares show the group mean with the $1\sigma$ scatter.
\textit{Right:} distributions of $\delta z \equiv (z_{\rm spec}-z_{\rm exp})/(1+z_{\rm exp})$ for each NB slice; the dashed black line marks $\delta z=0$, and the solid red line indicates the mean offset $\mu$ (labelled in each panel). 

Alt text: Comparison of measured PFS spectroscopic redshifts with redshifts expected from the NB selections, together with distributions of their normalised differences. Most NB subsamples lie close to the one-to-one relation and have offsets near zero, while the O3E sample at $z_{\rm exp}\simeq0.94$ shows a negative offset.
}
    \label{fig:specz}
\end{figure*}

\begin{figure*}[h!]
\begin{center}
\includegraphics[width=1.0\linewidth]{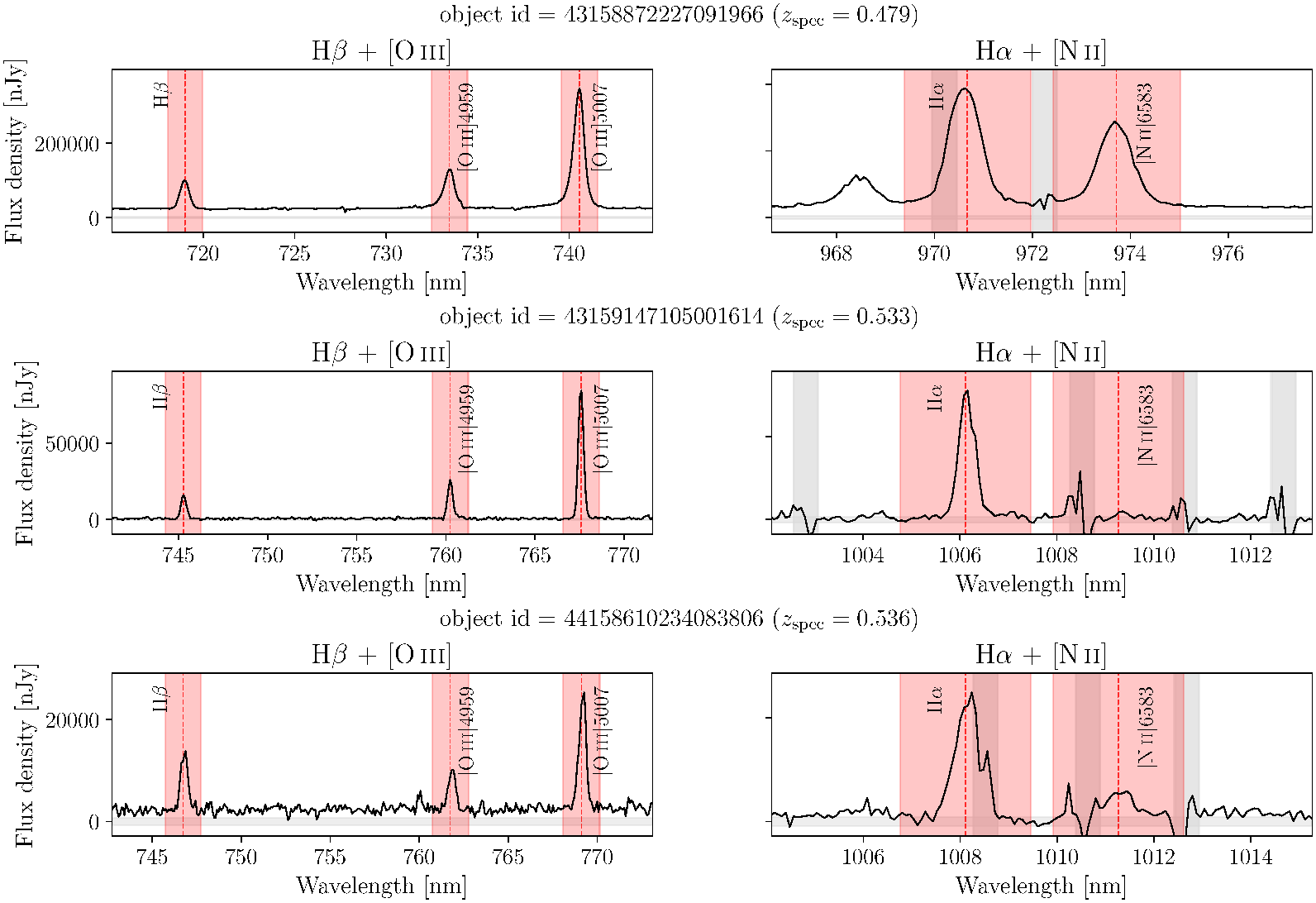}
\end{center}
\caption{ 
Example PFS spectra for three representative objects, shown as zoomed-in views of the spectra presented in figure~\ref{fig:full_spectrum_examples}.
For each object, the H$\beta$~$\lambda4861$, [O\thinspace{\sc iii}]~$\lambda\lambda4959, 5007$, H$\alpha$~$\lambda6563$, and [N\thinspace{\sc ii}]~$\lambda6583$ emission-line regions are shown. 
The expected positions of these emission lines, based on the spectroscopic redshift of each object, are marked by vertical red lines. 
grey vertical shaded regions indicate masked OH skylines, and red shaded regions denote the integration windows used for the line-flux measurements. 
Light grey horizontal bands indicate the typical local $\pm1\sigma$ noise levels estimated from nearby line-free regions. 

Alt text: Zoomed PFS spectra around H$\beta$+[O\thinspace{\sc iii}] and H$\alpha$+[N\thinspace{\sc ii}] for three representative objects. The principal emission lines appear near their expected wavelengths with varying strengths, while shaded regions indicate the adopted line-integration windows, masked OH skylines, and local noise levels. 
 }\label{fig:spec}
\end{figure*}

To obtain homogeneous and reproducible redshift measurements for our targets, we apply an automated line-based procedure as summarised below. 
The automated redshift estimation begins with emission-line peak detection in the reduced 1D spectra.
We estimate the local noise using a robust sliding-window estimator with a window size of 31 pixels, and subtract the continuum estimated with a 61-pixel median filter.
The residual spectrum is further smoothed using a median filter and converted into a signal-to-noise spectrum using the locally estimated noise.
Emission-line candidates are then identified using the \texttt{find\_peaks} routine in the Python SciPy package, applied to the smoothed S/N spectrum with a detection threshold of $S/N > 2$, a minimum peak separation of three pixels, and an allowed peak width of 1--50 pixels.
To mitigate spurious detections near the instrumental response edges, we mask asymmetric wavelength margins of 25 nm on the short-wavelength side and 5 nm on the long-wavelength side.
We also reduce false positives caused by sky-line residuals by excluding candidate peaks within $\pm 0.25$ nm of strong OH lines 
(selected from the Rousselot list with relative intensities larger than 50 and $\lambda \gtrsim 700$ nm; \citealt{Rousselot2000}).

\begin{figure}
\begin{center}
\includegraphics[width=1\linewidth]{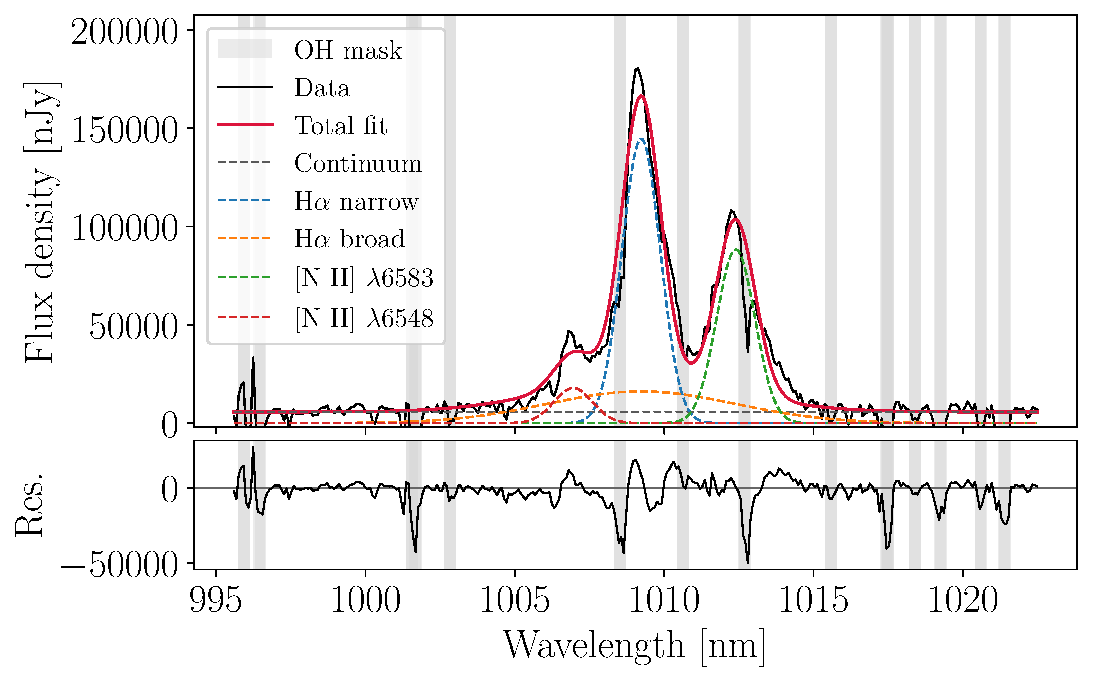}
\end{center}
\caption{ 
PFS spectrum around the H$\alpha$ region for the object (object\_id = 43158743378075529 in HSC-SSP PDR2),
together with the best-fitting multi-component model.
The top panel shows the observed spectrum (black), the total model (red),
and individual components including the continuum, narrow and broad H$\alpha$,
and the [N\thinspace{\sc ii}] $\lambda\lambda6548,6583$ doublet (dashed curves).
grey shaded regions indicate masked OH skylines.
The bottom panel shows the residuals after subtracting the best-fitting model. 

Alt text: PFS spectrum around H$\alpha$ for the broad-line object, decomposed into continuum, narrow and broad H$\alpha$, and [N\thinspace{\sc ii}] components. A prominent broad H$\alpha$ component is required in addition to the narrow emission lines, and the lower panel shows the residuals from the best-fitting model.
}\label{fig:broadH}
\end{figure}

Candidate redshifts are generated by anchoring on the strongest plausible line expected from the NB selection.
For O3Es, we adopt [O\thinspace{\sc iii}]~$\lambda5007$ as the primary anchor, while for HAEs we allow H$\alpha$ and [O\thinspace{\sc iii}]~$\lambda5007$ as anchor candidates.
For each anchor peak, we search for additional lines ([O\thinspace{\sc iii}]~$\lambda4959$, H$\beta$, [O\thinspace{\sc ii}]~$\lambda3727$, H$\alpha$, and [N\thinspace{\sc ii}]~$\lambda6583$) within $\pm 0.8$ nm of the expected observed wavelengths.
We evaluate solutions over $0 \le z_{\rm spec} \le 1.2$ and prioritise  the redshift that maximises (i) the number of consistently matched lines and (ii) the peak significance of the anchor.
We first restrict the peak search to windows centred on the expected lines at $z_{\rm exp}$ with a tolerance of $|z-z_{\rm exp}| \le 0.08$, and then perform a wider search over the full spectral range if no reliable solution is found.

For [O\thinspace{\sc iii}] emitters, we explicitly check the [O\thinspace{\sc iii}] doublet consistency, requiring the presence of $\lambda4959$ at the expected position and that the $\lambda5007$ component is stronger.
To recover rare cases where $\lambda4959$ is lost to strong OH features, we allow an OH-loss safeguard: if only one line is missing, the anchor is detected with $S/N \ge 5$, and $|z_{\rm spec}-z_{\rm exp}| \le 0.03$, the solution is retained.

For each object we record $z_{\rm spec}$, an empirical redshift uncertainty estimated from the scatter of line-based redshifts, the number of matched lines $N_{\rm lines}$, and the maximum line significance $S/N_{\rm max}$. Here, $N_{\rm lines}$ denotes the number of emission-line features matched to a common redshift solution. 
We assign the highest spectral quality flag satisfying the following criteria, based on $(N_{\rm lines}, S/N_{\rm max})$: $Q=4$ for $N_{\rm lines}\ge 4$ and $S/N_{\rm max}\ge 3.0$, $Q=3$ for $N_{\rm lines}\ge 3$ and $S/N_{\rm max}\ge 2.5$, $Q=2$ for $N_{\rm lines}\ge 2$ and $S/N_{\rm max}\ge 2.0$, and $Q=1$ for $N_{\rm lines}\ge 1$ and $S/N_{\rm max}\ge 1.0$. 
The $Q=1$ category represents the lowest-confidence redshift solutions and should be regarded as tentative.

We summarise the spectroscopic outcomes for the high-EW AGN candidates in table~\ref{table1}.
We define ``Confirmed'' targets as objects with a reliable spectroscopic redshift solution,
corresponding to a spectral quality flag of $Q\ge2$. 
We additionally include two manually identified $z\sim4$ interlopers as confirmed sources.
Thus, the 583 confirmed sources consist of 581 sources with $Q\ge2$ and two manually identified $z\sim4$ interlopers.
Among the confirmed targets, we classify objects whose $z_{\rm spec}$ falls outside the redshift range expected from the narrowband filter bandpass as ``Redshift mismatch'' systems.
Targets without a reliable spectroscopic redshift solution are classified as ``Unconfirmed''; parenthesised counts in table~\ref{table1} indicate a small number of additional targets rejected after manual inspection (e.g., severe sky-line residuals).

We validate our spectroscopic redshifts by comparing $z_{\rm spec}$ measured from the PFS spectra with the redshift expected from the NB selection, $z_{\rm exp}$, which is set by the central wavelength of each NB filter and the assumed emission line.
As shown in figure~\ref{fig:specz}, most objects follow the one-to-one relation, and the $\delta z$ distributions, where $\delta z \equiv (z_{\rm spec}-z_{\rm exp})/(1+z_{\rm exp})$, are narrowly peaked around zero, indicating that our line identification and redshift measurements are robust for the majority of the sample. 
A notable exception is the O3Es at $z_{\rm exp}\simeq0.94$, which show a larger systematic offset ($\mu\simeq-0.146$). 
Most of the mismatched sources are identified as H$\alpha$ emitters at $z\simeq0.48$, suggesting contamination and/or selection effects in this slice. 

As an external check, we also compare the PFS spectroscopic redshifts with the spectroscopic redshifts compiled in the HSC catalog ($z^{\rm HSC}_{\rm spec}$). 
Among the 60 objects with available $z^{\rm HSC}_{\rm spec}$ measurements, all are classified as Q4 and all satisfy $|z^{\rm PFS}_{\rm spec}-z^{\rm HSC}_{\rm spec}|/(1+z^{\rm HSC}_{\rm spec})<0.01$, corresponding to a success rate of 100\% for the Q4 sample with available $z^{\rm HSC}_{\rm spec}$ measurements. 
No $z^{\rm HSC}_{\rm spec}$ measurements are available for the Q1--Q3 objects.

\begin{figure}
\begin{center}
\includegraphics[width=0.9\linewidth]{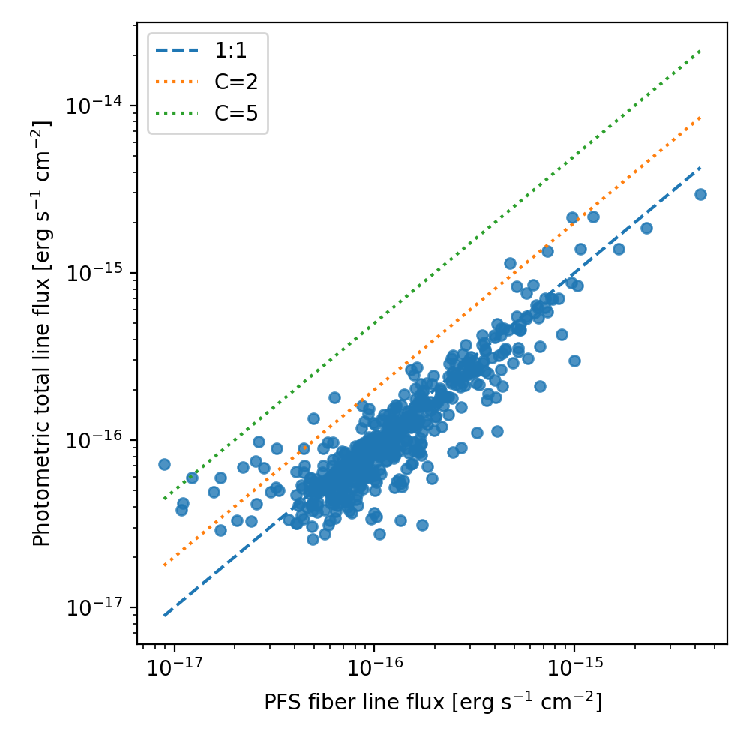}
\end{center}
\caption{
Comparison between photometric total line fluxes estimated from HSC-SSP photometry and PFS fibre line fluxes.
The dashed, dotted, and dot-dashed lines indicate one-to-one, factor-of-two, and factor-of-five relations, respectively. 

Alt text: Log--log comparison of total emission-line flux estimated from HSC-SSP photometry with PFS fibre line flux. The two measurements are correlated, with most objects lying within approximately a factor of two
of the one-to-one relation, although some objects show larger deviations.
}
\label{fig:apcorr}
\end{figure}

\subsection{Emission-line flux measurements}

We measure emission-line fluxes from the reduced PFS 1D spectra after fixing the spectroscopic redshift $z_{\rm spec}$.
The two manually identified $z\sim4$ interlopers are excluded from the rest-frame optical emission-line analysis described below because the relevant diagnostic lines fall outside the PFS wavelength coverage.
We focus on seven key lines used in this work, namely
[Ne\thinspace{\sc v}] $\lambda3426$,
[O\thinspace{\sc ii}] $\lambda3727$, 
He\thinspace{\sc ii} $\lambda4686$, 
H$\beta$,
[O\thinspace{\sc iii}] $\lambda5007$,
H$\alpha$,
and [N\thinspace{\sc ii}] $\lambda6583$. 

For each line, we compute the observed central wavelength 
$\lambda_{\rm obs}=\lambda_{\rm rest}(1+z_{\rm spec})$ 
and integrate the continuum-subtracted flux within a symmetric velocity window of 
$\pm v$, where we adopt $v=400$ km s$^{-1}$.  
This corresponds to a half-width
$\Delta\lambda = \lambda_{\rm obs} (v/c)$.
The line flux is measured by trapezoidal integration within
$\lambda_{\rm obs}\pm \Delta\lambda$.
Spectral pixels affected by strong OH skylines are masked, and the fluxes in the masked regions are linearly interpolated.
We adopt direct integration rather than Gaussian or Lorentzian profile fitting, as it provides a simple and robust flux measurement without assuming an emission-line profile, particularly for weak lines such as [N\thinspace{\sc ii}].

Local baselines are estimated in a window of $\pm 3\Delta\lambda$ around each line by fitting a linear model to the off-line pixels.
To ensure robust continuum determination against residual sky features and weak emission-line wings, the fit is performed with $2\sigma$ sigma-clipping, combined with an additional rejection of positive outliers identified from rolling-median peak detection and dilation in wavelength space.
For H$\alpha$ and [N\thinspace{\sc ii}] $\lambda6583$, we adopt a common baseline derived from a wider local fit that simultaneously excludes both line cores to minimise cross-contamination in the blended region.

We estimate line-flux uncertainties primarily from the local continuum residuals.
When sufficient off-line pixels are available (typically $\ge5$ pixels in the continuum sidebands after OH line masking), we derive a robust 1$\sigma$ noise estimate from the median absolute deviation of the continuum sidebands and propagate it through the trapezoidal integration.
If the sidebands are insufficient, we instead estimate the noise from a broader $\pm 5\Delta\lambda$ neighbourhood excluding the line window and OH-masked pixels.
As an additional safeguard against spurious single-pixel features, we require that the continuum-subtracted spectrum within the integration window contains at least three consecutive pixels with a per-pixel significance exceeding $S/N=1.5$; otherwise, the line is treated as undetected. 
The measured fluxes and their uncertainties are used for the emission-line ratio analysis, including the BPT classification and the identification of high-ionisation features such as He\thinspace{\sc ii} and [Ne\thinspace{\sc v}]. 
We regard He\thinspace{\sc ii} and [Ne\thinspace{\sc v}] as detected when their measured line fluxes have $S/N>3$. 
Figure~\ref{fig:spec} shows zoomed views of the H$\beta$+[O\thinspace{\sc iii}] and H$\alpha$+[N\thinspace{\sc ii}] emission-line regions for the three representative objects shown in figure~\ref{fig:full_spectrum_examples}. The panels illustrate the local continuum estimation, OH-line masking, and the integration windows adopted for the line-flux measurements.

\begin{figure*}[h!]
\begin{center}
\includegraphics[width=1\linewidth]{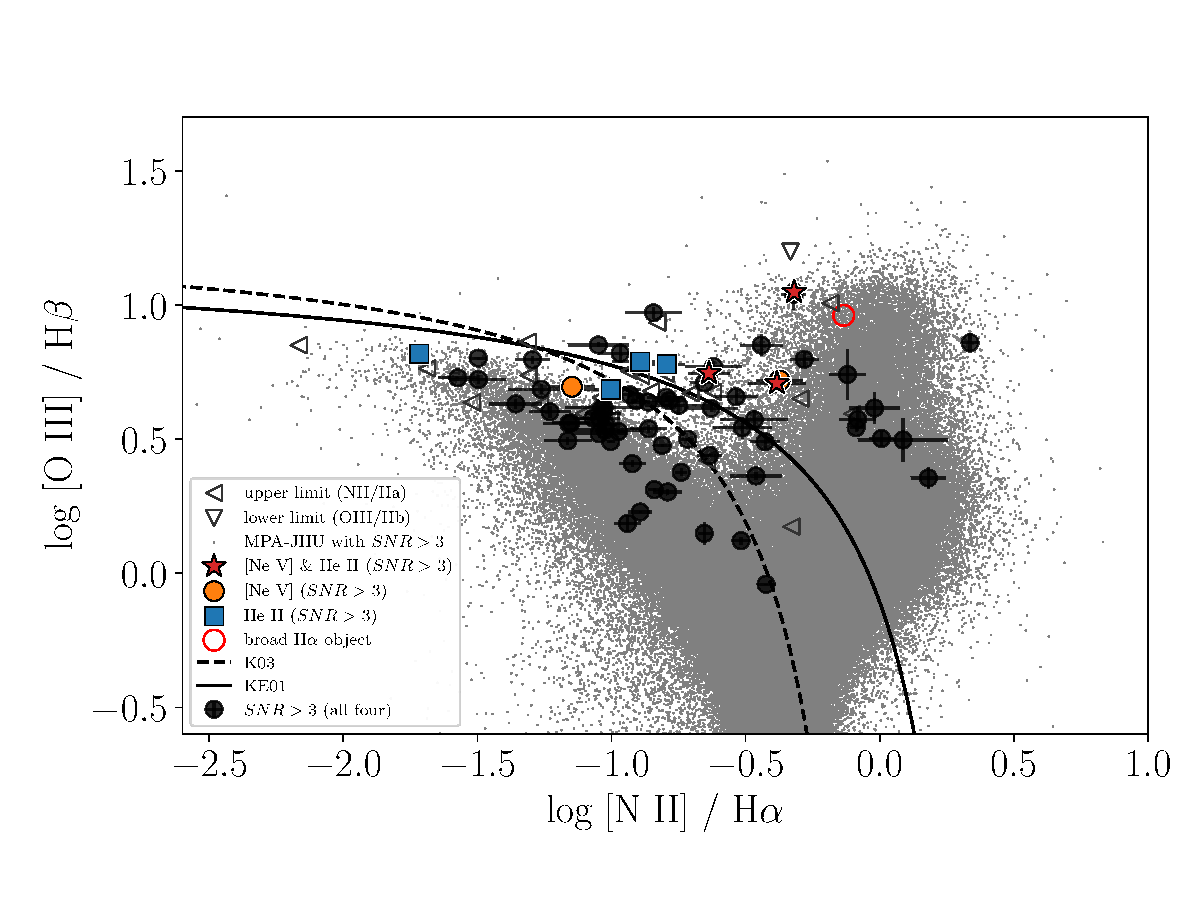}
\end{center}
\caption{ 
    BPT diagram of the NB-selected PFS sample.
    Black circles show objects with $S/N>3$ in all four lines.
    Open symbols with arrows indicate upper/lower limits when one of the lines
    falls below the threshold.
    The grey dots show SDSS galaxies from the MPA--JHU catalog
    with the same $S/N>3$ requirement.
    The dashed and solid curves denote the \citet{Kauffmann2003} and \citet{Kewley2001} demarcations, respectively.
    Objects with detected He\thinspace{\sc ii} and/or [Ne\thinspace{\sc v}]
    are highlighted.
    The broad-H$\alpha$ object is marked as open red circle separately. For this object, the emission-line ratios were calculated using only the narrow emission-line component. 
    
    Alt text: BPT diagram comparing the NB-selected PFS sample with the local SDSS galaxy distribution. The PFS sources span the star-forming, intermediate valley, and AGN regions, while several He\thinspace{\sc ii}- and/or [Ne\thinspace{\sc v}]-detected objects occupy the valley or even the classical star-forming region.
    }
    \label{fig:bpt}
\end{figure*}

One object in our PFS sample (object\_id = 43158743378075529 in HSC-SSP PDR2) shows evidence for a broad H$\alpha$ component (see figure~\ref{fig:broadH}). For this source, we do not adopt the standard box-integration approach. 
Since the broad and narrow H$\alpha$ components are blended in this object, the standard box-integration approach cannot reliably recover the narrow-line flux. Therefore, only for this source, we perform a dedicated multi-component fit to the H$\alpha$+[N\thinspace{\sc ii}] complex.
We simultaneously fit a constant continuum and four Gaussian components corresponding to a narrow H$\alpha$, a broad H$\alpha$, and the [N\thinspace{\sc ii}] $\lambda\lambda6548,6583$ doublet.
The velocity dispersion of the narrow H$\alpha$ component is tied to that of the [N\thinspace{\sc ii}] lines to ensure a stable decomposition of the blended region, while the fluxes of the two [N\thinspace{\sc ii}] lines are allowed to vary independently. We allow the fit redshift to vary within the uncertainty of the spectroscopic solution and mask wavelength ranges affected by strong OH residuals.
 
To assess whether the inclusion of a broad component is supported by the data, 
we compare this model with a narrow-line-only model that includes narrow H$\alpha$, 
[N\thinspace{\sc ii}] $\lambda\lambda6548,6583$, and a constant continuum.
We evaluate the models using the Akaike Information Criterion (AIC) and 
Bayesian Information Criterion (BIC), which balance goodness of fit against model complexity.
Following commonly used interpretations of the AIC and BIC, 
$\Delta{\rm AIC}>10$ and $\Delta{\rm BIC}>6$ indicate strong evidence in favour of the model with the lower AIC and BIC, respectively.
Defining 
$\Delta {\rm AIC} \equiv {\rm AIC}_{\rm narrow}-{\rm AIC}_{\rm broad}$ and
$\Delta {\rm BIC} \equiv {\rm BIC}_{\rm narrow}-{\rm BIC}_{\rm broad}$, 
we obtain $\Delta{\rm AIC}=13.7$ and $\Delta{\rm BIC}=6.7$, 
indicating that the model including a broad H$\alpha$ component is preferred.
We also confirmed that the broad component is recovered over a range of initial parameter values, 
yielding a consistent broad-line width of ${\rm FWHM}\simeq2800~{\rm km~s^{-1}}$.
The broad-line flux $F_{\rm H\alpha, b}$ is derived from the best-fit amplitude and width of the broad Gaussian component.
For the other diagnostic lines in this object ([O\thinspace{\sc iii}] and H$\beta$), 
we measure fluxes using the same velocity-window integration adopted for the main sample.

Because PFS is a fibre-fed spectroscopic instrument, the finite fibre aperture may lead to flux losses relative to the total galaxy emission.
We assess the impact of aperture effects by comparing the PFS fibre line fluxes with photometric total line-flux estimates derived from HSC-SSP measurements. 
Because the continua are generally too weak to be robustly detected in the PFS spectra for the majority of our high-EW sample, we use the emission-line fluxes, rather than continuum fluxes, to assess the aperture effect. 
We define a photometric-to-spectroscopic flux ratio as
\begin{equation}
C_{\rm aper}
=
\frac{F_{\rm phot}}
     {F_{\rm PFS}},
\end{equation}
where $F_{\rm phot}$ denotes the photometric estimate of the total emission-line flux, while $F_{\rm PFS}$ represents the line flux measured from the PFS fibre spectrum.
Figure~\ref{fig:apcorr} shows the comparison between the photometric and spectroscopic fluxes.
The distribution is centred close to unity, with a median value of $C_{\rm aper}=0.87$. 
The median value slightly below unity indicates that, on average, the PFS line fluxes are modestly larger than the photometric estimates. 
This small offset may reflect systematic uncertainties in the photometric line-flux estimates and/or the relative flux calibration, rather than to aperture effects alone. 
Most objects lie within a factor of approximately two of the one-to-one relation, indicating that large aperture losses are not typical for the majority of the sample. 
Some objects exhibit larger values ($C_{\rm aper}\gtrsim3$), which may reflect a combination of true aperture losses, uncertainties in photometric equivalent-width estimates, and residual systematics associated with narrowband transmission effects.
Because the object-by-object photometric flux estimates themselves carry non-negligible uncertainties, applying individual aperture corrections could introduce additional scatter into the measured line fluxes.
We therefore do not apply object-by-object aperture corrections in the main analysis and instead regard the comparison as a consistency check on the absolute flux scale.
For the broad-line source (figure \ref{fig:broadH}), we obtain $C_{\rm aper}=1.84$, suggesting a moderate aperture loss.
If applied, this would increase the broad H$\alpha$ luminosity by the same factor and quantities depending on it accordingly.
We therefore regard this value as an estimate of the systematic uncertainty associated with aperture effects, which will be discussed in section~\ref{subsec:broad}.

\section{Results}

A primary goal of this work is to investigate the ionisation properties of NB-selected emitters using emission-line diagnostics enabled by the PFS spectroscopy. 
We thus examine the location of the sample on the BPT diagram. 
We restrict the primary BPT analysis to the 70 sources with $S/N>3$ detections in all four diagnostic lines, H$\alpha$, H$\beta$, [O\thinspace{\sc iii}], and [N\thinspace{\sc ii}]. 

Figure~\ref{fig:bpt} shows the BPT diagram for the NB-selected PFS sample.
In addition to these 70 sources, we show 22 sources for which one of the four lines falls below the detection threshold as upper or lower limits, giving 92 sources in total in the diagram. 
For comparison, we also overlay the locus of SDSS galaxies from the MPA--JHU catalogue. 
The majority of the PFS emitters occupy the star-forming and composite regions, while a subset extends into the AGN-dominated regime above the \citet{Kewley2001} demarcations. 
We additionally highlight objects with significant high-ionisation lines
He\thinspace{\sc ii}~$\lambda4686$ and/or [Ne\thinspace{\sc v}]~$\lambda3426$.
The production of He$^{+}$ and Ne$^{4+}$ requires photons with energies above 54.4 and 97.1~eV, respectively, making these lines useful tracers of hard ionising spectra associated with AGNs \citep{Shirazi2012,Feltre2016}.
These objects preferentially lie in or near the AGN region, although a subset lies in the star-forming region. 
The single object exhibiting a clear broad H$\alpha$ component is also shown, and its BPT position is consistent with AGN-like line ratios.

In our NB-selected PFS sample, 24 out of 70 objects ($34$\%) are classified as AGNs based on the \citet{Kewley2001} demarcation, following the definition adopted in section~2.2. 
This fraction is substantially higher than the 5.1 \% AGN fraction found in the local high-EW emitter population discussed in section~2.2, although differences in sample selection and BPT-line detectability should be taken into account when interpreting this comparison. 
Among the 22 sources shown as limits, 14 can still be robustly classified with respect to the \citet{Kewley2001} demarcation. 
Including these sources yields a similar AGN fraction of 25/84 ($30$ \%).  
Among the PFS sources, the AGN fraction is higher for O3Es (17/33; 52\%) than for HAEs (7/37; 19\%).
This trend is qualitatively consistent with that seen in the local high-EW emitter population (section~2.2).
The origin of this elevated AGN fraction is discussed further in section~\ref{sec:highagn}.

A large fraction of the PFS emitters lie in or near the valley region.
Such valley objects are also seen in the SDSS/MPA--JHU comparison sample (see figure~\ref{fig:bpt_mpa_jhu}).  
We detect He\thinspace{\sc ii} and/or [Ne\thinspace{\sc v}] in nine emitters, three of which remain in the star-forming region of the BPT diagram ($N_{\rm HeII/SF}=2$, $N_{\rm NeV/SF}=1$).
The detection of these high-ionisation lines indicates that some objects classified as star-forming in the classical BPT diagram may nevertheless host hard ionising sources (e.g., \citealt{Izotov2008}).

\begin{figure*}
\begin{center}
\includegraphics[width=1\linewidth]{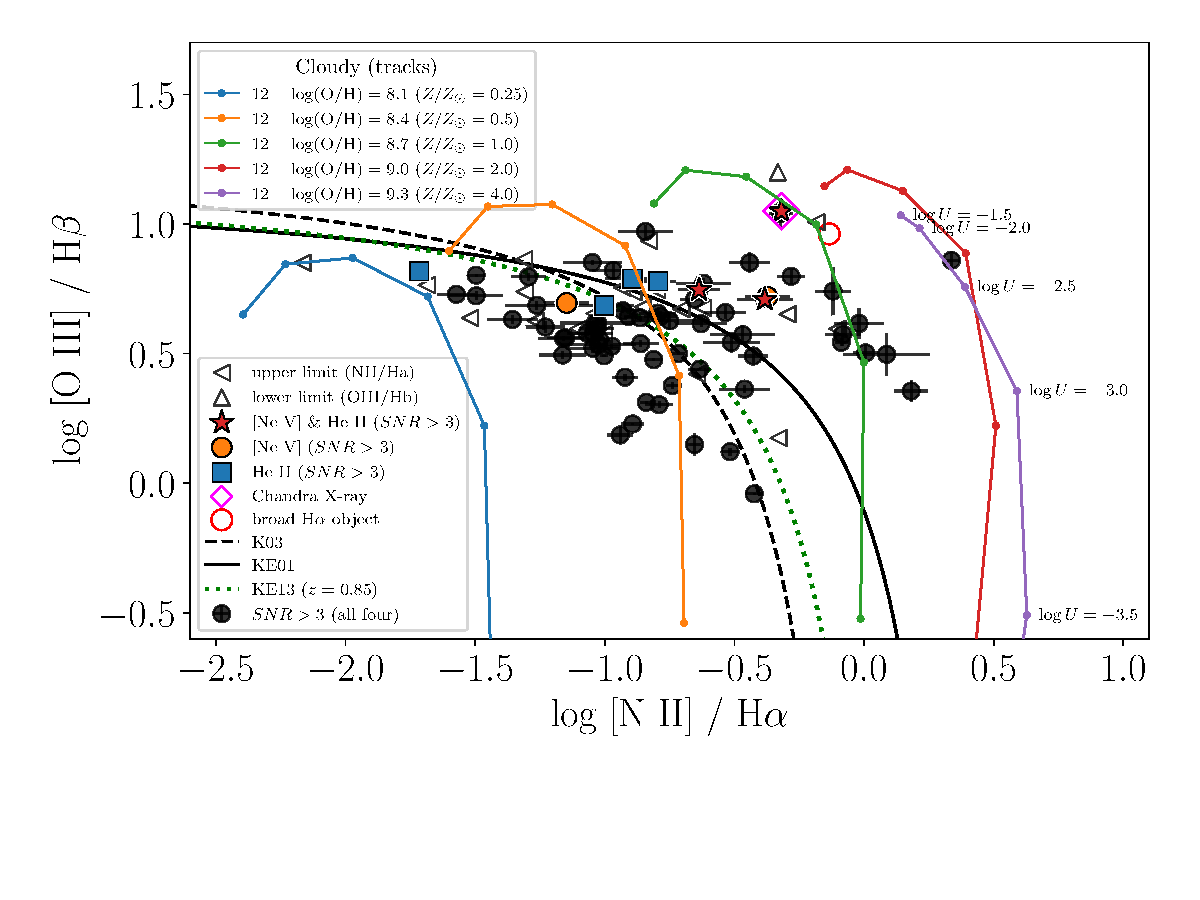}
\end{center}
\caption{ 
BPT diagram for the NB-selected PFS sample.
Black symbols show objects with reliable emission-line measurements
([O\thinspace{\sc iii}]~$\lambda5007$, H$\beta$, H$\alpha$, and [N\thinspace{\sc ii}]~$\lambda6583$)
above the adopted signal-to-noise thresholds.
The single broad H$\alpha$ object is indicated separately (open red circle).
The magenta diamond marks the object with a counterpart in the Chandra COSMOS-Legacy survey. 
Overlaid coloured curves show AGN photoionisation model tracks generated with \textsc{Cloudy},
spanning $Z/Z_{\odot} = 0.25-4$ and $\log U = -4.0$ to $-1.5$.
Nitrogen abundances are tied to metallicity following the \citet{Groves2006} prescription.
The green dashed curve shows the redshift-dependent AGN/star-forming separation proposed by \citet{Kewley2013} for $z=0.85$, approximately corresponding to the highest redshift probed by our four-line BPT sample. 
For clarity, the SDSS reference locus is not shown in this figure.

Alt text: BPT diagram of the PFS sample overlaid with CLOUDY AGN photoionisation tracks covering a range of metallicities and ionisation parameters. Many objects in the BPT valley overlap the sub-solar model tracks, while the high-ionisation-line detections extend towards the lower-metallicity portion of the model grid.
}
    \label{fig:bpt_cloudy}
\end{figure*}

\section{Discussion}

\subsection{Photoionisation models and the origin of valley objects}

A key result of this study is that a non-negligible fraction of our NB-selected PFS emitters  occupy the valley region between the star-forming and AGN branches in the BPT diagram. 
While such locations may arise from composite ionisation or from low-metallicity AGN, several of our valley objects extend into the AGN region of the diagram. 
To investigate whether these systems can be explained by AGN photoionisation under low-metallicity conditions, we compare the observations with tailored AGN photoionisation models.

We generate tailored AGN narrow-line region (NLR) grids with \textsc{Cloudy} C23 \citep{Ferland2017},
adopting a constant-density gas with hydrogen density $n_{\rm H}=10^{3} {\rm cm^{-3}}$
(\texttt{hden}~3) and including ISM-type dust grains.
The dust-to-gas ratio is scaled linearly with metallicity 
(\texttt{grains abundance}  $\propto Z/Z_{\odot}$), consistently with the scaling applied to metals
(\texttt{metals} $Z$ \texttt{linear}).
For the ionising continuum, we use the built-in \texttt{AGN} spectral shape,
parameterised by a Big Blue Bump temperature $T=10^{6}$~K and spectral indices
$\alpha_{\rm ox}=-1.6$, $\alpha_{\rm UV}=-0.5$, and $\alpha_{\rm X}=-1.0$.
We vary the total metallicity over $Z/Z_{\odot} =$ 0.25, 0.5, 1, 2, and 4,
and sample the ionisation parameter over $\log U = -4.0$ to $-1.5$ in 0.5-dex steps.
All models adopt the GASS10 abundance set with $12+\log({\rm O/H})_{\odot}=8.69$,
and tie the nitrogen abundance to metallicity following the \citet{Groves2006} prescription.
Calculations are iterated to convergence and stopped at $T=3000$~K.

Figure~\ref{fig:bpt_cloudy} presents the BPT diagram for our PFS sample
overlaid with the resulting \textsc{Cloudy} AGN tracks.
The locations of many valley objects are broadly consistent with sub-solar AGN photoionisation models, under the assumption
that the observed emission lines are predominantly powered by AGN activity.
In particular, their distribution is consistent with
$Z/Z_{\odot}\approx0.5$--1.0
(equivalently $12+\log({\rm O/H})\approx8.4$--8.7),
suggesting that these systems may host relatively metal-poor narrow-line regions. 

At the same time, star-forming galaxies at intermediate and high redshifts are known to be offset from the local SDSS star-forming
sequence in the BPT plane, likely owing to harder ionising spectra and higher ionisation parameters than those found in local star-forming galaxies (e.g., \citealt{Steidel2014, Shapley2015, Sanders2016, Strom2017, Shapley2019}).
For reference, figure~\ref{fig:bpt_cloudy} also shows the redshift-dependent AGN/star-forming separation proposed by \citet{Kewley2013} at $z=0.85$, approximately corresponding to the highest redshift probed by our four-line BPT sample. 
The resulting boundary differs only slightly from the local \citet{Kewley2001} relation, suggesting that redshift evolution of the star-forming sequence alone is unlikely to account for the large number of valley objects observed in our sample.

Therefore, the observed distribution is broadly consistent with a combination of composite systems containing both star formation and AGN activity, and low-metallicity AGNs with hard ionising spectra.

\subsection{Morphological constraints from source extendedness}

We examine the morphological properties of our sample using the ratio
of the source size to the local PSF size, to investigate whether the
sample shows structural properties distinct from star-forming galaxies
and AGNs.

We quantify the extendedness of each object using the size ratio,
\begin{equation}
r \equiv \frac{\mathrm{source~size}}{\mathrm{PSF~size}}.
\end{equation}
The source size is derived from the second moments measured by the HSC
pipeline (the \texttt{sdssshape} algorithm). Specifically, we define the
source size as
\begin{equation}
\mathrm{source~size}
=
\left(I_{11}I_{22}-I_{12}^{2}\right)^{1/4},
\end{equation}
where $I_{11}$, $I_{22}$, and $I_{12}$ are the adaptive second moments
of the object.
The PSF size is defined analogously as
\begin{equation}
\mathrm{PSF~size}
=
\left(I_{11}^{\rm PSF}I_{22}^{\rm PSF}
-\left(I_{12}^{\rm PSF}\right)^2\right)^{1/4},
\end{equation}
where $I_{11}^{\rm PSF}$, $I_{22}^{\rm PSF}$, and $I_{12}^{\rm PSF}$
are the second moments of the PSF model evaluated at the position of
each object.
The measurements are based on the HSC NB images corresponding to the
selection band of each object (e.g., NB718, NB816, NB921, NB973, or
NB1010).
Because these bands include the targeted emission lines, the measured
sizes are expected to reflect both continuum and line-emitting components.
By construction, point sources are expected to have $r \sim 1$, while
extended sources exhibit $r > 1$.
We note that the size ratio may in principle depend on redshift through
cosmological angular-size effects and galaxy size evolution.
However, within the BPT-classified subsample with reliable size measurements,
we find no significant correlation between the size ratio and redshift
(Spearman rank test: $N=66$, $\rho=-0.11$, $p=0.39$), indicating that such effects are unlikely to drive the observed distributions. 

\begin{figure}
\begin{center}
\includegraphics[width=1\linewidth]{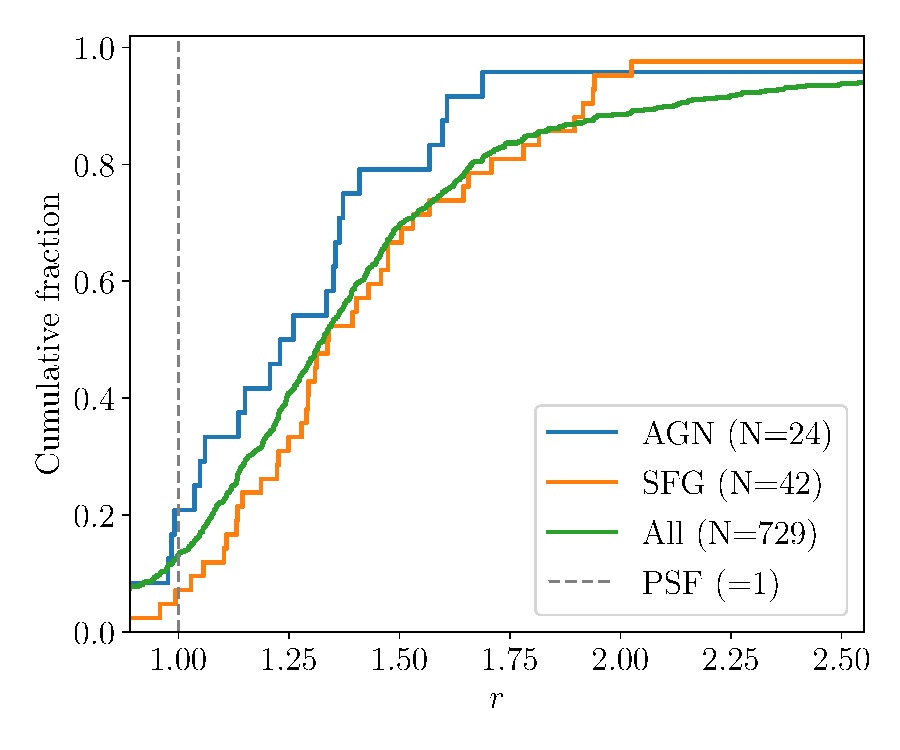}
\end{center}
\caption{
Cumulative distributions of the size ratio $r$ for the AGN, SFG, and all targets. The dashed line indicates $r=1$, corresponding to the PSF size. 
The horizontal axis range is restricted to $0.9 < r < 2.5$ to highlight the region around the PSF. 

Alt text: Cumulative distributions of source-to-PSF size ratio for AGN, star-forming galaxies, and the full target sample. AGN are more concentrated towards ratios close to unity, indicating more point-like morphologies, while star-forming galaxies extend to larger sizes and the full target sample lies between the two distributions.
}\label{fig:size_ratio_hist}
\end{figure} 

Figure~\ref{fig:size_ratio_hist} shows the distributions of the size ratio for SFGs, AGN, and the target sample. 
Here, SFGs and AGN are defined based on their locations in the BPT diagram (see figure \ref{fig:bpt_cloudy}), with AGN lying above the \citet{Kewley2001} line and SFGs below it. 
We note that objects below the \citet{Kewley2001} line are not necessarily purely star-forming and may include composite systems or AGNs not identified by the classical BPT criterion; for simplicity, we refer to this population as SFGs throughout this analysis. 
Only objects with reliable size measurements are included in this analysis, corresponding to 24/24, 42/46, and 729/827 of the AGN, SFG, and target samples, respectively.
The SFG population exhibits a broader distribution extending towards larger values of $r$, reflecting their spatially extended nature. In contrast, AGN tend to be more concentrated towards $r\sim1$. 
This may indicate intrinsically more compact light distributions; however, the bright nuclear emission associated with AGN can also reduce the detectability of faint extended host-galaxy light, potentially biasing the observed size ratios towards smaller values. The target sample shows an intermediate distribution between these two populations. 

We further quantify these trends using the fraction of point-like sources. 
We classify an object as point-like if its measured size is consistent with the PSF within a tolerance, adopting a threshold of $r < 1.1$. 
We find point-source fractions of $12\% \pm 5\%$ for SFGs, $33\% \pm 10\%$ for AGN, and $22\% \pm 2\%$ for the target sample. 
The spectroscopic subsample (i.e., the combined AGN and SFG samples) yields a similar value of $20\% \pm 5\%$ to that of the target sample, indicating that it is not strongly biased relative to the parent sample. 
The higher point-source fraction of the target sample compared to SFGs suggests that purely extended, star-formation-dominated emission is unlikely to explain the full population. 
At the same time, the fraction remains lower than that of AGN, implying that a significant contribution from host-galaxy emission is still present.

The origin of the relatively compact morphologies of the AGN sample remains unclear. One possible interpretation is that some AGN in our sample may represent systems at an early evolutionary stage, in which the host galaxies themselves have not yet grown to the sizes typically seen in more evolved systems. 

These results support a scenario in which nuclear activity contributes to the observed emission-line properties of the sample, while not dominating in all cases. 
This is consistent with an interpretation in which at least part of the population occupies an intermediate regime, where both AGN photoionisation and star formation contribute to the observed line ratios.

As an independent check of whether the difference in source compactness is associated with broader differences in host-galaxy morphology, we cross-matched the sources in our PFS sample with the HST/ACS-based ZEST morphology catalog \citep{Scarlata2007} using a matching radius of 1\arcsec. 
Among the BPT-classified objects with available ZEST classifications, 23/40 (58 \%) SFGs and 7/12 (58 \%) AGN are classified as disc galaxies. Early-type systems are found only among the AGN population (2/12), whereas none are present among the SFGs. 
Thus, the ZEST classifications do not reveal a clear difference in the broad morphological mix between the two populations, suggesting that the difference in HSC-based compactness cannot be simply attributed to a different fraction of disc and early-type galaxies.

\subsection{High-ionisation emission lines as AGN diagnostics}

Some objects with significant detections of He\thinspace{\sc ii}~$\lambda4686$ and/or [Ne\thinspace{\sc v}]~$\lambda3426$ are located in the star-forming or composite regions of the classical BPT diagram. 
A subset of these objects is broadly consistent with the lower-metallicity portion of the \textsc{Cloudy} AGN grids ($Z/Z_{\odot}\sim0.25$--0.5), although the number of such sources is small.
Previous studies have demonstrated that high-ionisation lines such as He\thinspace{\sc ii} and [Ne\thinspace{\sc v}] can serve as useful diagnostics of AGN-like ionising sources in low-metallicity systems (e.g., \citealt{Groves2006, Izotov2008}). 
As an additional check, we cross-match the sources in our PFS sample with the Chandra COSMOS-Legacy catalog \citep{Civano2016,Marchesi2016} using a matching radius of 1\arcsec. 
We find a single X-ray counterpart in our sample (figure \ref{fig:bpt_cloudy}). 
This source is also detected in both He\thinspace{\sc ii} and [Ne\thinspace{\sc v}] and is located in the AGN region of the BPT diagram. 
Using the spectroscopic redshift and the observed hard-band ($2-10$ keV) X-ray flux, we estimate an X-ray luminosity of $\log (L_{\rm X}/{\rm erg s^{-1}})=43.12$, well above the commonly adopted AGN threshold of $L_{\rm X}=10^{42}\ {\rm erg s^{-1}}$ \citep{Lanzuisi2018}. 
This independently confirms the presence of an accreting supermassive black hole in this source. 
Our results suggest that high-ionisation lines such as He\thinspace{\sc ii} and [Ne\thinspace{\sc v}] provide a useful complementary probe of AGN activity among high-EW NB-selected emitters, particularly in cases where classical BPT classifications alone may be ambiguous.

\subsection{Interpreting the high AGN fraction}\label{sec:highagn}

\begin{figure}
\begin{center}
\includegraphics[width=1\linewidth]{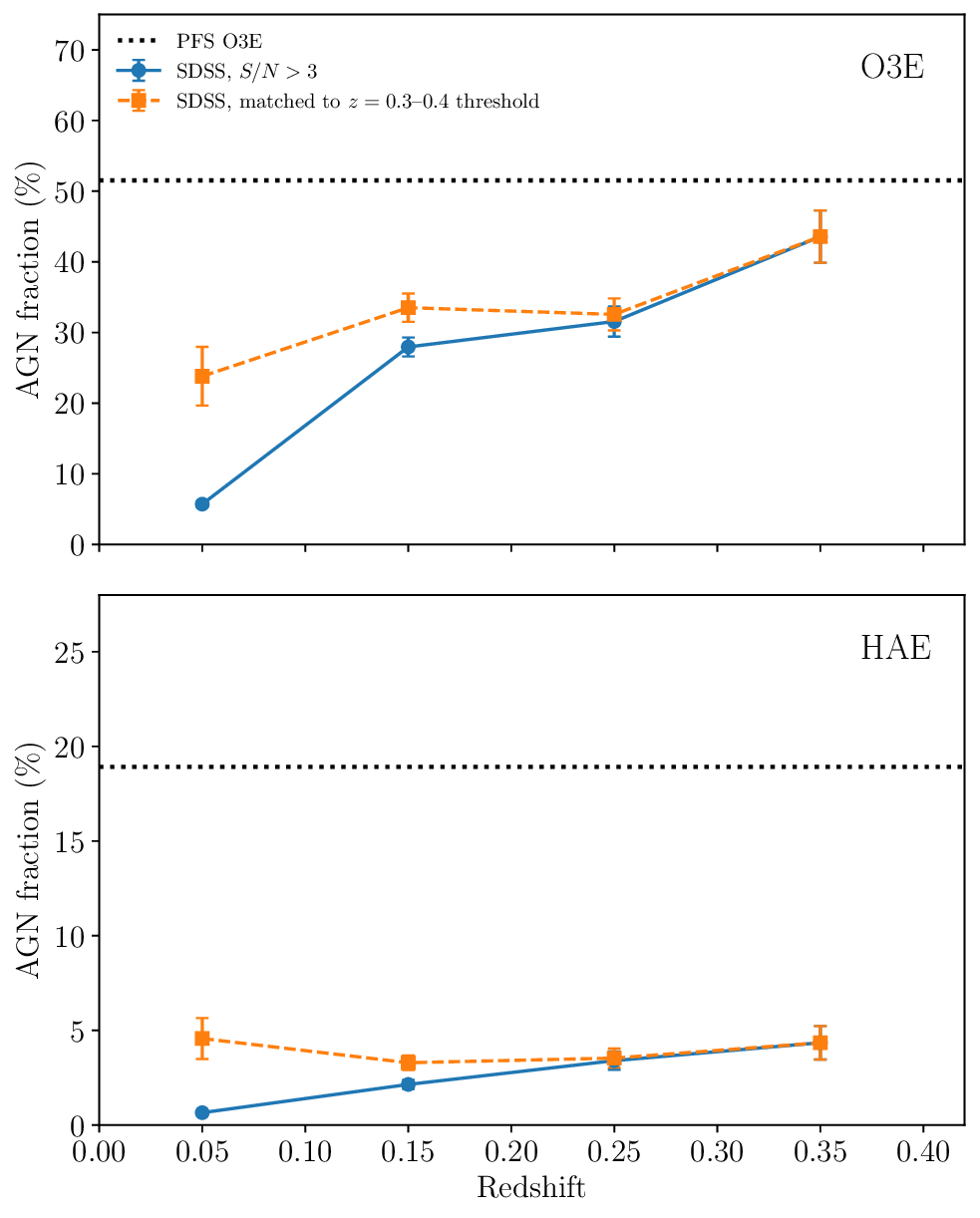}
\end{center}
\caption{
AGN fraction as a function of redshift for high-EW O3Es (top panel) and HAEs (bottom panel) in the SDSS/MPA--JHU sample. The blue curves show the AGN fraction measured from the original SDSS sample requiring S/N $>$ 3 detections in all four BPT emission lines. The orange curves show the AGN fraction after applying a common luminosity threshold corresponding to the minimum four-line detection limit in the highest-redshift SDSS bin ($0.3<z<0.4$), thereby reducing luminosity-dependent selection effects. Horizontal dotted lines indicate the AGN fractions measured in the PFS O3E (52 \%) and HAE (19 \%) samples. 

Alt text: AGN fraction versus redshift for high-EW O3Es and HAEs in SDSS. The apparent AGN fraction increases with redshift in the original four-line-selected sample, but this trend is substantially reduced after imposing a common luminosity threshold; the corresponding PFS AGN fractions remain higher than the SDSS values. 
}\label{fig:agnfraction}
\end{figure} 

Our PFS sample exhibits a substantially higher AGN fraction than that observed in local high-EW emission-line galaxies. Among the 70 sources with significant detections in all four BPT emission lines, 24 objects (34 \%) are classified as AGNs, compared to an AGN fraction of only 5.1 \% in the local high-EW emitter sample selected from the SDSS/MPA--JHU catalog.

A possible explanation is that the elevated AGN fraction simply reflects our selection of high-EW emitters. However, high equivalent width alone is unlikely to explain the observed difference. 
For example, \citet{Cardamone2009} studied Green Pea galaxies at $0.112<z<0.360$ and found that only 10 out of 103 narrow-line objects ($\sim$10\%) are classified as AGNs based on the BPT diagram, although an additional 13 objects were identified as composite/transition systems. 
Thus, even galaxy populations with extremely large emission-line equivalent widths are not necessarily dominated by AGN activity. The elevated AGN fraction observed in our sample therefore cannot be attributed solely to the selection of high-EW emitters.

To investigate the origin of the elevated AGN fraction, we examined the AGN fraction of SDSS high-EW emitters as a function of redshift (figure~\ref{fig:agnfraction}). The SDSS comparison is limited to $z\lesssim0.4$ because H$\alpha$ and [N\thinspace{\sc ii}] move beyond the SDSS spectral coverage at higher redshift. In the original SDSS sample, the AGN fraction increases strongly towards higher redshift for both O3Es and HAEs. However, much of this apparent increase disappears when a common luminosity threshold corresponding to the four-line detection limit in the highest-redshift SDSS bin ($0.3<z<0.4$) is applied. This result indicates that luminosity-dependent selection effects associated with the four-line BPT requirement contribute significantly to the observed increase in AGN fraction.
Indeed, the fraction of SDSS galaxies satisfying the four-line BPT criterion decreases from approximately 76 \% at $0.02<z<0.05$ to only 11 \% at $0.30<z<0.40$. Consequently, only the most luminous emission-line systems remain in the BPT-classifiable sample at higher redshift, naturally increasing the measured AGN fraction.
Nevertheless, some residual increase remains for the O3E population. Although the statistical significance of this trend is limited, it suggests that selection effects alone may not fully account for the observed behaviour.

A similar selection bias may also be present within the PFS sample itself. 
Among the 581 spectroscopically confirmed sources included in the rest-frame optical emission-line analysis, only 70 objects have significant detections in all four BPT lines. 
The principal limitation is the low detection rate of [N\thinspace{\sc ii}], which is detected in only 14 \% of the parent sample. Consequently, the measured AGN fraction should be interpreted as the AGN fraction among BPT-classifiable emitters rather than among all PFS-confirmed emitters.
However, the magnitude of this bias cannot currently be quantified. 
Most sources lacking four-line detections remain unclassified even when upper and lower limits are considered, preventing a robust estimate of the AGN fraction in the full parent sample. 
Therefore, while the SDSS comparison demonstrates that the four-line BPT selection can significantly enhance the observed AGN fraction, the magnitude of the corresponding bias in the PFS sample remains uncertain.
In addition, the PFS sample extends to substantially higher redshifts ($z\sim0.4$--1.0) than the SDSS comparison sample. 
Since the incidence of AGN activity is generally known to increase towards higher redshift, a contribution from cosmic evolution of AGN activity cannot be excluded. 
Therefore, while selection effects are likely to play an important role, intrinsic redshift evolution and/or differences in the underlying galaxy population may also contribute to the elevated AGN fraction observed in the PFS sample.

\subsection{Broad H$\alpha$ properties}\label{subsec:broad}

One object in our sample exhibits a clear broad H$\alpha$ component,
indicating the presence of a Type~1 AGN.
Using the broad H$\alpha$ luminosity and velocity width, we estimate
the black hole mass using a standard single-epoch virial estimator
\citep{GreeneHo2005}:

\begin{equation}
M_{\rm BH} =
2\times10^{6}\,M_\odot
\left(\frac{L_{\rm H\alpha,b}}{10^{42}\ {\rm erg\,s^{-1}}}\right)^{0.55}
\left(\frac{{\rm FWHM}_{\rm H\alpha,b}}{10^{3}\ {\rm km\,s^{-1}}}\right)^{2.06}.
\end{equation}

The bolometric luminosity is estimated assuming
\begin{equation}
L_{\rm bol} = 130\,L_{\rm H\alpha,b},
\end{equation}
following \citet{SternLaor2012}.
The Eddington luminosity is given by
\begin{equation}
L_{\rm Edd} =
1.26\times10^{38}
\left(\frac{M_{\rm BH}}{M_\odot}\right)
\ {\rm erg\,s^{-1}},
\end{equation}
and the Eddington ratio is defined as
\begin{equation}
\lambda_{\rm Edd} = \frac{L_{\rm bol}}{L_{\rm Edd}}.
\end{equation}

Using these relations, we derive a black hole mass of
$M_{\rm BH} = 8.8\times10^{6}\ M_{\odot}$ and an Eddington ratio of $\lambda_{\rm Edd} = 0.035$. 
These values place the source in the low-mass AGN regime with moderate accretion activity. 

We also assess the impact of aperture losses using the comparison between photometric total line fluxes and PFS fibre line fluxes described in section~4.2.
For this broad-line source, we obtain an aperture factor of $C_{\rm aper}=1.84$, somewhat larger than the median value for the full sample.
If applied directly, this correction would increase the broad H$\alpha$ luminosity by a factor of 1.84.
The corresponding virial black-hole mass and Eddington ratio would increase by factors of $1.84^{0.55}\simeq1.40$ and $1.84^{0.45}\simeq1.31$, respectively.
Because the photometric flux estimates themselves carry additional uncertainties, we do not apply this correction to the fiducial values quoted above; instead, these values provide an estimate of the maximum systematic uncertainty associated with aperture effects.

We note that, when compared with our \textsc{Cloudy} NLR grids, this object
lies close to the near-solar metallicity track ($12+\log({\rm O/H}) \simeq 8.7$; figure \ref{fig:bpt_cloudy}).
Given the model dependence of the NLR metallicity estimates and the fact that
this source is treated separately from the narrow-line sample, we do not
include it in the low-metallicity AGN subset discussed above.

Although only one object in our current sample shows a clear broad-line
component, this source provides a useful reference for the AGN population
identified via narrow-line diagnostics.
Future larger PFS samples will enable statistical constraints on the broad-line AGN population at similar redshifts (e.g., \citealt{Takada2014,Greene2022}).

\subsection{Constraints on outflow signatures from stacked spectra}

We investigate the presence of outflow signatures using the [O\thinspace{\sc iii}] emission-line profiles. 
Visual inspection of individual sources does not reveal any obvious signatures of asymmetric or broad line components. 
To characterise the average kinematic properties, we construct stacked spectra separately for the AGN and SFG subsamples. 
Before stacking, each spectrum is shifted to the rest frame and normalised by the median continuum flux density over the rest-frame $505$--$515$ nm range.
The normalised spectra are then interpolated onto a common rest-frame wavelength grid and median-stacked at each wavelength, while the associated uncertainty is estimated from the standard error of the median. 
For the AGN subsample, the object exhibiting clear broad emission-line components (see section~\ref{subsec:broad}) is excluded from the stacking in order to focus on the average properties of narrow-line regions. 

\begin{figure}
\begin{center}
\includegraphics[width=1\linewidth]{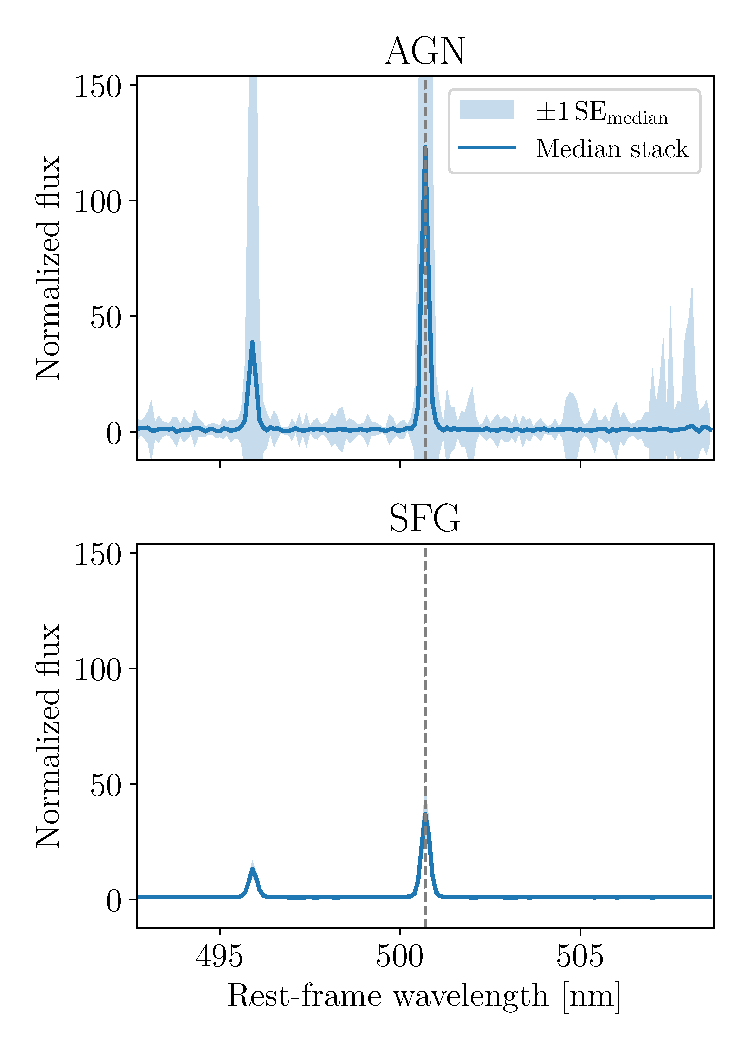}
\end{center}
\caption{
Stacked [O\thinspace{\sc iii}] emission-line profiles for the AGN (top) and SFG (bottom) subsamples.
Before stacking, each spectrum is shifted to the rest frame and normalised by the median continuum flux density over the rest-frame $505$--$515$ nm range.
The normalised spectra are median-stacked on a common rest-frame wavelength grid.
The shaded regions represent the approximate standard error of the median, estimated as $1.253 \sigma/\sqrt{N}$, where $\sigma$ is the standard deviation of the normalised flux densities and $N$ is the number of spectra contributing at each wavelength.
The uncertainty for the SFG stack is very small and is therefore not clearly visible. 
The vertical dashed line marks the rest-frame wavelength of [O\thinspace{\sc iii}]~$\lambda5007$. 

Alt text: Median-stacked [O\thinspace{\sc iii}] emission-line profiles for the AGN and SFG subsamples. The AGN stack has stronger [O\thinspace{\sc iii}] emission relative to the normalised continuum and larger uncertainty, but both stacks are approximately symmetric around [O\thinspace{\sc iii}] $\lambda5007$ without prominent broad or asymmetric wings.
}\label{fig:outflow}
\end{figure} 

Figure \ref{fig:outflow} shows the resulting stacked [O\thinspace{\sc iii}] emission-line profiles for the AGN and SFG samples. 
The [O\thinspace{\sc iii}] emission relative to the local continuum is stronger in the AGN stack than in the SFG stack, consistent with a harder ionising radiation field in the AGN subsample. 
The larger uncertainty in the AGN stack compared to the SFG stack is likely driven by the smaller sample size and the greater object-to-object diversity among AGN. 
Both profiles appear largely symmetric around the systemic wavelength, with no clear evidence for a prominent asymmetric wing component that would indicate strong outflows. 
To quantify this, we fitted the stacked [O\thinspace{\sc iii}] profiles with both single- and two-component Gaussian models. 
For the AGN stack, a single Gaussian provides an adequate description of the line profile, and the addition of a broad component is not supported by the BIC. 
Although the SFG stack is marginally better fitted by a two-component model, the amplitude of the broad component is less than 1 \% of that of the narrow component. 
This suggests that, on average, outflow-dominated kinematics do not play a major role in these systems. 
The absence of a significant broad or asymmetric [O {\sc iii}] component further suggests that strong AGN-driven outflows are not prevalent in this sample. 

We also visually inspected the stacked spectra to search for continuum features associated with the host galaxies. 
However, the continuum remains weak and no clear stellar absorption features are detected in the current stacks. 
A more detailed investigation of the continuum properties will require higher signal-to-noise data and is left for future work.

\section{Summary}
\label{sec:summary}

We have conducted PFS spectroscopic follow-up observations of high-EW NB selected targets to assess the range of physical properties exhibited by high-EW sources and to test whether such systems preferentially include low-metallicity galaxies and/or AGNs.

From the PFS spectra, we measured emission-line fluxes using velocity-window integration with local continuum subtraction, masking wavelength regions affected by strong OH residuals. For one object showing a clear broad H$\alpha$ component, we performed a dedicated multi-component fit to the H$\alpha$+[N\thinspace{\sc ii}] complex to obtain reliable narrow-line fluxes for placement on the BPT diagram.

Using the measured line ratios, we constructed BPT diagnostics for the subset of sources with significant detections in the four classical lines. 
We find that a non-negligible fraction of the sample lies in or near the valley between the star-forming and AGN branches. 
We also identify high-ionisation detections (e.g., He\thinspace{\sc ii} and [Ne\thinspace{\sc v}]) in nine emitters, indicating the presence of hard ionising spectra within the NB-selected population.  

Comparison to the \textsc{Cloudy} model suggests that the valley population is consistent with sub-solar metallicities. 
Some He\thinspace{\sc ii}- or [Ne\thinspace{\sc v}]-detected sources are found in the star-forming region with the lowest-metallicity region of the model grids, corresponding to $Z/Z_{\odot}\approx 0.25$--$0.5$ (or $12+\log({\rm O/H})\approx 8.1$--$8.4$). 
These results suggest that high-EW NB selection can efficiently identify systems with hard ionising sources, some of which may be candidates for metal-poor AGN. 

We examined the structural properties of the sample using the size ratio relative to the PSF. We find that the point-source fraction of our sample ($22\%\pm2\%$) lies between those of typical star-forming galaxies ($12\%\pm5\%$) and AGNs ($33\%\pm10\%$), indicating structural properties intermediate between the two populations. 

For the broad-line object, the broad H$\alpha$ luminosity and FWHM imply a black hole mass of $M_{\rm BH}=8.8\times10^{6}\,M_{\odot}$ and an Eddington ratio of $\lambda_{\rm Edd}=0.035$. This object lies close to the near-solar metallicity track in our adopted \textsc{Cloudy} grids ($12+\log({\rm O/H})\simeq 8.7$), and we treat it separately from the low-metallicity subset defined by the narrow-line high-ionisation detections. 

We also examined the [O\thinspace{\sc iii}] emission-line profiles using stacked spectra of the AGN and SFG subsamples. 
The resulting line profiles are largely symmetric, with no clear evidence for prominent asymmetric wings that would indicate strong outflows. 
This suggests that strong outflow signatures are not prevalent on average in the stacked subsamples. 
These results may be consistent with a scenario in which at least some of these objects represent relatively early or less evolved phases of AGN activity. 

Future larger PFS samples will enable tighter statistical constraints on the incidence and physical nature of high-EW AGN candidates, including the relative roles of metallicity, ionisation parameter, and mixed ionising sources in shaping their positions on diagnostic diagrams.

\section*{Funding}
This work was supported by the Japan Society for the Promotion of Science (JSPS) KAKENHI (JP22K14075 and JP24K00684 (H.U.); JP25H00663 (H.U. and Y.L.); JP23K25911 and JP25H00671 (T.N.); JP24K17084 (Y.L.); JP25K07361 (M.O.); and JP26K17200 (Y.S.)). 
A.R.A acknowledges partial support from Conselho Nacional de Desenvolvimento Científico e Tecnologico (CNPq) through grant 313739/2023-4.

\section*{Acknowledgements}

The instrument {\okina}Ōnohi{\okina}ula Prime Focus Spectrograph (PFS), including both hardware and software, was developed by the PFS collaboration consisting of over 25 institutes across multiple countries, where the technical
activities were conducted by (in alphabetical order) Academia Sinica Institute of Astronomy and Astrophysics (Taiwan), California Institute of Technology, Johns Hopkins University, Kavli Institute for the Physics and Mathematics of the Universe in the University of Tokyo (Kavli IPMU), Laboratoire d'Astrophysique de Marseille,
Laborat\'orio Nacional de Astrof\'isica (Brazil),
Max-Planck-Institut f\"ur Astrophysik,
Max-Planck-Institut f\"ur extraterrestrische Physik,
NASA Jet Propulsion Laboratory,
National Astronomical Observatory of Japan (NAOJ),
Princeton University, and Universidade de S\~ao Paulo under the oversight by Project Office hosted by Kavli IPMU (later NAOJ).
There were also essential commitments from academic and industrial partners such as Durham University (United Kingdom) and Bertin Technologies (France).

This work is based in part on data collected at the Subaru Telescope, which is operated by the National Astronomical Observatory of Japan. 
We are honored and grateful for the opportunity of observing the Universe from Maunakea, which has the cultural, historical, and natural significance in Hawaii. 

We appreciate the development and operation of PFS Science Platform by Subaru Telescope and Astronomy Data Center at NAOJ, which enables access to both PFS and HSC data and various analyses on the server side. 

The Hyper Suprime-Cam (HSC) collaboration includes the astronomical communities of Japan and Taiwan, and Princeton University.  The HSC instrumentation and software were developed by the National Astronomical Observatory of Japan (NAOJ), the Kavli Institute for the Physics and Mathematics of the Universe (Kavli IPMU), the University of Tokyo, the High Energy Accelerator Research Organization (KEK), the Academia Sinica Institute for Astronomy and Astrophysics in Taiwan (ASIAA), and Princeton University.  Funding was contributed by the FIRST programme from the Japanese Cabinet Office, the Ministry of Education, Culture, Sports, Science and Technology (MEXT), the Japan Society for the Promotion of Science (JSPS), Japan Science and Technology Agency  (JST), the Toray Science  Foundation, NAOJ, Kavli IPMU, KEK, ASIAA, and Princeton University.
This paper is based in part on data collected at the Subaru Telescope and retrieved from the HSC data archive system, which is operated by Subaru Telescope and Astronomy Data Center (ADC) at NAOJ. Data analysis was in part carried out with the cooperation of Center for Computational Astrophysics (CfCA) at NAOJ.  We are honored and grateful for the opportunity of observing the Universe from Maunakea, which has the cultural, historical and natural significance in Hawaii.

This paper makes use of software developed for Vera C. Rubin Observatory. We thank the Rubin Observatory for making their code available as free software at http://pipelines.lsst.io/. 

The Pan-STARRS1 Surveys (PS1) and the PS1 public science archive have been made possible through contributions by the Institute for Astronomy, the University of Hawaii, the Pan-STARRS Project Office, the Max Planck Society and its participating institutes, the Max Planck Institute for Astronomy, Heidelberg, and the Max Planck Institute for Extraterrestrial Physics, Garching, The Johns Hopkins University, Durham University, the University of Edinburgh, the Queen’s University Belfast, the Harvard-Smithsonian Center for Astrophysics, the Las Cumbres Observatory Global Telescope Network Incorporated, the National Central University of Taiwan, the Space Telescope Science Institute, the National Aeronautics and Space Administration under grant No. NNX08AR22G issued through the Planetary Science Division of the NASA Science Mission Directorate, the National Science Foundation grant No. AST-1238877, the University of Maryland, Eotvos Lorand University (ELTE), the Los Alamos National Laboratory, and the Gordon and Betty Moore Foundation.

ChatGPT (OpenAI; GPT-5 series) was used to assist with English wording and with the development, review, and debugging of Python code. All suggestions, code, and resulting outputs were reviewed and verified by the authors.

\appendix

\section{Colour Selection for the NB1010 Sample}\label{sec:nb1010_colour}

\begin{figure*}
\begin{center}
\includegraphics[width=0.6\linewidth]{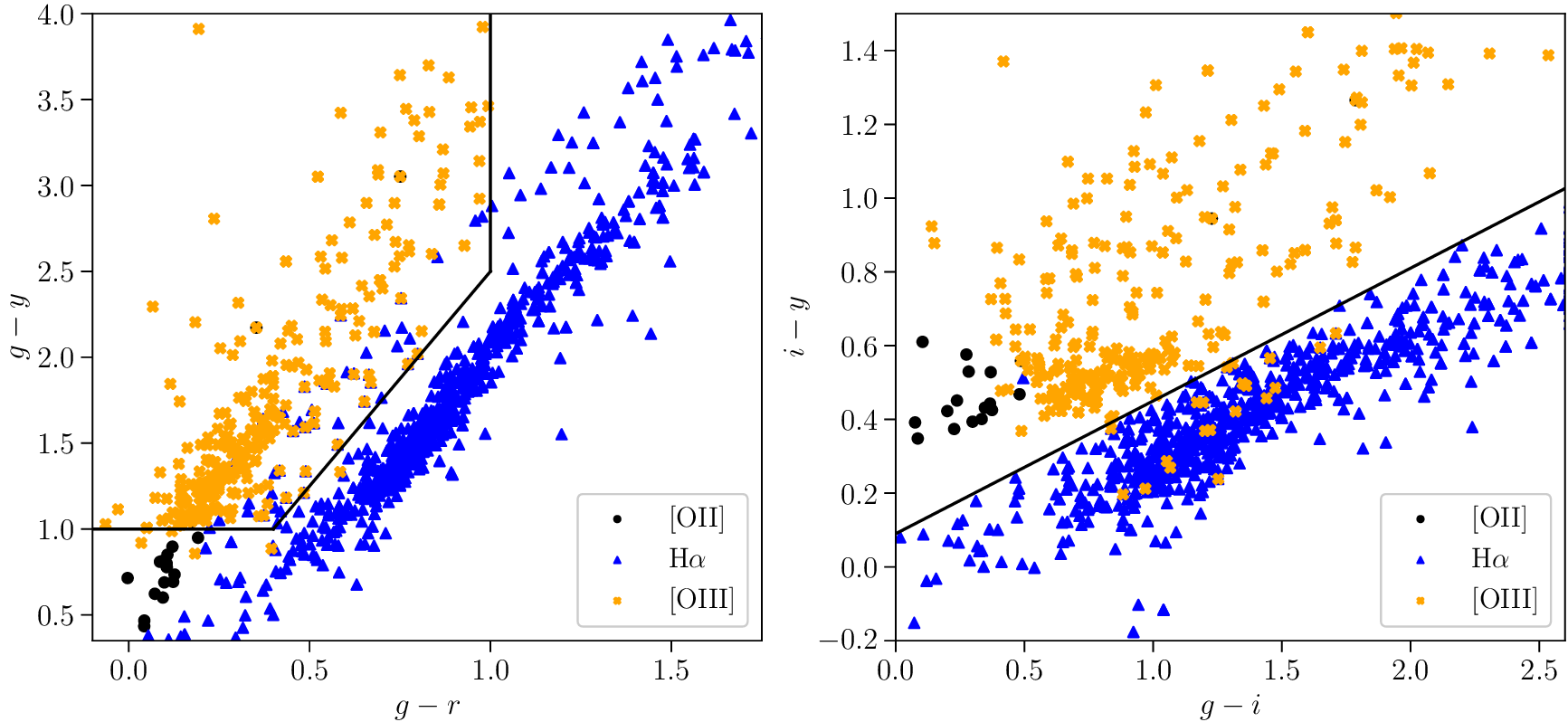}
\end{center}
\caption{Colour--colour diagrams for spectroscopically confirmed emitters in the NB1010 sample. 
Left: $(g-r)$ versus $(g-y)$. Right: $(g-i)$ versus $(i-y)$. 
Black circles, blue triangles, and orange squares indicate [O\thinspace{\sc ii}], H$\alpha$, and [O\thinspace{\sc iii}] emitters, respectively. 
The solid lines show the adopted colour-selection boundaries used to identify candidate emitters. 

Alt text: Two colour--colour diagrams for spectroscopically confirmed NB1010 emitters. [O\thinspace{\sc ii}], H$\alpha$, and [O\thinspace{\sc iii}] emitters occupy distinct but partially overlapping regions, with the adopted linear boundaries separating most H$\alpha$ and [O\thinspace{\sc iii}]  candidates from the other emitter populations.
}\label{fig:colour_colour_NB1010}
\end{figure*}

In the NB1010 selection, we apply colour criteria to separate candidate emitters in colour--colour space. 
Figure~\ref{fig:colour_colour_NB1010} illustrates the distributions of spectroscopically identified [O\thinspace{\sc ii}], H$\alpha$, and [O\thinspace{\sc iii}] emitters on the $(g-r)$--$(g-y)$ and $(g-i)$--$(i-y)$ planes, together with the adopted selection boundaries. 
Based on these distributions, [O\thinspace{\sc iii}] candidates were selected using
\begin{subequations}
\begin{align}
g - y &> 1.0, \\
g - y &> 2.5\,(g - r), \\
g - r &< 1.0,
\end{align}
\end{subequations}
while H$\alpha$ candidates were selected using
\begin{equation}
i - y < 0.36\,(g - i) + 0.09.
\end{equation}
These criteria are used as an auxiliary colour selection for the NB1010 sample.

\bibliographystyle{aasjournal}
\bibliography{refs}

@ARTICLE{Kauffmann2003,
       author = {{Kauffmann}, Guinevere and {Heckman}, Timothy M. and
                 {Tremonti}, Christy and {Brinchmann}, Jarle and
                 {Charlot}, St{\'e}phane and {White}, Simon D. M. and
                 {Ridgway}, Susan E. and {Brinkmann}, Jon and
                 {Fukugita}, Masataka and {Hall}, Patrick B. and
                 {Ivezi{\'c}}, {\v Z}eljko and {Richards}, Gordon T. and
                 {Schneider}, Donald P.},
        title = "{The host galaxies of active galactic nuclei}",
      journal = {\mnras},
         year = 2003,
        month = dec,
       volume = {346},
       number = {4},
        pages = {1055--1077},
          doi = {10.1111/j.1365-2966.2003.07154.x},
       eprint = {astro-ph/0304239},
 archivePrefix = {arXiv}
}

@ARTICLE{Aihara2022,
       author = {{Aihara}, Hiroaki and {AlSayyad}, Yusra and {Ando}, Makoto and {Armstrong}, Robert and {Bosch}, James and {Egami}, Eiichi and {Furusawa}, Hisanori and {Furusawa}, Junko and {Harasawa}, Sumiko and {Harikane}, Yuichi and {Hsieh}, Bau-Ching and {Ikeda}, Hiroyuki and {Ito}, Kei and {Iwata}, Ikuru and {Kodama}, Tadayuki and {Koike}, Michitaro and {Kokubo}, Mitsuru and {Komiyama}, Yutaka and {Li}, Xiangchong and {Liang}, Yongming and {Lin}, Yen-Ting and {Lupton}, Robert H. and {Lust}, Nate B. and {MacArthur}, Lauren A. and {Mawatari}, Ken and {Mineo}, Sogo and {Miyatake}, Hironao and {Miyazaki}, Satoshi and {More}, Surhud and {Morishima}, Takahiro and {Murayama}, Hitoshi and {Nakajima}, Kimihiko and {Nakata}, Fumiaki and {Nishizawa}, Atsushi J. and {Oguri}, Masamune and {Okabe}, Nobuhiro and {Okura}, Yuki and {Ono}, Yoshiaki and {Osato}, Ken and {Ouchi}, Masami and {Pan}, Yen-Chen and {Plazas Malag{\'o}n}, Andr{\'e}s A. and {Price}, Paul A. and {Reed}, Sophie L. and {Rykoff}, Eli S. and {Shibuya}, Takatoshi and {Simunovic}, Mirko and {Strauss}, Michael A. and {Sugimori}, Kanako and {Suto}, Yasushi and {Suzuki}, Nao and {Takada}, Masahiro and {Takagi}, Yuhei and {Takata}, Tadafumi and {Takita}, Satoshi and {Tanaka}, Masayuki and {Tang}, Shenli and {Taranu}, Dan S. and {Terai}, Tsuyoshi and {Toba}, Yoshiki and {Turner}, Edwin L. and {Uchiyama}, Hisakazu and {Vijarnwannaluk}, Bovornpratch and {Waters}, Christopher Z. and {Yamada}, Yoshihiko and {Yamamoto}, Naoaki and {Yamashita}, Takuji},
        title = "{Third data release of the Hyper Suprime-Cam Subaru Strategic Program}",
      journal = {\pasj},
         year = 2022,
        month = apr,
       volume = {74},
       number = {2},
        pages = {247-272},
          doi = {10.1093/pasj/psab122},
archivePrefix = {arXiv},
       eprint = {2108.13045},
 primaryClass = {astro-ph.IM},
       adsurl = {https://ui.adsabs.harvard.edu/abs/2022PASJ...74..247A}
}

@ARTICLE{Aihara2019,
       author = {{Aihara}, Hiroaki and {AlSayyad}, Yusra and {Ando}, Makoto and {Armstrong}, Robert and {Bosch}, James and {Egami}, Eiichi and {Furusawa}, Hisanori and {Furusawa}, Junko and {Goulding}, Andy and {Harikane}, Yuichi and {Hikage}, Chiaki and {Ho}, Paul T.~P. and {Hsieh}, Bau-Ching and {Huang}, Song and {Ikeda}, Hiroyuki and {Imanishi}, Masatoshi and {Ito}, Kei and {Iwata}, Ikuru and {Jaelani}, Anton T. and {Kakuma}, Ryota and {Kawana}, Kojiro and {Kikuta}, Satoshi and {Kobayashi}, Umi and {Koike}, Michitaro and {Komiyama}, Yutaka and {Li}, Xiangchong and {Liang}, Yongming and {Lin}, Yen-Ting and {Luo}, Wentao and {Lupton}, Robert and {Lust}, Nate B. and {MacArthur}, Lauren A. and {Matsuoka}, Yoshiki and {Mineo}, Sogo and {Miyatake}, Hironao and {Miyazaki}, Satoshi and {More}, Surhud and {Murata}, Ryoma and {Namiki}, Shigeru V. and {Nishizawa}, Atsushi J. and {Oguri}, Masamune and {Okabe}, Nobuhiro and {Okamoto}, Sakurako and {Okura}, Yuki and {Ono}, Yoshiaki and {Onodera}, Masato and {Onoue}, Masafusa and {Osato}, Ken and {Ouchi}, Masami and {Shibuya}, Takatoshi and {Strauss}, Michael A. and {Sugiyama}, Naoshi and {Suto}, Yasushi and {Takada}, Masahiro and {Takagi}, Yuhei and {Takata}, Tadafumi and {Takita}, Satoshi and {Tanaka}, Masayuki and {Terai}, Tsuyoshi and {Toba}, Yoshiki and {Uchiyama}, Hisakazu and {Utsumi}, Yousuke and {Wang}, Shiang-Yu and {Wang}, Wenting and {Yamada}, Yoshihiko},
        title = "{Second data release of the Hyper Suprime-Cam Subaru Strategic Program}",
      journal = {\pasj},
         year = 2019,
        month = dec,
       volume = {71},
       number = {6},
          eid = {114},
        pages = {114},
          doi = {10.1093/pasj/psz103},
archivePrefix = {arXiv},
       eprint = {1905.12221},
 primaryClass = {astro-ph.IM},
       adsurl = {https://ui.adsabs.harvard.edu/abs/2019PASJ...71..114A}
}

@ARTICLE{HaringRix2004,
       author = {{H{\"a}ring}, Nadine and {Rix}, Hans-Walter},
        title = "{On the Black Hole Mass-Bulge Mass Relation}",
      journal = {\apjl},
         year = 2004,
        month = apr,
       volume = {604},
       number = {2},
        pages = {L89-L92},
          doi = {10.1086/383567},
archivePrefix = {arXiv},
       eprint = {astro-ph/0402376},
 primaryClass = {astro-ph},
       adsurl = {https://ui.adsabs.harvard.edu/abs/2004ApJ...604L..89H}
}

@ARTICLE{KormendyHo2013,
       author = {{Kormendy}, John and {Ho}, Luis C.},
        title = "{Coevolution (Or Not) of Supermassive Black Holes and Host Galaxies}",
      journal = {\araa},
         year = 2013,
        month = aug,
       volume = {51},
       number = {1},
        pages = {511-653},
          doi = {10.1146/annurev-astro-082708-101811},
archivePrefix = {arXiv},
       eprint = {1304.7762},
 primaryClass = {astro-ph.CO},
       adsurl = {https://ui.adsabs.harvard.edu/abs/2013ARA&A..51..511K}
}

@ARTICLE{GrahamSahu2023,
       author = {{Graham}, Alister W. and {Sahu}, Nandini},
        title = "{Appreciating mergers for understanding the non-linear M$_{bh}$-M$_{*,spheroid}$ and M$_{bh}$-M$_{*, galaxy}$ relations, updated herein, and the implications for the (reduced) role of AGN feedback}",
      journal = {\mnras},
         year = 2023,
        month = jan,
       volume = {518},
       number = {2},
        pages = {2177-2200},
          doi = {10.1093/mnras/stac2019},
archivePrefix = {arXiv},
       eprint = {2209.14526},
 primaryClass = {astro-ph.GA},
       adsurl = {https://ui.adsabs.harvard.edu/abs/2023MNRAS.518.2177G}
}

@ARTICLE{Cardamone2009,
       author = {{Cardamone}, Carolin and {Schawinski}, Kevin and {Sarzi}, Marc and {Bamford}, Steven P. and {Bennert}, Nicola and {Urry}, C.~M. and {Lintott}, Chris and {Keel}, William C. and {Parejko}, John and {Nichol}, Robert C. and {Thomas}, Daniel and {Andreescu}, Dan and {Murray}, Phil and {Raddick}, M. Jordan and {Slosar}, An{\v{z}}e and {Szalay}, Alex and {Vandenberg}, Jan},
        title = "{Galaxy Zoo Green Peas: discovery of a class of compact extremely star-forming galaxies}",
      journal = {\mnras},
         year = 2009,
        month = nov,
       volume = {399},
       number = {3},
        pages = {1191-1205},
          doi = {10.1111/j.1365-2966.2009.15383.x},
archivePrefix = {arXiv},
       eprint = {0907.4155},
 primaryClass = {astro-ph.CO},
       adsurl = {https://ui.adsabs.harvard.edu/abs/2009MNRAS.399.1191C}
}

@ARTICLE{Lanzuisi2018,
       author = {{Lanzuisi}, G. and {Civano}, F. and {Marchesi}, S. and {Comastri}, A. and {Brusa}, M. and {Gilli}, R. and {Vignali}, C. and {Zamorani}, G. and {Brightman}, M. and {Griffiths}, R.~E. and {Koekemoer}, A.~M.},
        title = "{The Chandra COSMOS Legacy Survey: Compton thick AGN at high redshift}",
      journal = {\mnras},
         year = 2018,
        month = oct,
       volume = {480},
       number = {2},
        pages = {2578-2592},
          doi = {10.1093/mnras/sty2025},
archivePrefix = {arXiv},
       eprint = {1803.08547},
 primaryClass = {astro-ph.GA},
       adsurl = {https://ui.adsabs.harvard.edu/abs/2018MNRAS.480.2578L}
}

@ARTICLE{Scarlata2007,
       author = {{Scarlata}, C. and {Carollo}, C.~M. and {Lilly}, S. and {Sargent}, M.~T. and {Feldmann}, R. and {Kampczyk}, P. and {Porciani}, C. and {Koekemoer}, A. and {Scoville}, N. and {Kneib}, J.-P. and {Leauthaud}, A. and {Massey}, R. and {Rhodes}, J. and {Tasca}, L. and {Capak}, P. and {Maier}, C. and {McCracken}, H.~J. and {Mobasher}, B. and {Renzini}, A. and {Taniguchi}, Y. and {Thompson}, D. and {Sheth}, K. and {Ajiki}, M. and {Aussel}, H. and {Murayama}, T. and {Sanders}, D.~B. and {Sasaki}, S. and {Shioya}, Y. and {Takahashi}, M.},
        title = "{COSMOS Morphological Classification with the Zurich Estimator of Structural Types (ZEST) and the Evolution Since z = 1 of the Luminosity Function of Early, Disk, and Irregular Galaxies}",
      journal = {\apjs},
         year = 2007,
        month = sep,
       volume = {172},
       number = {1},
        pages = {406-433},
          doi = {10.1086/516582},
       adsurl = {https://ui.adsabs.harvard.edu/abs/2007ApJS..172..406S}
}

@ARTICLE{Civano2016,
       author = {{Civano}, F. and {Marchesi}, S. and {Comastri}, A. and {Urry}, M.~C. and {Elvis}, M. and {Cappelluti}, N. and {Puccetti}, S. and {Brusa}, M. and {Zamorani}, G. and {Hasinger}, G. and {Aldcroft}, T. and {Alexander}, D.~M. and {Allevato}, V. and {Brunner}, H. and {Capak}, P. and {Finoguenov}, A. and {Fiore}, F. and {Fruscione}, A. and {Gilli}, R. and {Glotfelty}, K. and {Griffiths}, R.~E. and {Hao}, H. and {Harrison}, F.~A. and {Jahnke}, K. and {Kartaltepe}, J. and {Karim}, A. and {LaMassa}, S.~M. and {Lanzuisi}, G. and {Miyaji}, T. and {Ranalli}, P. and {Salvato}, M. and {Sargent}, M. and {Scoville}, N.~J. and {Schawinski}, K. and {Schinnerer}, E. and {Silverman}, J. and {Smolcic}, V. and {Stern}, D. and {Toft}, S. and {Trakhtenbrot}, B. and {Treister}, E. and {Vignali}, C.},
        title = "{The Chandra Cosmos Legacy Survey: Overview and Point Source Catalog}",
      journal = {\apj},
         year = 2016,
        month = mar,
       volume = {819},
       number = {1},
          eid = {62},
        pages = {62},
          doi = {10.3847/0004-637X/819/1/62},
archivePrefix = {arXiv},
       eprint = {1601.00941},
 primaryClass = {astro-ph.GA},
       adsurl = {https://ui.adsabs.harvard.edu/abs/2016ApJ...819...62C}
}

@ARTICLE{Marchesi2016,
       author = {{Marchesi}, S. and {Civano}, F. and {Elvis}, M. and {Salvato}, M. and {Brusa}, M. and {Comastri}, A. and {Gilli}, R. and {Hasinger}, G. and {Lanzuisi}, G. and {Miyaji}, T. and {Treister}, E. and {Urry}, C.~M. and {Vignali}, C. and {Zamorani}, G. and {Allevato}, V. and {Cappelluti}, N. and {Cardamone}, C. and {Finoguenov}, A. and {Griffiths}, R.~E. and {Karim}, A. and {Laigle}, C. and {LaMassa}, S.~M. and {Jahnke}, K. and {Ranalli}, P. and {Schawinski}, K. and {Schinnerer}, E. and {Silverman}, J.~D. and {Smolcic}, V. and {Suh}, H. and {Trakhtenbrot}, B.},
        title = "{The Chandra COSMOS Legacy survey: optical/IR identifications}",
      journal = {\apj},
         year = 2016,
        month = jan,
       volume = {817},
       number = {1},
          eid = {34},
        pages = {34},
          doi = {10.3847/0004-637X/817/1/34},
archivePrefix = {arXiv},
       eprint = {1512.01105},
 primaryClass = {astro-ph.GA},
       adsurl = {https://ui.adsabs.harvard.edu/abs/2016ApJ...817...34M}
}

@ARTICLE{Steidel2014,
       author = {{Steidel}, Charles C. and {Rudie}, Gwen C. and {Strom}, Allison L. and {Pettini}, Max and {Reddy}, Naveen A. and {Shapley}, Alice E. and {Trainor}, Ryan F. and {Erb}, Dawn K. and {Turner}, Monica L. and {Konidaris}, Nicholas P. and {Kulas}, Kristin R. and {Mace}, Gregory and {Matthews}, Keith and {McLean}, Ian S.},
        title = "{Strong Nebular Line Ratios in the Spectra of z \raisebox{-0.5ex}\textasciitilde 2-3 Star Forming Galaxies: First Results from KBSS-MOSFIRE}",
      journal = {\apj},
         year = 2014,
        month = nov,
       volume = {795},
       number = {2},
          eid = {165},
        pages = {165},
          doi = {10.1088/0004-637X/795/2/165},
archivePrefix = {arXiv},
       eprint = {1405.5473},
 primaryClass = {astro-ph.GA},
       adsurl = {https://ui.adsabs.harvard.edu/abs/2014ApJ...795..165S}
}

@ARTICLE{Shapley2015,
       author = {{Shapley}, Alice E. and {Reddy}, Naveen A. and {Kriek}, Mariska and {Freeman}, William R. and {Sanders}, Ryan L. and {Siana}, Brian and {Coil}, Alison L. and {Mobasher}, Bahram and {Shivaei}, Irene and {Price}, Sedona H. and {de Groot}, Laura},
        title = "{The MOSDEF Survey: Excitation Properties of z {\ensuremath{\sim}} 2.3 Star-forming Galaxies}",
      journal = {\apj},
         year = 2015,
        month = mar,
       volume = {801},
       number = {2},
          eid = {88},
        pages = {88},
          doi = {10.1088/0004-637X/801/2/88},
archivePrefix = {arXiv},
       eprint = {1409.7071},
 primaryClass = {astro-ph.GA},
       adsurl = {https://ui.adsabs.harvard.edu/abs/2015ApJ...801...88S}
}

@ARTICLE{Sanders2016,
       author = {{Sanders}, Ryan L. and {Shapley}, Alice E. and {Kriek}, Mariska and {Reddy}, Naveen A. and {Freeman}, William R. and {Coil}, Alison L. and {Siana}, Brian and {Mobasher}, Bahram and {Shivaei}, Irene and {Price}, Sedona H. and {de Groot}, Laura},
        title = "{The MOSDEF Survey: Electron Density and Ionization Parameter at z \raisebox{-0.5ex}\textasciitilde 2.3}",
      journal = {\apj},
         year = 2016,
        month = jan,
       volume = {816},
       number = {1},
          eid = {23},
        pages = {23},
          doi = {10.3847/0004-637X/816/1/23},
archivePrefix = {arXiv},
       eprint = {1509.03636},
 primaryClass = {astro-ph.GA},
       adsurl = {https://ui.adsabs.harvard.edu/abs/2016ApJ...816...23S}
}

@ARTICLE{Strom2017,
       author = {{Strom}, Allison L. and {Steidel}, Charles C. and {Rudie}, Gwen C. and {Trainor}, Ryan F. and {Pettini}, Max and {Reddy}, Naveen A.},
        title = "{Nebular Emission Line Ratios in z ≃ 2-3 Star-forming Galaxies with KBSS-MOSFIRE: Exploring the Impact of Ionization, Excitation, and Nitrogen-to-Oxygen Ratio}",
      journal = {\apj},
         year = 2017,
        month = feb,
       volume = {836},
       number = {2},
          eid = {164},
        pages = {164},
          doi = {10.3847/1538-4357/836/2/164},
archivePrefix = {arXiv},
       eprint = {1608.02587},
 primaryClass = {astro-ph.GA},
       adsurl = {https://ui.adsabs.harvard.edu/abs/2017ApJ...836..164S}
}

@ARTICLE{Shapley2019,
       author = {{Shapley}, Alice E. and {Sanders}, Ryan L. and {Shao}, Peng and {Reddy}, Naveen A. and {Kriek}, Mariska and {Coil}, Alison L. and {Mobasher}, Bahram and {Siana}, Brian and {Shivaei}, Irene and {Freeman}, William R. and {Azadi}, Mojegan and {Price}, Sedona H. and {Leung}, Gene C.~K. and {Fetherolf}, Tara and {de Groot}, Laura and {Zick}, Tom and {Fornasini}, Francesca M. and {Barro}, Guillermo},
        title = "{The MOSDEF Survey: Sulfur Emission-line Ratios Provide New Insights into Evolving Interstellar Medium Conditions at High Redshift}",
      journal = {\apjl},
         year = 2019,
        month = aug,
       volume = {881},
       number = {2},
          eid = {L35},
        pages = {L35},
          doi = {10.3847/2041-8213/ab385a},
archivePrefix = {arXiv},
       eprint = {1907.07189},
 primaryClass = {astro-ph.GA},
       adsurl = {https://ui.adsabs.harvard.edu/abs/2019ApJ...881L..35S}
}

@ARTICLE{Shirazi2012,
       author = {{Shirazi}, Maryam and {Brinchmann}, Jarle},
        title = "{Strongly star forming galaxies in the local Universe with nebular He II{\ensuremath{\lambda}}4686 emission}",
      journal = {\mnras},
         year = 2012,
        month = apr,
       volume = {421},
       number = {2},
        pages = {1043-1063},
          doi = {10.1111/j.1365-2966.2012.20439.x},
archivePrefix = {arXiv},
       eprint = {1201.1290},
 primaryClass = {astro-ph.CO},
       adsurl = {https://ui.adsabs.harvard.edu/abs/2012MNRAS.421.1043S}
}

@ARTICLE{Feltre2016,
       author = {{Feltre}, A. and {Charlot}, S. and {Gutkin}, J.},
        title = "{Nuclear activity versus star formation: emission-line diagnostics at ultraviolet and optical wavelengths}",
      journal = {\mnras},
         year = 2016,
        month = mar,
       volume = {456},
       number = {3},
        pages = {3354-3374},
          doi = {10.1093/mnras/stv2794},
archivePrefix = {arXiv},
       eprint = {1511.08217},
 primaryClass = {astro-ph.GA},
       adsurl = {https://ui.adsabs.harvard.edu/abs/2016MNRAS.456.3354F}
}

@ARTICLE{Brinchmann2004,
       author = {{Brinchmann}, J. and {Charlot}, S. and {White}, S.~D.~M. and {Tremonti}, C. and {Kauffmann}, G. and {Heckman}, T. and {Brinkmann}, J.},
        title = "{The physical properties of star-forming galaxies in the low-redshift Universe}",
      journal = {\mnras},
         year = 2004,
        month = jul,
       volume = {351},
       number = {4},
        pages = {1151-1179},
          doi = {10.1111/j.1365-2966.2004.07881.x},
archivePrefix = {arXiv},
       eprint = {astro-ph/0311060},
 primaryClass = {astro-ph},
       adsurl = {https://ui.adsabs.harvard.edu/abs/2004MNRAS.351.1151B}
}

@ARTICLE{Xu2018,
       author = {{Xu}, Fei and {Bian}, Fuyan and {Shen}, Yue and {Zuo}, Wenwen and {Fan}, Xiaohui and {Zhu}, Zonghong},
        title = "{The evolution of chemical abundance in quasar broad line region}",
      journal = {\mnras},
         year = 2018,
        month = oct,
       volume = {480},
       number = {1},
        pages = {345-357},
          doi = {10.1093/mnras/sty1763},
archivePrefix = {arXiv},
       eprint = {1807.01978},
 primaryClass = {astro-ph.GA},
       adsurl = {https://ui.adsabs.harvard.edu/abs/2018MNRAS.480..345X}
}

@ARTICLE{Morokuma2016,
       author = {{Morokuma}, Tomoki and {Tominaga}, Nozomu and {Tanaka}, Masaomi and {Yasuda}, Naoki and {Furusawa}, Hisanori and {Taniguchi}, Yuki and {Kato}, Takahiro and {Jiang}, Ji-an and {Nagao}, Tohru and {Kuncarayakti}, Hanindyo and {Morokuma-Matsui}, Kana and {Ikeda}, Hiroyuki and {Blinnikov}, Sergei and {Nomoto}, Ken'ichi and {Kokubo}, Mitsuru and {Doi}, Mamoru},
        title = "{An effective selection method for low-mass active black holes and first spectroscopic identification}",
      journal = {\pasj},
         year = 2016,
        month = jun,
       volume = {68},
       number = {3},
          eid = {40},
        pages = {40},
          doi = {10.1093/pasj/psw033},
archivePrefix = {arXiv},
       eprint = {1603.02302},
 primaryClass = {astro-ph.GA},
       adsurl = {https://ui.adsabs.harvard.edu/abs/2016PASJ...68...40M}
}

@ARTICLE{Greene2020,
       author = {{Greene}, Jenny E. and {Strader}, Jay and {Ho}, Luis C.},
        title = "{Intermediate-Mass Black Holes}",
      journal = {\araa},
         year = 2020,
        month = aug,
       volume = {58},
        pages = {257-312},
          doi = {10.1146/annurev-astro-032620-021835},
archivePrefix = {arXiv},
       eprint = {1911.09678},
 primaryClass = {astro-ph.GA},
       adsurl = {https://ui.adsabs.harvard.edu/abs/2020ARA&A..58..257G}
}

@ARTICLE{McConnellMa2013,
       author = {{McConnell}, Nicholas J. and {Ma}, Chung-Pei},
        title = "{Revisiting the Scaling Relations of Black Hole Masses and Host Galaxy Properties}",
      journal = {\apj},
         year = 2013,
        month = feb,
       volume = {764},
       number = {2},
          eid = {184},
        pages = {184},
          doi = {10.1088/0004-637X/764/2/184},
archivePrefix = {arXiv},
       eprint = {1211.2816},
 primaryClass = {astro-ph.CO},
       adsurl = {https://ui.adsabs.harvard.edu/abs/2013ApJ...764..184M}
}

@ARTICLE{Ding2023,
       author = {{Ding}, Xuheng and {Onoue}, Masafusa and {Silverman}, John D. and {Matsuoka}, Yoshiki and {Izumi}, Takuma and {Strauss}, Michael A. and {Jahnke}, Knud and {Phillips}, Camryn L. and {Li}, Junyao and {Volonteri}, Marta and {Haiman}, Zoltan and {Andika}, Irham Taufik and {Aoki}, Kentaro and {Baba}, Shunsuke and {Bieri}, Rebekka and {Bosman}, Sarah E.~I. and {Bottrell}, Connor and {Eilers}, Anna-Christina and {Fujimoto}, Seiji and {Habouzit}, Melanie and {Imanishi}, Masatoshi and {Inayoshi}, Kohei and {Iwasawa}, Kazushi and {Kashikawa}, Nobunari and {Kawaguchi}, Toshihiro and {Kohno}, Kotaro and {Lee}, Chien-Hsiu and {Lupi}, Alessandro and {Lyu}, Jianwei and {Nagao}, Tohru and {Overzier}, Roderik and {Schindler}, Jan-Torge and {Schramm}, Malte and {Shimasaku}, Kazuhiro and {Toba}, Yoshiki and {Trakhtenbrot}, Benny and {Trebitsch}, Maxime and {Treu}, Tommaso and {Umehata}, Hideki and {Venemans}, Bram P. and {Vestergaard}, Marianne and {Walter}, Fabian and {Wang}, Feige and {Yang}, Jinyi},
        title = "{Detection of stellar light from quasar host galaxies at redshifts above 6}",
      journal = {\nat},
         year = 2023,
        month = sep,
       volume = {621},
       number = {7977},
        pages = {51-55},
          doi = {10.1038/s41586-023-06345-5},
archivePrefix = {arXiv},
       eprint = {2211.14329},
 primaryClass = {astro-ph.GA},
       adsurl = {https://ui.adsabs.harvard.edu/abs/2023Natur.621...51D}
}

@ARTICLE{Stone2024,
       author = {{Stone}, Meredith A. and {Lyu}, Jianwei and {Rieke}, George H. and {Alberts}, Stacey and {Hainline}, Kevin N.},
        title = "{Undermassive Host Galaxies of Five z {\ensuremath{\sim}} 6 Luminous Quasars Detected with JWST}",
      journal = {\apj},
         year = 2024,
        month = mar,
       volume = {964},
       number = {1},
          eid = {90},
        pages = {90},
          doi = {10.3847/1538-4357/ad2a57},
archivePrefix = {arXiv},
       eprint = {2310.18395},
 primaryClass = {astro-ph.GA},
       adsurl = {https://ui.adsabs.harvard.edu/abs/2024ApJ...964...90S}
}

@ARTICLE{Pacucci2024,
       author = {{Pacucci}, Fabio and {Loeb}, Abraham},
        title = "{The Redshift Evolution of the M $_{{\ensuremath{\bullet}}}$─M $_{{\ensuremath{\star}}}$ Relation for JWST's Supermassive Black Holes at z > 4}",
      journal = {\apj},
         year = 2024,
        month = apr,
       volume = {964},
       number = {2},
          eid = {154},
        pages = {154},
          doi = {10.3847/1538-4357/ad3044},
archivePrefix = {arXiv},
       eprint = {2401.04159},
 primaryClass = {astro-ph.GA},
       adsurl = {https://ui.adsabs.harvard.edu/abs/2024ApJ...964..154P}
}

@ARTICLE{Lauer2007,
       author = {{Lauer}, Tod R. and {Tremaine}, Scott and {Richstone}, Douglas and {Faber}, S.~M.},
        title = "{Selection Bias in Observing the Cosmological Evolution of the M$_{{\ensuremath{\bullet}}}$-{\ensuremath{\sigma}} and M$_{{\ensuremath{\bullet}}}$-L Relationships}",
      journal = {\apj},
         year = 2007,
        month = nov,
       volume = {670},
       number = {1},
        pages = {249-260},
          doi = {10.1086/522083},
archivePrefix = {arXiv},
       eprint = {0705.4103},
 primaryClass = {astro-ph},
       adsurl = {https://ui.adsabs.harvard.edu/abs/2007ApJ...670..249L}
}

@ARTICLE{Greene2004,
       author = {{Greene}, Jenny E. and {Ho}, Luis C.},
        title = "{Active Galactic Nuclei with Candidate Intermediate-Mass Black Holes}",
      journal = {\apj},
         year = 2004,
        month = aug,
       volume = {610},
       number = {2},
        pages = {722-736},
          doi = {10.1086/421719},
archivePrefix = {arXiv},
       eprint = {astro-ph/0404110},
 primaryClass = {astro-ph},
       adsurl = {https://ui.adsabs.harvard.edu/abs/2004ApJ...610..722G}
}

@ARTICLE{Miller2015,
       author = {{Miller}, Brendan P. and {Gallo}, Elena and {Greene}, Jenny E. and {Kelly}, Brandon C. and {Treu}, Tommaso and {Woo}, Jong-Hak and {Baldassare}, Vivienne},
        title = "{X-Ray Constraints on the Local Supermassive Black Hole Occupation Fraction}",
      journal = {\apj},
         year = 2015,
        month = jan,
       volume = {799},
       number = {1},
          eid = {98},
        pages = {98},
          doi = {10.1088/0004-637X/799/1/98},
archivePrefix = {arXiv},
       eprint = {1403.4246},
 primaryClass = {astro-ph.GA},
       adsurl = {https://ui.adsabs.harvard.edu/abs/2015ApJ...799...98M}
}

@ARTICLE{Satyapal2014,
       author = {{Satyapal}, S. and {Secrest}, N.~J. and {McAlpine}, W. and {Ellison}, S.~L. and {Fischer}, J. and {Rosenberg}, J.~L.},
        title = "{Discovery of a Population of Bulgeless Galaxies with Extremely Red Mid-IR Colors: Obscured AGN Activity in the Low-mass Regime?}",
      journal = {\apj},
         year = 2014,
        month = apr,
       volume = {784},
       number = {2},
          eid = {113},
        pages = {113},
          doi = {10.1088/0004-637X/784/2/113},
archivePrefix = {arXiv},
       eprint = {1401.5483},
 primaryClass = {astro-ph.GA},
       adsurl = {https://ui.adsabs.harvard.edu/abs/2014ApJ...784..113S}
}

@ARTICLE{Nyland2017,
       author = {{Nyland}, Kristina and {Davis}, Timothy A. and {Nguyen}, Dieu D. and {Seth}, Anil and {Wrobel}, Joan M. and {Kamble}, Atish and {Lacy}, Mark and {Alatalo}, Katherine and {Karovska}, Margarita and {Maksym}, W. Peter and {Mukherjee}, Dipanjan and {Young}, Lisa M.},
        title = "{A Multi-wavelength Study of the Turbulent Central Engine of the Low-mass AGN Hosted by NGC 404}",
      journal = {\apj},
         year = 2017,
        month = aug,
       volume = {845},
       number = {1},
          eid = {50},
        pages = {50},
          doi = {10.3847/1538-4357/aa7ecf},
archivePrefix = {arXiv},
       eprint = {1707.02303},
 primaryClass = {astro-ph.GA},
       adsurl = {https://ui.adsabs.harvard.edu/abs/2017ApJ...845...50N}
}

@ARTICLE{Baldassare2018,
       author = {{Baldassare}, Vivienne F. and {Geha}, Marla and {Greene}, Jenny},
        title = "{Identifying AGNs in Low-mass Galaxies via Long-term Optical Variability}",
      journal = {\apj},
         year = 2018,
        month = dec,
       volume = {868},
       number = {2},
          eid = {152},
        pages = {152},
          doi = {10.3847/1538-4357/aae6cf},
archivePrefix = {arXiv},
       eprint = {1808.09578},
 primaryClass = {astro-ph.GA},
       adsurl = {https://ui.adsabs.harvard.edu/abs/2018ApJ...868..152B}
}

@ARTICLE{Reines2013,
       author = {{Reines}, Amy E. and {Greene}, Jenny E. and {Geha}, Marla},
        title = "{Dwarf Galaxies with Optical Signatures of Active Massive Black Holes}",
      journal = {\apj},
         year = 2013,
        month = oct,
       volume = {775},
       number = {2},
          eid = {116},
        pages = {116},
          doi = {10.1088/0004-637X/775/2/116},
archivePrefix = {arXiv},
       eprint = {1308.0328},
 primaryClass = {astro-ph.CO},
       adsurl = {https://ui.adsabs.harvard.edu/abs/2013ApJ...775..116R}
}

@ARTICLE{Matsuoka2011,
       author = {{Matsuoka}, K. and {Nagao}, T. and {Marconi}, A. and {Maiolino}, R. and {Taniguchi}, Y.},
        title = "{The mass-metallicity relation of SDSS quasars}",
      journal = {\aap},
         year = 2011,
        month = mar,
       volume = {527},
          eid = {A100},
        pages = {A100},
          doi = {10.1051/0004-6361/201015584},
archivePrefix = {arXiv},
       eprint = {1011.5811},
 primaryClass = {astro-ph.CO},
       adsurl = {https://ui.adsabs.harvard.edu/abs/2011A&A...527A.100M}
}

@ARTICLE{Groves2006,
       author = {{Groves}, Brent A. and {Heckman}, Timothy M. and {Kauffmann}, Guinevere},
        title = "{Emission-line diagnostics of low-metallicity active galactic nuclei}",
      journal = {\mnras},
         year = 2006,
        month = oct,
       volume = {371},
       number = {4},
        pages = {1559-1569},
          doi = {10.1111/j.1365-2966.2006.10812.x},
archivePrefix = {arXiv},
       eprint = {astro-ph/0607311},
 primaryClass = {astro-ph},
       adsurl = {https://ui.adsabs.harvard.edu/abs/2006MNRAS.371.1559G}
}

@ARTICLE{Kawasaki2017,
       author = {{Kawasaki}, Kota and {Nagao}, Tohru and {Toba}, Yoshiki and {Terao}, Koki and {Matsuoka}, Kenta},
        title = "{Active Galactic Nuclei with a Low-metallicity Narrow-line Region}",
      journal = {\apj},
         year = 2017,
        month = jun,
       volume = {842},
       number = {1},
          eid = {44},
        pages = {44},
          doi = {10.3847/1538-4357/aa70e1},
archivePrefix = {arXiv},
       eprint = {1707.08731},
 primaryClass = {astro-ph.GA},
       adsurl = {https://ui.adsabs.harvard.edu/abs/2017ApJ...842...44K}
}

@ARTICLE{Izotov2008,
       author = {{Izotov}, Yuri I. and {Thuan}, Trinh X.},
        title = "{Active Galactic Nuclei in Four Metal-poor Dwarf Emission-Line Galaxies}",
      journal = {\apj},
         year = 2008,
        month = nov,
       volume = {687},
       number = {1},
        pages = {133-140},
          doi = {10.1086/591660},
archivePrefix = {arXiv},
       eprint = {0807.2029},
 primaryClass = {astro-ph},
       adsurl = {https://ui.adsabs.harvard.edu/abs/2008ApJ...687..133I}
}

@ARTICLE{Hayashi2018,
       author = {{Hayashi}, Masao and {Tanaka}, Masayuki and {Shimakawa}, Rhythm and {Furusawa}, Hisanori and {Momose}, Rieko and {Koyama}, Yusei and {Silverman}, John D. and {Kodama}, Tadayuki and {Komiyama}, Yutaka and {Leauthaud}, Alexie and {Lin}, Yen-Ting and {Miyazaki}, Satoshi and {Nagao}, Tohru and {Nishizawa}, Atsushi J. and {Ouchi}, Masami and {Shibuya}, Takatoshi and {Tadaki}, Ken-ichi and {Yabe}, Kiyoto},
        title = "{A 16 deg$^{2}$ survey of emission-line galaxies at z < 1.5 in HSC-SSP Public Data Release 1}",
      journal = {\pasj},
         year = 2018,
        month = jan,
       volume = {70},
          eid = {S17},
        pages = {S17},
          doi = {10.1093/pasj/psx088},
archivePrefix = {arXiv},
       eprint = {1704.05978},
 primaryClass = {astro-ph.GA},
       adsurl = {https://ui.adsabs.harvard.edu/abs/2018PASJ...70S..17H}
}

@ARTICLE{Hayashi2020,
       author = {{Hayashi}, Masao and {Shimakawa}, Rhythm and {Tanaka}, Masayuki and {Onodera}, Masato and {Koyama}, Yusei and {Inoue}, Akio K. and {Komiyama}, Yutaka and {Lee}, Chien-Hsiu and {Lin}, Yen-Ting and {Yabe}, Kiyoto},
        title = "{A 16 deg$^{2}$ survey of emission-line galaxies at z < 1.6 from HSC-SSP PDR2 and CHORUS}",
      journal = {\pasj},
         year = 2020,
        month = oct,
       volume = {72},
       number = {5},
          eid = {86},
        pages = {86},
          doi = {10.1093/pasj/psaa076},
archivePrefix = {arXiv},
       eprint = {2007.07413},
 primaryClass = {astro-ph.GA},
       adsurl = {https://ui.adsabs.harvard.edu/abs/2020PASJ...72...86H}
}

@ARTICLE{Aihara2018,
       author = {{Aihara}, Hiroaki and {Arimoto}, Nobuo and {Armstrong}, Robert and {Arnouts}, St{\'e}phane and {Bahcall}, Neta A. and {Bickerton}, Steven and {Bosch}, James and {Bundy}, Kevin and {Capak}, Peter L. and {Chan}, James H.~H. and {Chiba}, Masashi and {Coupon}, Jean and {Egami}, Eiichi and {Enoki}, Motohiro and {Finet}, Francois and {Fujimori}, Hiroki and {Fujimoto}, Seiji and {Furusawa}, Hisanori and {Furusawa}, Junko and {Goto}, Tomotsugu and {Goulding}, Andy and {Greco}, Johnny P. and {Greene}, Jenny E. and {Gunn}, James E. and {Hamana}, Takashi and {Harikane}, Yuichi and {Hashimoto}, Yasuhiro and {Hattori}, Takashi and {Hayashi}, Masao and {Hayashi}, Yusuke and {He{\l}miniak}, Krzysztof G. and {Higuchi}, Ryo and {Hikage}, Chiaki and {Ho}, Paul T.~P. and {Hsieh}, Bau-Ching and {Huang}, Kuiyun and {Huang}, Song and {Ikeda}, Hiroyuki and {Imanishi}, Masatoshi and {Inoue}, Akio K. and {Iwasawa}, Kazushi and {Iwata}, Ikuru and {Jaelani}, Anton T. and {Jian}, Hung-Yu and {Kamata}, Yukiko and {Karoji}, Hiroshi and {Kashikawa}, Nobunari and {Katayama}, Nobuhiko and {Kawanomoto}, Satoshi and {Kayo}, Issha and {Koda}, Jin and {Koike}, Michitaro and {Kojima}, Takashi and {Komiyama}, Yutaka and {Konno}, Akira and {Koshida}, Shintaro and {Koyama}, Yusei and {Kusakabe}, Haruka and {Leauthaud}, Alexie and {Lee}, Chien-Hsiu and {Lin}, Lihwai and {Lin}, Yen-Ting and {Lupton}, Robert H. and {Mandelbaum}, Rachel and {Matsuoka}, Yoshiki and {Medezinski}, Elinor and {Mineo}, Sogo and {Miyama}, Shoken and {Miyatake}, Hironao and {Miyazaki}, Satoshi and {Momose}, Rieko and {More}, Anupreeta and {More}, Surhud and {Moritani}, Yuki and {Moriya}, Takashi J. and {Morokuma}, Tomoki and {Mukae}, Shiro and {Murata}, Ryoma and {Murayama}, Hitoshi and {Nagao}, Tohru and {Nakata}, Fumiaki and {Niida}, Mana and {Niikura}, Hiroko and {Nishizawa}, Atsushi J. and {Obuchi}, Yoshiyuki and {Oguri}, Masamune and {Oishi}, Yukie and {Okabe}, Nobuhiro and {Okamoto}, Sakurako and {Okura}, Yuki and {Ono}, Yoshiaki and {Onodera}, Masato and {Onoue}, Masafusa and {Osato}, Ken and {Ouchi}, Masami and {Price}, Paul A. and {Pyo}, Tae-Soo and {Sako}, Masao and {Sawicki}, Marcin and {Shibuya}, Takatoshi and {Shimasaku}, Kazuhiro and {Shimono}, Atsushi and {Shirasaki}, Masato and {Silverman}, John D. and {Simet}, Melanie and {Speagle}, Joshua and {Spergel}, David N. and {Strauss}, Michael A. and {Sugahara}, Yuma and {Sugiyama}, Naoshi and {Suto}, Yasushi and {Suyu}, Sherry H. and {Suzuki}, Nao and {Tait}, Philip J. and {Takada}, Masahiro and {Takata}, Tadafumi and {Tamura}, Naoyuki and {Tanaka}, Manobu M. and {Tanaka}, Masaomi and {Tanaka}, Masayuki and {Tanaka}, Yoko and {Terai}, Tsuyoshi and {Terashima}, Yuichi and {Toba}, Yoshiki and {Tominaga}, Nozomu and {Toshikawa}, Jun and {Turner}, Edwin L. and {Uchida}, Tomohisa and {Uchiyama}, Hisakazu and {Umetsu}, Keiichi and {Uraguchi}, Fumihiro and {Urata}, Yuji and {Usuda}, Tomonori and {Utsumi}, Yousuke and {Wang}, Shiang-Yu and {Wang}, Wei-Hao and {Wong}, Kenneth C. and {Yabe}, Kiyoto and {Yamada}, Yoshihiko and {Yamanoi}, Hitomi and {Yasuda}, Naoki and {Yeh}, Sherry and {Yonehara}, Atsunori and {Yuma}, Suraphong},
        title = "{The Hyper Suprime-Cam SSP Survey: Overview and survey design}",
      journal = {\pasj},
         year = 2018,
        month = jan,
       volume = {70},
          eid = {S4},
        pages = {S4},
          doi = {10.1093/pasj/psx066},
archivePrefix = {arXiv},
       eprint = {1704.05858},
 primaryClass = {astro-ph.IM},
       adsurl = {https://ui.adsabs.harvard.edu/abs/2018PASJ...70S...4A}
}

@ARTICLE{Inoue2020,
       author = {{Inoue}, Akio K. and {Yamanaka}, Satoshi and {Ouchi}, Masami and {Iwata}, Ikuru and {Shimasaku}, Kazuhiro and {Taniguchi}, Yoshiaki and {Nagao}, Tohru and {Kashikawa}, Nobunari and {Ono}, Yoshiaki and {Mawatari}, Ken and {Shibuya}, Takatoshi and {Hayashi}, Masao and {Ikeda}, Hiroyuki and {Zhang}, Haibin and {Liang}, Yongming and {Lee}, Chien-Hsiu and {Hilmi}, Miftahul and {Kikuta}, Satoshi and {Kusakabe}, Haruka and {Furusawa}, Hisanori and {Hayashino}, Tomoki and {Kajisawa}, Masaru and {Matsuda}, Yuichi and {Nakajima}, Kimihiko and {Momose}, Rieko and {Harikane}, Yuichi and {Saito}, Tomoki and {Kodama}, Tadayuki and {Kikuchihara}, Shotaro and {Iye}, Masanori and {Goto}, Tomotsugu},
        title = "{CHORUS. I. Cosmic HydrOgen Reionization Unveiled with Subaru: Overview}",
      journal = {\pasj},
         year = 2020,
        month = dec,
       volume = {72},
       number = {6},
          eid = {101},
        pages = {101},
          doi = {10.1093/pasj/psaa100},
archivePrefix = {arXiv},
       eprint = {2011.07211},
 primaryClass = {astro-ph.GA},
       adsurl = {https://ui.adsabs.harvard.edu/abs/2020PASJ...72..101I}
}

@ARTICLE{Takada2014,
       author = {{Takada}, Masahiro and {Ellis}, Richard S. and {Chiba}, Masashi and {Greene}, Jenny E. and {Aihara}, Hiroaki and {Arimoto}, Nobuo and {Bundy}, Kevin and {Cohen}, Judith and {Dor{\'e}}, Olivier and {Graves}, Genevieve and {Gunn}, James E. and {Heckman}, Timothy and {Hirata}, Christopher M. and {Ho}, Paul and {Kneib}, Jean-Paul and {Le F{\`e}vre}, Olivier and {Lin}, Lihwai and {More}, Surhud and {Murayama}, Hitoshi and {Nagao}, Tohru and {Ouchi}, Masami and {Seiffert}, Michael and {Silverman}, John D. and {Sodr{\'e}}, Laerte and {Spergel}, David N. and {Strauss}, Michael A. and {Sugai}, Hajime and {Suto}, Yasushi and {Takami}, Hideki and {Wyse}, Rosemary},
        title = "{Extragalactic science, cosmology, and Galactic archaeology with the Subaru Prime Focus Spectrograph}",
      journal = {\pasj},
         year = 2014,
        month = feb,
       volume = {66},
       number = {1},
          eid = {R1},
        pages = {R1},
          doi = {10.1093/pasj/pst019},
archivePrefix = {arXiv},
       eprint = {1206.0737},
 primaryClass = {astro-ph.CO},
       adsurl = {https://ui.adsabs.harvard.edu/abs/2014PASJ...66R...1T}
}

@INPROCEEDINGS{Tamura2016,
       author = {{Tamura}, Naoyuki and {Takato}, Naruhisa and {Shimono}, Atsushi and {Moritani}, Yuki and {Yabe}, Kiyoto and {Ishizuka}, Yuki and {Ueda}, Akitoshi and {Kamata}, Yukiko and {Aghazarian}, Hrand and {Arnouts}, St{\'e}phane and {Barban}, Gabriel and {Barkhouser}, Robert H. and {Borges}, Renato C. and {Braun}, David F. and {Carr}, Michael A. and {Chabaud}, Pierre-Yves and {Chang}, Yin-Chang and {Chen}, Hsin-Yo and {Chiba}, Masashi and {Chou}, Richard C.~Y. and {Chu}, You-Hua and {Cohen}, Judith and {de Almeida}, Rodrigo P. and {de Oliveira}, Antonio C. and {de Oliveira}, Ligia S. and {Dekany}, Richard G. and {Dohlen}, Kjetil and {dos Santos}, Jesulino B. and {dos Santos}, Leandro H. and {Ellis}, Richard and {Fabricius}, Maximilian and {Ferrand}, Didier and {Ferreira}, D{\'e}cio and {Golebiowski}, Mirek and {Greene}, Jenny E. and {Gross}, Johannes and {Gunn}, James E. and {Hammond}, Randolph and {Harding}, Albert and {Hart}, Murdock and {Heckman}, Timothy M. and {Hirata}, Christopher M. and {Ho}, Paul and {Hope}, Stephen C. and {Hovland}, Larry and {Hsu}, Shu-Fu and {Hu}, Yen-Shan and {Huang}, Ping-Jie and {Jaquet}, Marc and {Jing}, Yipeng and {Karr}, Jennifer and {Kimura}, Masahiko and {King}, Matthew E. and {Komatsu}, Eiichiro and {Le Brun}, Vincent and {Le F{\`e}vre}, Olivier and {Le Fur}, Arnaud and {Le Mignant}, David and {Ling}, Hung-Hsu and {Loomis}, Craig P. and {Lupton}, Robert H. and {Madec}, Fabrice and {Mao}, Peter and {Marrara}, Lucas S. and {Mendes de Oliveira}, Claudia and {Minowa}, Yosuke and {Morantz}, Chaz and {Murayama}, Hitoshi and {Murray}, Graham J. and {Ohyama}, Youichi and {Orndorff}, Joseph and {Pascal}, Sandrine and {Pereira}, Jefferson M. and {Reiley}, Daniel and {Reinecke}, Martin and {Ritter}, Andreas and {Roberts}, Mitsuko and {Schwochert}, Mark A. and {Seiffert}, Michael D. and {Smee}, Stephen A. and {Sodre}, Laerte and {Spergel}, David N. and {Steinkraus}, Aaron J. and {Strauss}, Michael A. and {Surace}, Christian and {Suto}, Yasushi and {Suzuki}, Nao and {Swinbank}, John and {Tait}, Philip J. and {Takada}, Masahiro and {Tamura}, Tomonori and {Tanaka}, Yoko and {Tresse}, Laurence and {Verducci}, Orlando and {Vibert}, Didier and {Vidal}, Clement and {Wang}, Shiang-Yu and {Wen}, Chih-Yi and {Yan}, Chi-Hung and {Yasuda}, Naoki},
        title = "{Prime Focus Spectrograph (PFS) for the Subaru telescope: overview, recent progress, and future perspectives}",
    booktitle = {Ground-based and Airborne Instrumentation for Astronomy VI},
         year = 2016,
       editor = {{Evans}, Christopher J. and {Simard}, Luc and {Takami}, Hideki},
       series = {Society of Photo-Optical Instrumentation Engineers (SPIE) Conference Series},
       volume = {9908},
        month = aug,
          eid = {99081M},
        pages = {99081M},
          doi = {10.1117/12.2232103},
archivePrefix = {arXiv},
       eprint = {1608.01075},
 primaryClass = {astro-ph.IM},
       adsurl = {https://ui.adsabs.harvard.edu/abs/2016SPIE.9908E..1MT}
}

@ARTICLE{Kawanomoto2018,
       author = {{Kawanomoto}, Satoshi and {Uraguchi}, Fumihiro and {Komiyama}, Yutaka and {Miyazaki}, Satoshi and {Furusawa}, Hisanori and {Finet}, Fran{\c{c}}ois and {Hattori}, Takashi and {Wang}, Shiang-Yu and {Yasuda}, Naoki and {Suzuki}, Naotaka},
        title = "{Hyper Suprime-Cam: Filters}",
      journal = {\pasj},
         year = 2018,
        month = aug,
       volume = {70},
       number = {4},
          eid = {66},
        pages = {66},
          doi = {10.1093/pasj/psy056},
       adsurl = {https://ui.adsabs.harvard.edu/abs/2018PASJ...70...66K}
}

@ARTICLE{Komiyama2018,
       author = {{Komiyama}, Yutaka and {Obuchi}, Yoshiyuki and {Nakaya}, Hidehiko and {Kamata}, Yukiko and {Kawanomoto}, Satoshi and {Utsumi}, Yousuke and {Miyazaki}, Satoshi and {Uraguchi}, Fumihiro and {Furusawa}, Hisanori and {Morokuma}, Tomoki and {Uchida}, Tomohisa and {Miyatake}, Hironao and {Mineo}, Sogo and {Fujimori}, Hiroki and {Aihara}, Hiroaki and {Karoji}, Hiroshi and {Gunn}, James E. and {Wang}, Shiang-Yu},
        title = "{Hyper Suprime-Cam: Camera dewar design}",
      journal = {\pasj},
         year = 2018,
        month = jan,
       volume = {70},
          eid = {S2},
        pages = {S2},
          doi = {10.1093/pasj/psx069},
       adsurl = {https://ui.adsabs.harvard.edu/abs/2018PASJ...70S...2K}
}

@ARTICLE{Furusawa2018,
       author = {{Furusawa}, Hisanori and {Koike}, Michitaro and {Takata}, Tadafumi and {Okura}, Yuki and {Miyatake}, Hironao and {Lupton}, Robert H. and {Bickerton}, Steven and {Price}, Paul A. and {Bosch}, James and {Yasuda}, Naoki and {Mineo}, Sogo and {Yamada}, Yoshihiko and {Miyazaki}, Satoshi and {Nakata}, Fumiaki and {Koshida}, Shintaro and {Komiyama}, Yutaka and {Utsumi}, Yousuke and {Kawanomoto}, Satoshi and {Jeschke}, Eric and {Noumaru}, Junichi and {Schubert}, Kiaina and {Iwata}, Ikuru and {Finet}, Francois and {Fujiyoshi}, Takuya and {Tajitsu}, Akito and {Terai}, Tsuyoshi and {Lee}, Chien-Hsiu},
        title = "{The on-site quality-assurance system for Hyper Suprime-Cam: OSQAH}",
      journal = {\pasj},
         year = 2018,
        month = jan,
       volume = {70},
          eid = {S3},
        pages = {S3},
          doi = {10.1093/pasj/psx079},
       adsurl = {https://ui.adsabs.harvard.edu/abs/2018PASJ...70S...3F}
}

@ARTICLE{Bosch2018,
       author = {{Bosch}, James and {Armstrong}, Robert and {Bickerton}, Steven and {Furusawa}, Hisanori and {Ikeda}, Hiroyuki and {Koike}, Michitaro and {Lupton}, Robert and {Mineo}, Sogo and {Price}, Paul and {Takata}, Tadafumi and {Tanaka}, Masayuki and {Yasuda}, Naoki and {AlSayyad}, Yusra and {Becker}, Andrew C. and {Coulton}, William and {Coupon}, Jean and {Garmilla}, Jose and {Huang}, Song and {Krughoff}, K. Simon and {Lang}, Dustin and {Leauthaud}, Alexie and {Lim}, Kian-Tat and {Lust}, Nate B. and {MacArthur}, Lauren A. and {Mandelbaum}, Rachel and {Miyatake}, Hironao and {Miyazaki}, Satoshi and {Murata}, Ryoma and {More}, Surhud and {Okura}, Yuki and {Owen}, Russell and {Swinbank}, John D. and {Strauss}, Michael A. and {Yamada}, Yoshihiko and {Yamanoi}, Hitomi},
        title = "{The Hyper Suprime-Cam software pipeline}",
      journal = {\pasj},
         year = 2018,
        month = jan,
       volume = {70},
          eid = {S5},
        pages = {S5},
          doi = {10.1093/pasj/psx080},
archivePrefix = {arXiv},
       eprint = {1705.06766},
 primaryClass = {astro-ph.IM},
       adsurl = {https://ui.adsabs.harvard.edu/abs/2018PASJ...70S...5B}
}

@ARTICLE{Schlafly2012,
       author = {{Schlafly}, E.~F. and {Finkbeiner}, D.~P. and {Juri{\'c}}, M. and {Magnier}, E.~A. and {Burgett}, W.~S. and {Chambers}, K.~C. and {Grav}, T. and {Hodapp}, K.~W. and {Kaiser}, N. and {Kudritzki}, R.-P. and {Martin}, N.~F. and {Morgan}, J.~S. and {Price}, P.~A. and {Rix}, H.-W. and {Stubbs}, C.~W. and {Tonry}, J.~L. and {Wainscoat}, R.~J.},
        title = "{Photometric Calibration of the First 1.5 Years of the Pan-STARRS1 Survey}",
      journal = {\apj},
         year = 2012,
        month = sep,
       volume = {756},
       number = {2},
          eid = {158},
        pages = {158},
          doi = {10.1088/0004-637X/756/2/158},
archivePrefix = {arXiv},
       eprint = {1201.2208},
 primaryClass = {astro-ph.IM},
       adsurl = {https://ui.adsabs.harvard.edu/abs/2012ApJ...756..158S}
}

@ARTICLE{Tonry2012,
       author = {{Tonry}, J.~L. and {Stubbs}, C.~W. and {Lykke}, K.~R. and {Doherty}, P. and {Shivvers}, I.~S. and {Burgett}, W.~S. and {Chambers}, K.~C. and {Hodapp}, K.~W. and {Kaiser}, N. and {Kudritzki}, R.-P. and {Magnier}, E.~A. and {Morgan}, J.~S. and {Price}, P.~A. and {Wainscoat}, R.~J.},
        title = "{The Pan-STARRS1 Photometric System}",
      journal = {\apj},
         year = 2012,
        month = may,
       volume = {750},
       number = {2},
          eid = {99},
        pages = {99},
          doi = {10.1088/0004-637X/750/2/99},
archivePrefix = {arXiv},
       eprint = {1203.0297},
 primaryClass = {astro-ph.IM},
       adsurl = {https://ui.adsabs.harvard.edu/abs/2012ApJ...750...99T}
}

@ARTICLE{Magnier2013,
       author = {{Magnier}, E.~A. and {Schlafly}, E. and {Finkbeiner}, D. and {Juric}, M. and {Tonry}, J.~L. and {Burgett}, W.~S. and {Chambers}, K.~C. and {Flewelling}, H.~A. and {Kaiser}, N. and {Kudritzki}, R.-P. and {Morgan}, J.~S. and {Price}, P.~A. and {Sweeney}, W.~E. and {Stubbs}, C.~W.},
        title = "{The Pan-STARRS 1 Photometric Reference Ladder, Release 12.01}",
      journal = {\apjs},
         year = 2013,
        month = apr,
       volume = {205},
       number = {2},
          eid = {20},
        pages = {20},
          doi = {10.1088/0067-0049/205/2/20},
archivePrefix = {arXiv},
       eprint = {1303.3634},
 primaryClass = {astro-ph.IM},
       adsurl = {https://ui.adsabs.harvard.edu/abs/2013ApJS..205...20M}
}

@ARTICLE{Tanaka2015,
       author = {{Tanaka}, Masayuki},
        title = "{Photometric Redshift with Bayesian Priors on Physical Properties of Galaxies}",
      journal = {\apj},
         year = 2015,
        month = mar,
       volume = {801},
       number = {1},
          eid = {20},
        pages = {20},
          doi = {10.1088/0004-637X/801/1/20},
archivePrefix = {arXiv},
       eprint = {1501.02047},
 primaryClass = {astro-ph.GA},
       adsurl = {https://ui.adsabs.harvard.edu/abs/2015ApJ...801...20T}
}

@ARTICLE{Tanaka2018,
       author = {{Tanaka}, Masayuki and {Coupon}, Jean and {Hsieh}, Bau-Ching and {Mineo}, Sogo and {Nishizawa}, Atsushi J. and {Speagle}, Joshua and {Furusawa}, Hisanori and {Miyazaki}, Satoshi and {Murayama}, Hitoshi},
        title = "{Photometric redshifts for Hyper Suprime-Cam Subaru Strategic Program Data Release 1}",
      journal = {\pasj},
         year = 2018,
        month = jan,
       volume = {70},
          eid = {S9},
        pages = {S9},
          doi = {10.1093/pasj/psx077},
archivePrefix = {arXiv},
       eprint = {1704.05988},
 primaryClass = {astro-ph.GA},
       adsurl = {https://ui.adsabs.harvard.edu/abs/2018PASJ...70S...9T}
}

@ARTICLE{HsiehYee2014,
       author = {{Hsieh}, B.~C. and {Yee}, H.~K.~C.},
        title = "{Estimating Luminosities and Stellar Masses of Galaxies Photometrically without Determining Redshifts}",
      journal = {\apj},
         year = 2014,
        month = sep,
       volume = {792},
       number = {2},
          eid = {102},
        pages = {102},
          doi = {10.1088/0004-637X/792/2/102},
archivePrefix = {arXiv},
       eprint = {1407.5151},
 primaryClass = {astro-ph.GA},
       adsurl = {https://ui.adsabs.harvard.edu/abs/2014ApJ...792..102H}
}

@ARTICLE{Laigle2016,
       author = {{Laigle}, C. and {McCracken}, H.~J. and {Ilbert}, O. and {Hsieh}, B.~C. and {Davidzon}, I. and {Capak}, P. and {Hasinger}, G. and {Silverman}, J.~D. and {Pichon}, C. and {Coupon}, J. and {Aussel}, H. and {Le Borgne}, D. and {Caputi}, K. and {Cassata}, P. and {Chang}, Y.-Y. and {Civano}, F. and {Dunlop}, J. and {Fynbo}, J. and {Kartaltepe}, J.~S. and {Koekemoer}, A. and {Le F{\`e}vre}, O. and {Le Floc'h}, E. and {Leauthaud}, A. and {Lilly}, S. and {Lin}, L. and {Marchesi}, S. and {Milvang-Jensen}, B. and {Salvato}, M. and {Sanders}, D.~B. and {Scoville}, N. and {Smolcic}, V. and {Stockmann}, M. and {Taniguchi}, Y. and {Tasca}, L. and {Toft}, S. and {Vaccari}, Mattia and {Zabl}, J.},
        title = "{The COSMOS2015 Catalog: Exploring the 1 < z < 6 Universe with Half a Million Galaxies}",
      journal = {\apjs},
         year = 2016,
        month = jun,
       volume = {224},
       number = {2},
          eid = {24},
        pages = {24},
          doi = {10.3847/0067-0049/224/2/24},
archivePrefix = {arXiv},
       eprint = {1604.02350},
 primaryClass = {astro-ph.GA},
       adsurl = {https://ui.adsabs.harvard.edu/abs/2016ApJS..224...24L}
}

@ARTICLE{Kewley2001,
       author = {{Kewley}, L.~J. and {Dopita}, M.~A. and {Sutherland}, R.~S. and {Heisler}, C.~A. and {Trevena}, J.},
        title = "{Theoretical Modeling of Starburst Galaxies}",
      journal = {\apj},
         year = 2001,
        month = jul,
       volume = {556},
       number = {1},
        pages = {121-140},
          doi = {10.1086/321545},
archivePrefix = {arXiv},
       eprint = {astro-ph/0106324},
 primaryClass = {astro-ph},
       adsurl = {https://ui.adsabs.harvard.edu/abs/2001ApJ...556..121K}
}

@ARTICLE{Kewley2013,
       author = {{Kewley}, Lisa J. and {Dopita}, Michael A. and {Leitherer}, Claus and {Dav{\'e}}, Romeel and {Yuan}, Tiantian and {Allen}, Mark and {Groves}, Brent and {Sutherland}, Ralph},
        title = "{Theoretical Evolution of Optical Strong Lines across Cosmic Time}",
      journal = {\apj},
         year = 2013,
        month = sep,
       volume = {774},
       number = {2},
          eid = {100},
        pages = {100},
          doi = {10.1088/0004-637X/774/2/100},
archivePrefix = {arXiv},
       eprint = {1307.0508},
 primaryClass = {astro-ph.CO},
       adsurl = {https://ui.adsabs.harvard.edu/abs/2013ApJ...774..100K}
}

@ARTICLE{Sugai2015,
       author = {{Sugai}, Hajime and {Tamura}, Naoyuki and {Karoji}, Hiroshi and {Shimono}, Atsushi and {Takato}, Naruhisa and {Kimura}, Masahiko and {Ohyama}, Youichi and {Ueda}, Akitoshi and {Aghazarian}, Hrand and {de Arruda}, Marcio Vital and {Barkhouser}, Robert H. and {Bennett}, Charles L. and {Bickerton}, Steve and {Bozier}, Alexandre and {Braun}, David F. and {Bui}, Khanh and {Capocasale}, Christopher M. and {Carr}, Michael A. and {Castilho}, Bruno and {Chang}, Yin-Chang and {Chen}, Hsin-Yo and {Chou}, Richard C.~Y. and {Dawson}, Olivia R. and {Dekany}, Richard G. and {Ek}, Eric M. and {Ellis}, Richard S. and {English}, Robin J. and {Ferrand}, Didier and {Ferreira}, D{\'e}cio and {Fisher}, Charles D. and {Golebiowski}, Mirek and {Gunn}, James E. and {Hart}, Murdock and {Heckman}, Timothy M. and {Ho}, Paul T.~P. and {Hope}, Stephen and {Hovland}, Larry E. and {Hsu}, Shu-Fu and {Hu}, Yen-Shan and {Huang}, Pin Jie and {Jaquet}, Marc and {Karr}, Jennifer E. and {Kempenaar}, Jason G. and {King}, Matthew E. and {le F{\`e}vre}, Olivier and {Mignant}, David Le and {Ling}, Hung-Hsu and {Loomis}, Craig and {Lupton}, Robert H. and {Madec}, Fabrice and {Mao}, Peter and {Souza Marrara}, Lucas and {M{\'e}nard}, Brice and {Morantz}, Chaz and {Murayama}, Hitoshi and {Murray}, Graham J. and {Cesar de Oliveira}, Antonio and {Mendes de Oliveira}, Claudia and {Souza de Oliveira}, Ligia and {Orndorff}, Joe D. and {de Paiva Vila{\c{c}}a}, Rodrigo and {Partos}, Eamon J. and {Pascal}, Sandrine and {Pegot-Ogier}, Thomas and {Reiley}, Daniel J. and {Riddle}, Reed and {Santos}, Leandro and {dos Santos}, Jesulino Bispo and {Schwochert}, Mark A. and {Seiffert}, Michael D. and {Smee}, Stephen A. and {Smith}, Roger M. and {Steinkraus}, Ronald E. and {Sodr{\'e}}, Laerte and {Spergel}, David N. and {Surace}, Christian and {Tresse}, Laurence and {Vidal}, Cl{\'e}ment and {Vives}, Sebastien and {Wang}, Shiang-Yu and {Wen}, Chih-Yi and {Wu}, Amy C. and {Wyse}, Rosie and {Yan}, Chi-Hung},
        title = "{Prime Focus Spectrograph for the Subaru telescope: massively multiplexed optical and near-infrared fiber spectrograph}",
      journal = {Journal of Astronomical Telescopes, Instruments, and Systems},
         year = 2015,
        month = jul,
       volume = {1},
          eid = {035001},
        pages = {035001},
          doi = {10.1117/1.JATIS.1.3.035001},
archivePrefix = {arXiv},
       eprint = {1507.00725},
 primaryClass = {astro-ph.IM},
       adsurl = {https://ui.adsabs.harvard.edu/abs/2015JATIS...1c5001S}
}

@INPROCEEDINGS{Tamura2022,
       author = {{Tamura}, Naoyuki and {Moritani}, Yuki and {Yabe}, Kiyoto and {Ishizuka}, Yuki and {Kamata}, Yukiko and {Allaoui}, Ali and {Arai}, Akira and {Arnouts}, St{\'e}phane and {Barkhouser}, Robert H. and {Barette}, Rudy and {Blanchard}, Patrick and {Bergeron}, Eddie and {Caplar}, Neven and {Chabaud}, Pierre-Yves and {Chang}, Yin-Chang and {Chen}, Hsin-Yo and {Chou}, Chueh-Yi and {Chu}, You-Hua and {Cohen}, Judith G. and {da Costa}, Richardo L. and {Crauchet}, Thibaut and {de Almeida}, Rodrigo P. and {de Oliveira}, Antonio C. and {de Oliveira}, Ligia S. and {Dohlen}, Kjetil and {dos Santos}, Leandro H. and {Ellis}, Richard S. and {Fabricius}, Maximilian and {Ferreira}, D{\'e}cio and {Furusawa}, Hisanori and {Givans}, Jahmour J. and {Garci{\'a}-Carpio}, Javier and {Golebiowski}, Mirek and {Gray}, Aidan and {Gunn}, James E. and {Hamano}, Satoshi and {Hammond}, Randolph P. and {Harding}, Albert and {Hayashi}, Kota and {He}, Wanqiu and {Heckman}, Timothy M. and {Hope}, Stephen C. and {Hsu}, Shu-Fu and {Hu}, Yen-Shan and {Huang}, Pin Jie and {Ishigaki}, Miho N. and {Jeschke}, Eric and {Jing}, Yipeng and {Kado-Fong}, Erin and {Karr}, Jennifer L. and {Kawanomoto}, Satoshi and {Kimura}, Masahiko and {Koike}, Michitaro and {Komatsu}, Eiichiro and {Koshida}, Shintaro and {Le Brun}, Vincent and {Le Fur}, Arnaud and {Le Mignant}, David and {Lhoussaine}, Romain and {Lin}, Yen-Ting and {Ling}, Hung-Hsu and {Loomis}, Craig P. and {Lupton}, Robert H. and {Madec}, Fabrice and {Marchesini}, Danilo and {Marguerite}, Edouard and {Marrara}, Lucas S. and {Medvedev}, Dmitry and {Mineo}, Sogo and {Miyazaki}, Satoshi and {Morishima}, Takahiro and {Murata}, Kazumi and {Murayama}, Hitoshi and {Murray}, Graham J. and {Okita}, Hirofumi and {Onodera}, Masato and {Peebles}, Joshua and {Price}, Paul and {Pyo}, Tae-Soo and {Ramos}, Lucio and {Reiley}, Daniel J. and {Reinecke}, Martin and {Roberts}, Mitsuko and {Rosa}, Josimar A. and {Rousselle}, Julien P. and {Sarkis}, Mira and {Seiffert}, Michael D. and {Schubert}, Kiaina and {Siddiqui}, Hassan and {Smee}, Stephen A. and {Sodr{\'e}}, Laerte and {Strauss}, Michael A. and {Surace}, Christian and {Taghizadeh Popp}, Manuchehr and {Tait}, Philip J. and {Takada}, Masahiro and {Takagi}, Yuhei and {Tanaka}, Masayuki and {Tanaka}, Yoko and {Thakar}, Aniruddha R. and {Vibert}, Didier and {Wang}, Shiang-Yu and {Wen}, Chih-Yi and {Werner}, Suzanne and {Wung}, Matthew and {Lemson}, Gerald and {Mitschang}, Arik and {Yasuda}, Naoki and {Yoshida}, Hiroshige and {Yan}, Chi-Hung and {Yoshida}, Michitoshi and {Yamashita}, Takuji},
        title = "{Prime Focus Spectrograph (PFS) for the Subaru Telescope: its start of the last development phase}",
    booktitle = {Ground-based and Airborne Instrumentation for Astronomy IX},
         year = 2022,
       editor = {{Evans}, Christopher J. and {Bryant}, Julia J. and {Motohara}, Kentaro},
       series = {Society of Photo-Optical Instrumentation Engineers (SPIE) Conference Series},
       volume = {12184},
        month = aug,
          eid = {1218410},
        pages = {1218410},
          doi = {10.1117/12.2628152},
       adsurl = {https://ui.adsabs.harvard.edu/abs/2022SPIE12184E..10T}
}

@INPROCEEDINGS{Tamura2024,
       author = {{Tamura}, Naoyuki and {Yabe}, Kiyoto and {Koshida}, Shintaro and {Moritani}, Yuki and {Tanaka}, Masayuki and {Ishigaki}, Miho N. and {Ishizuka}, Yuki and {Kamata}, Yukiko and {Allaoui}, Ali and {Arai}, Akira and {Arnouts}, St{\'e}phane and {Barette}, Rudy and {Barkhouser}, Robert H. and {Bergeron}, Eddie and {Blanchard}, Patrick and {Caplar}, Neven and {Carle}, Michael and {Chabaud}, Pierre-Yves and {Chang}, Yin-Chang and {Chen}, Hsin-Yo and {Chou}, Richard C.~Y. and {Cohen}, Judith G. and {Costa}, Ricardo and {Crauchet}, Thibaut and {de Almeida}, Rodorigo P. and {de Oliveira}, Antonio C. and {de Oliveira}, Ligia S. and {Dohlen}, Kjetil and {dos Santos}, Leandro H. and {Dobos}, L{\'a}szl{\'o} and {Ellis}, Richard S. and {Ertel}, Steve and {Fabricius}, Maximilian and {Ferreira}, D{\'e}cio and {Furusawa}, Hisanori and {Gee}, Wilfred T. and {Garci{\'a}-Carpio}, Javier and {Gerasimov}, Roman and {Golebiowski}, Mirek and {Gray}, Aidan and {Gunn}, James E. and {Hahn}, ChangHoon and {Hamano}, Satoshi and {Hammond}, Randolph P. and {Harding}, Albert and {Hattori}, Takashi and {Hayashi}, Kota and {He}, Wanqiu and {Heckman}, Timothy M. and {Hope}, Stephen C. and {Hsu}, Shu-Fu and {Huang}, Pin-Jie and {Jaquet}, Marc and {Jeschke}, Eric and {Jespersen}, Christian K. and {Jing}, Yipeng and {Kackley}, Russell and {Karr}, Jennifer L. and {Kawanomoto}, Satoshi and {Kimura}, Masahiko and {Kirby}, Evan N. and {Koike}, Michitaro and {Komatsu}, Eiichiro and {Koyama}, Yusei and {Le Brun}, Vincent and {Le Fur}, Arnaud and {Le Mignant}, David and {Lemson}, Gerald and {Lin}, Yen-Ting and {Ling}, Hung-Hsu and {Loomis}, Craig P. and {Lupton}, Robert H. and {Madec}, Fabrice and {Marchesini}, Danilo and {Marrara}, Lucas S. and {Medvedev}, Dmitry and {Mineo}, Sogo and {Mitschang}, Arik and {Miyazaki}, Satoshi and {Morihana}, Kumiko and {Morishima}, Takahiro and {Murayama}, Hitoshi and {Murray}, Graham J. and {Okamoto}, Sakurako and {Okita}, Hirofumi and {Onodera}, Masato and {Passegger}, Vera M. and {Peebles}, Joshua and {Price}, Paul A. and {Pyo}, Tae-Soo and {Ramos}, Lucio and {Reiley}, Daniel J. and {Reinecke}, Martin and {Roberts}, Mitsuko and {Rosa}, Josemar A. and {Rousselle}, Julien P. and {Rubio}, Kody H. and {Schubert}, Kiaina and {Seiffert}, Michael D. and {Siegel}, Jared and {Smee}, Stephen A. and {Sodr{\'e}}, Laerte and {Strauss}, Michael A. and {Sunayama}, Tomomi and {Surace}, Christian and {Takada}, Masahiro and {Takagi}, Yuhei and {Tanaka}, Ichi and {Tanaka}, Yoko and {Thakar}, Aniruddha R. and {Vibert}, Didier and {Wang}, Shiang-Yu and {Wen}, Chih-Yi and {Werner}, Suzanne and {Wung}, Matthew and {Yan}, Chi-Hung and {Yasuda}, Naoki and {Yoshida}, Hiroshige},
        title = "{Prime Focus Spectrograph (PFS) for Subaru Telescope: progressing final steps to science operation}",
    booktitle = {Ground-based and Airborne Instrumentation for Astronomy X},
         year = 2024,
       editor = {{Bryant}, Julia J. and {Motohara}, Kentaro and {Vernet}, Jo{\"e}l. R.~D.},
       series = {Society of Photo-Optical Instrumentation Engineers (SPIE) Conference Series},
       volume = {13096},
        month = jul,
          eid = {1309605},
        pages = {1309605},
          doi = {10.1117/12.3015967},
       adsurl = {https://ui.adsabs.harvard.edu/abs/2024SPIE13096E..05T}
}

@INPROCEEDINGS{Shimono2016,
       author = {{Shimono}, Atsushi and {Tamura}, Naoyuki and {Takato}, Naruhisa and {Yasuda}, Naoki and {Suzuki}, Nao and {Loomis}, Craig P. and {Lupton}, Robert H. and {Moritani}, Yuki and {Yabe}, Kiyoto},
        title = "{The survey operation software system development for Prime Focus Spectrograph (PFS) on Subaru Telescope}",
    booktitle = {Software and Cyberinfrastructure for Astronomy IV},
         year = 2016,
       editor = {{Chiozzi}, Gianluca and {Guzman}, Juan C.},
       series = {Society of Photo-Optical Instrumentation Engineers (SPIE) Conference Series},
       volume = {9913},
        month = jul,
          eid = {99133B},
        pages = {99133B},
          doi = {10.1117/12.2232844},
archivePrefix = {arXiv},
       eprint = {1608.01163},
 primaryClass = {astro-ph.IM},
       adsurl = {https://ui.adsabs.harvard.edu/abs/2016SPIE.9913E..3BS}
}

@ARTICLE{Rousselot2000,
       author = {{Rousselot}, P. and {Lidman}, C. and {Cuby}, J.-G. and {Moreels}, G. and {Monnet}, G.},
        title = "{Night-sky spectral atlas of OH emission lines in the near-infrared}",
      journal = {\aap},
         year = 2000,
        month = feb,
       volume = {354},
        pages = {1134-1150},
       adsurl = {https://ui.adsabs.harvard.edu/abs/2000A&A...354.1134R}
}

@ARTICLE{Ferland2017,
       author = {{Ferland}, G.~J. and {Chatzikos}, M. and {Guzm{\'a}n}, F. and {Lykins}, M.~L. and {van Hoof}, P.~A.~M. and {Williams}, R.~J.~R. and {Abel}, N.~P. and {Badnell}, N.~R. and {Keenan}, F.~P. and {Porter}, R.~L. and {Stancil}, P.~C.},
        title = "{The 2017 Release Cloudy}",
      journal = {Revista Mexicana de Astronomía y Astrofísica},
         year = 2017,
        month = oct,
       volume = {53},
        pages = {385-438},
          doi = {10.48550/arXiv.1705.10877},
archivePrefix = {arXiv},
       eprint = {1705.10877},
 primaryClass = {astro-ph.GA},
       adsurl = {https://ui.adsabs.harvard.edu/abs/2017RMxAA..53..385F}
}

@ARTICLE{GreeneHo2005,
       author = {{Greene}, Jenny E. and {Ho}, Luis C.},
        title = "{Estimating Black Hole Masses in Active Galaxies Using the H{\ensuremath{\alpha}} Emission Line}",
      journal = {\apj},
         year = 2005,
        month = sep,
       volume = {630},
       number = {1},
        pages = {122-129},
          doi = {10.1086/431897},
archivePrefix = {arXiv},
       eprint = {astro-ph/0508335},
 primaryClass = {astro-ph},
       adsurl = {https://ui.adsabs.harvard.edu/abs/2005ApJ...630..122G}
}

@ARTICLE{SternLaor2012,
       author = {{Stern}, Jonathan and {Laor}, Ari},
        title = "{Type 1 AGN at low z- I. Emission properties}",
      journal = {\mnras},
         year = 2012,
        month = jun,
       volume = {423},
       number = {1},
        pages = {600-631},
          doi = {10.1111/j.1365-2966.2012.20901.x},
archivePrefix = {arXiv},
       eprint = {1203.3158},
 primaryClass = {astro-ph.CO},
       adsurl = {https://ui.adsabs.harvard.edu/abs/2012MNRAS.423..600S}
}

@ARTICLE{Greene2022,
       author = {{Greene}, Jenny and {Bezanson}, Rachel and {Ouchi}, Masami and {Silverman}, John and {the PFS Galaxy Evolution Working Group}},
        title = "{The Prime Focus Spectrograph Galaxy Evolution Survey}",
      journal = {arXiv e-prints},
         year = 2022,
        month = jun,
          eid = {arXiv:2206.14908},
        pages = {arXiv:2206.14908},
          doi = {10.48550/arXiv.2206.14908},
archivePrefix = {arXiv},
       eprint = {2206.14908},
 primaryClass = {astro-ph.GA},
       adsurl = {https://ui.adsabs.harvard.edu/abs/2022arXiv220614908G}
}

\end{document}